\documentclass[aps,prx,twocolumn,longbibliography,nofootinbib]{revtex4-2}
\usepackage[T1]{fontenc}
\usepackage{amsmath,amssymb,bm,graphicx,xcolor,booktabs,dcolumn}
\usepackage{wasysym}
\usepackage{enumitem}
\usepackage[colorlinks=true,linkcolor=blue!60!black,citecolor=blue!60!black,urlcolor=blue!60!black]{hyperref}
\usepackage{orcidlink}

\newcommand{\kk}{\mathbf{k}}
\newcommand{\QQ}{\mathbf{Q}}
\newcommand{\avg}[1]{\langle #1 \rangle}
\newcommand{\Z}{\mathbb{Z}}

\begin{document}

\title{Charge order before superconductivity in the doped kagome Dirac spin liquid}

\author{Yasir Iqbal\,\orcidlink{https://orcid.org/0000-0002-3387-0120}}
\email{yiqbal@physics.iitm.ac.in}
\affiliation{Department of Physics, Indian Institute of Technology Madras, Chennai 600036, India}

\date{\today}

\begin{abstract}
Doping a quantum spin liquid is expected to produce a superconductor.
We show that for the Dirac spin liquid on the kagome lattice the doped
charge orders first.  The charge is carried by a chargon doublet whose
two gauge components are degenerate on the square and triangular
lattices, and the kagome lattice, as the line graph of the honeycomb
lattice, splits them.  One component disperses and condenses in four
valleys.  The other has flat lowest bands and stays empty over a window
in the bare mass whose width, $(\sqrt6-1)t$, follows from the band
structure alone.  Within that window, in the mean-field chargon theory
constructed from the projective symmetry group of the spin liquid, the
doped state forms charge and loop-current crystals with the
electromagnetic U(1) unbroken and no superconductivity.  Pairing
requires a gapped descendant of the spin liquid as the parent and then
takes the form of a pair-density wave, with no uniform $d+id$ channel at
leading order and none at any order for the crystals at the $M$ point.
The predicted charge order is odd under the twofold rotation and has no
cubic invariant, which distinguishes it from the charge order of the
kagome metals.
\end{abstract}

\maketitle

\tableofcontents

\section{Introduction}
\label{sec:intro}

The fate of a quantum spin liquid once charge is restored to it is a
question nearly as old as the resonating-valence-bond proposal
\cite{AndersonRVB}.  The expectation that grew from that proposal was
that a Mott insulator whose spins form a liquid of singlets becomes a
superconductor once doped \cite{AndersonScience1987,KRS1987}.  This
paper works out what doping does to the U(1) Dirac spin liquid of the
kagome lattice, within a mean-field treatment that holds the parent
fixed, and finds charge order and loop-current order with the
electromagnetic U(1) unbroken, over a window that the band structure
fixes in closed form.  Superconductivity requires a different parent,
one of the gapped descendants of the spin liquid, and it then takes the
form of a pair-density wave.  The lattice geometry decides whether a
superconducting channel is available at all, and the interactions
select among the condensates that remain.

The chargon construction poses this question in a form controlled by
symmetry.  One represents the spin liquid by fermionic spinons with a
projective symmetry group (PSG) and carries the charge with bosonic
chargons \cite{KRS1987} that inherit the spinons' emergent gauge
structure.  The band minima of the chargon, the projective action of
the lattice symmetries on those valleys, and the resulting Landau theory
then dictate which conventional orders emerge upon doping or pressure
\cite{Chatterjee2016,Bonetti2026}.  Because the PSG differs from one
lattice to the next, the answer has to be worked out case by case.  On
the square lattice it yields $d$-wave superconductivity intertwined with
charge stripes \cite{Christos2023}.  On the triangular lattice it yields
a variety of density waves and finite-momentum superconductors
descending from the Dirac spin liquid \cite{Feuerpfeil2026}.  In both
cases the two components of the chargon gauge doublet are degenerate, so
the quartic couplings determine whether a given condensate is a
superconductor or a pure density wave, and both occur in those phase
diagrams \cite{Christos2023,Feuerpfeil2026}.  On kagome the two
components are split, because the kagome lattice is the line graph of
the honeycomb lattice.  The two lowest bands of one gauge component are
pinned flat at $-2t$, while the other component disperses with four
minima at $-(1+\sqrt6)t$.  Only the dispersive component can condense,
and it does so over a window in the bare mass of width $(\sqrt6-1)t$,
a separation of band bottoms and not an interacting gap.  Within this
window the doped Dirac spin liquid orders into charge crystals and
loop-current crystals without superconductivity.

The kagome lattice is well studied from both the insulating and the
metallic side.  On the insulating side, the ground state of the kagome
Heisenberg antiferromagnet remains unsettled among the $[0,\pi]$ Dirac
spin liquid (DSL, Fig.~\ref{fig:ansatz})
\cite{Hastings2000,Ran2007,Hermele2008}, a gapped $\Z_2$ spin liquid
\cite{YanHuseWhite2011,Depenbrock2012,Mei2017} and a thirty-six-site
valence-bond crystal \cite{MarstonZeng,SinghHuse2007}.  We take the DSL
as the parent.  It is the best variational description of the model
\cite{Iqbal2011,Iqbal2013}, and signatures of its Dirac cones have been
seen in density-matrix renormalization spectra \cite{HeZaletel2017},
tensor-network calculations \cite{Liao2017}, and the entanglement
response to flux insertion \cite{ZhuEnt2018}.  The leading gapped $\Z_2$
candidate is, moreover, itself one of its descendants
(Sec.~\ref{sec:z2}).  The central results do not depend on how that
debate is resolved, because the moduli-space theorem of
Sec.~\ref{sec:moduli} (through sextic order) and the entire
particle--hole composite catalog of Table~\ref{tab:ph} are identical for
the DSL and for all four of its $\Z_2$ descendants.  Herbertsmithite and
its analogs are the material realizations
\cite{Norman2016,Han2012,ZnBarlowite2025}.  The $\Z_2$ neighbors of the
DSL have been classified \cite{Lu2011}, and its chiral neighbor is an
established Kalmeyer--Laughlin phase
\cite{KalmeyerLaughlin1987,Gong2014,Bauer2014,Hu2015,Wietek2015,Bieri2016}.
The doped problem, however, has stood open since Ko, Lee, and Wen
proposed an exotic anyon superconductor from the doped kagome liquid
\cite{Ko2009}.  Strong-coupling numerics find insulating crystals of
spinless holons at light doping, with no superconductivity there
\cite{Jiang2017,Peng2021,XuGuYang2024}.  The one realized
electron-doping series, Li intercalation \cite{Kelly2016}, finds an
insulator at every composition studied, and Ga substitution leaves the
kagome plane undoped because its extra charge is anion-compensated
\cite{Puphal2019}.  A systematic symmetry analysis of what doping the
kagome DSL can produce has been missing.  On the metallic side, the
kagome metals $A$V$_3$Sb$_5$ and their relatives display $2\times2$
charge order, disputed time-reversal breaking, and pair-density-wave
superconductivity \cite{RMP2026}.  Their theoretical description has
been built on weak-coupling sublattice interference
\cite{KieselThomale2012,WuPairing2021,WuThomaleRaghu2023} and on Landau
and mean-field theories of the $2\times2$ charge and loop-current
orders
\cite{ParkYeBalents2021,Denner2021,Christensen2021,LinNandkishore2021,Feng2021,Tazai2023,Wagner2023,Fernandes2026},
and a link between the itinerant and strong-coupling pictures has been
asked for explicitly \cite{RMP2026}.

The two sides have so far been treated as separate problems.  On the
insulating side, the idea that competing phases descend from a single
low-energy structure, the algebraic spin liquid as the ``mother of many
competing orders'' \cite{HermeleSenthilFisher2005}, has been made
precise for the kagome lattice.  There the U(1) Dirac spin liquid is the
parent of its neighbors.  The magnetic and valence-bond orders around it
are components of a single set of low-energy operators, the fermion
bilinears together with the monopoles, whose symmetry quantum numbers
follow from the spinon band topology
\cite{Hermele2008,SongNC2019,SongPRX2020}.  The near-degeneracy that
makes kagome numerics so delicate is then a consequence of that
structure and not an accident of the model.  On the doped side, the
algebraic charge liquid \cite{Kaul2008} added charge to an algebraic
spin liquid through fermionic holons coupled to its gauge field.  No
counterpart of the classification exists, however, once charge is
allowed to fluctuate.  The operators that carry charge are absent from
the insulating theory by construction, so the superconducting,
pair-density-wave, charge-ordered and current-carrying phases that
descend from the parent cannot be read off from the monopole spectrum.
Yet these are the phases at issue in the doped spin liquid and in the
kagome metals.  The charge sector has its own multiplet, the projective
valley representation of the chargon, and a classification of its own
that is fixed by the projective symmetry group alone.

Here we construct that counterpart, the PSG-constrained chargon theory
of the kagome DSL, of its four $\Z_2$ descendants (with
translation-invariant spinon pairing), and of its chiral descendant
beyond the sector crossing of Sec.~\ref{sec:csl}, at and away from half
filling.  One chargon gauge component reproduces the holon band of
Ref.~\cite{Ko2009}, with four nondegenerate minima at $-(1+\sqrt6)t$
[Fig.~\ref{fig:bands}(a,b)].  The other carries the doubly degenerate
flat multiplet of that spectrum as its lowest states, pinned at $-2t$,
and the splitting $\Delta_{\rm flat}=(\sqrt6-1)t$ is the window quoted
above.  Because only one component condenses, the physical
electromagnetic U(1) stays unbroken, and the condensed state is a
chargon Higgs phase with no superfluid response.  Whether it is
metallic, semimetallic or insulating is determined by the spinon
spectrum in the condensate-renormalized link field.  At mean field all
three commensurate condensates are metals.  The emergent Gauss law moves
the spinons off half filling by the doped density, so the electron-like
quasiparticles form Fermi pockets of total Luttinger volume $x$ around
the two Dirac nodes.  An insulating gap at the commensurate lock-ins
must therefore come from physics beyond mean field (Sec.~\ref{sec:doped}).
With no superfluid response, the descendants proper, produced by
condensing the chargon, are distinct from the vestigial orders, which
are composites of the multiplet that order where superconductivity is
absent.  The distinction persists to finite temperature, where the
leading scale is the condensate stiffness and not a superconducting one.

For the condensing component, the projective symmetry action leaves one
quadratic and three quartic invariants, all in closed form, and these
reduce the whole quartic competition to a single angle
(Fig.~\ref{fig:wheel}).  The projected on-site repulsion sits on the
boundary between the $M$-star and $K/2$-star condensates.  This
direction, which we call the Hubbard ray, is protected to all orders in
the quartic and gauge couplings by an $O(4)\times U(1)$ symmetry of the
projected on-site quartic, a symmetry larger than the PSG requires.  The
$M$-star phase hosts a moduli space, the pair sphere of competing
charge crystals, which stays degenerate through sextic order and is
lifted at octic order, where the octic term derived for on-site
repulsion selects the uni-$M$ stripe.

Pairing requires the $\Z_2$ channels and is then a pair-density wave on
the phase-I and II$^-$ arcs of Fig.~\ref{fig:wheel}, which contain the
projected on-site and nearest-neighbor directions.  The pairing map of
the four $\Z_2$ descendants admits only one-dimensional uniform
channels, one per descendant, namely extended $s$ ($A_1$), the two $f$
waves ($B_1$, $B_2$), and $A_2$.  A uniform $d+id$ state is excluded at
leading order, because the uniform pair channel of every descendant is
one-dimensional and has no $E_2$ content.  It is excluded at every order
for the $M$-star condensates, whose crystal momenta forbid any uniform
pair composite.  Beyond a sector crossing, the chiral descendant becomes
a two-valley theory with a flat $\mathbb{CP}^1$ manifold at quartic
order.  Only intersite loop terms resolve this manifold, and every
on-site moment through cubic order is constant on it.  Because the
chiral state has no spinon pairing, it orders at quartic order as a
nonsuperconducting charge and current crystal.  The two-valley theory is
conditional, since we do not determine whether the doped state lies
beyond the crossing.  Placing it there requires identifying the chargon
links with the spinon links, and the symmetry-allowed intra-hexagon link
that this identification brings in closes the sector margin
(Sec.~\ref{sec:cslsectors}).

The composite orders realize symmetry channels that are new to the
kagome problem.  In the $M$-point irrep labels used for the kagome
metals, they include $C_2$-odd $F_3$ charge order with no cubic
invariant, $F_2'$ loop currents, a $K$-star loop-current crystal, and
$K/2$-star charge crystals at half the $K$ wavevector.  The DSL's own
operators below scaling dimension two carry only $\Gamma$- and $M$-point
momenta, and every $M$-triplet among them is $C_2$-even.  In the
insulating theory the $C_2$-odd $F_3$ channel occurs only among the
conserved currents, and the $K$ and $K/2$ stars do not occur.  No
chargon channel coincides with a mass or monopole channel, since the
spin-singlet $M$-triplets of the masses and monopoles are all
$\mathcal{T}$-even.  The common channels lie in the DSL's conserved
$M$-point currents, which contain both $F_3$ and $F_2'$
\cite{Hermele2008,SongPRX2020,SongNC2019}, and the mechanisms that
produce them differ between the two sectors.  The classification needs
no enhanced flavor symmetry of the charge sector.  The $O(4)\times U(1)$
of the Hubbard ray belongs to one coupling direction and plays no part
in it.

The charge and loop-current orders of the weak-coupling theory of the
kagome metals and of the present theory of the doped insulator
decompose into the same $F$ irreps of the $M$ star, so the
strong-coupling and weak-coupling descriptions of kagome charge order
can be compared directly.  They differ in the irrep of the charge order,
$F_1$ for the metals against $F_3$ here, and in the presence or absence
of the Landau cubic, while the $F_2'$ loop-current channel is common to
both.  The comparison is one of symmetry classification and does not
interpolate between the two limits.

Section~\ref{sec:experiment} turns these results into testable
consequences for doped herbertsmithite and for the kagome metals.  The
charge order is $C_2$-odd and transforms as the $F_3$ irrep of the $M$
point, so it admits no Landau cubic and no symmetry-mandated first-order
transition, whereas the $F_1$ order of the kagome metals has the opposite
parity, a difference accessible to scanned probes and to resonant x-ray
scattering at the Cu $L_3$ edge.  The superconductor supported by the
$\Z_2$ descendants is a pair-density wave with no uniform component at
the quartic level on the branches selected by on-site and
nearest-neighbor repulsion, and its vestigial phase has a flux period of
$h/4e$ or $h/6e$ depending on which crystal forms.  The loop currents
produce local fields of $3$ to $20$~G at $x=1/36$ and $18$ to $118$~G at
$x=1/6$ at the sites within $3$~\AA\ of the kagome plane, reduced by
factors of $1.4$ and $7.0$ once the local Gauss law is imposed
(Appendix~\ref{app:gausslaw}), which still leaves them one to two orders
of magnitude above the fields that zero-field $\mu$SR reports at the
charge-order transitions of the kagome metals, and polar Kerr and
anomalous Hall responses are forbidden by symmetry for the single-arm
stripe and allowed for the multi-arm crystals.  Charge order and
loop-current order set in together, at a single transition $T_{\rm CO}$
that our estimates place below a separate chargon condensation scale
$T_\rho$, provided the $\Gamma$-point flux orders with the condensate.

The paper proceeds as follows.  Section~\ref{sec:setup} sets up the
parton construction and shows that the chargon band structure splits
into two spectrally inequivalent sectors.  Section~\ref{sec:psg}
constructs the projective valley representation, the Landau--Higgs
functional, the moduli-space theorem, and the all-orders protection of
the Hubbard ray.  Section~\ref{sec:phases} maps the condensed phases,
including the loop-current stripe manifold of phase~I and the $K/2$
crystals of phase~II, and catalogs the composite order parameters.
Section~\ref{sec:z2} treats the four $\Z_2$ descendants and their
pairing channels, Sec.~\ref{sec:csl} treats the chiral descendant and
its two-valley theory, and Sec.~\ref{sec:doped} follows the doped case
from Gross--Pitaevskii to strong coupling.
Section~\ref{sec:experiment} compares the theory with experiment, and
Sec.~\ref{sec:conclusions} concludes.  The appendices collect gauge
conventions and verification methodology, proof details, the two-loop
check of the ray protection, the four projective symmetry groups, the
chiral-ansatz verification, and complete composite tables.  Where we
call a result exact, we mean a spectral identity of the free chargon
problem or a statement of representation theory and invariant algebra,
all checked numerically to the tolerances of
Appendix~\ref{app:conventions}.  Where a result depends on the
projection onto the condensation valleys or on the classical (mean-field
or Gross--Pitaevskii) truncation, we say so.

\section{Chargon theory of the kagome Dirac spin liquid}
\label{sec:setup}

\subsection{Parton construction and conventions}
\label{sec:parton}

We represent the spin-$\tfrac12$ degrees of freedom by fermionic spinons
$f_{{\bm i}\sigma}$, collected into the Nambu doublet
$\psi_{\bm i}=(f_{{\bm i}\uparrow},f^{\dagger}_{{\bm i}\downarrow})^{\!\top}$,
with the quadratic mean-field Hamiltonian
$H=-\sum_{ij}\psi^{\dagger}_{\bm i}u_{ij}\psi_{\bm j}$, where the link fields
$u_{ij}=u_{ji}^{\dagger}$ are $2\times2$ matrices in the SU(2) gauge space
\cite{Wen2002}.  A U(1) spin liquid is an ansatz whose invariant gauge
group is U(1); for the states considered here $u_{ij}$ commutes with
$\tau^{3}$.  A spin-independent spinon hopping
$-\sum_{ij\sigma}t_{ij}f^{\dagger}_{{\bm i}\sigma}f_{{\bm j}\sigma}$
corresponds to
$u_{ij}=\mathrm{diag}(t_{ij},-t_{ij}^{*})
=(\mathrm{Re}\,t_{ij})\,\tau^{3}+i\,(\mathrm{Im}\,t_{ij})\,\tau^{0}$,
with $\tau^{0}$ the unit matrix.  It is proportional to $\tau^{3}$ when
$t_{ij}$ is real, as in the time-reversal-symmetric DSL
[Eq.~\eqref{eq:uij}], and acquires a $\tau^{0}$ component in the chiral
ansatz of Sec.~\ref{sec:csl}.  In either case the residual gauge
transformations are rotations about $\tau^{3}$.

Charge fluctuations, relevant both for the doped system and for the
charge response of the insulator, are introduced through the SU(2) rotor
(chargon) decomposition of the electron
\cite{Chatterjee2016,Christos2023,Feuerpfeil2026,Bonetti2026},
which builds on the SU(2) slave-boson formulation of the $t$--$J$ model
\cite{WenLee1996,LeeNagaosaNgWen1998,LeeNagaosaWen2006},
\begin{equation}
\begin{pmatrix} c_{{\bm i}\uparrow} \\ c^{\dagger}_{{\bm i}\downarrow}\end{pmatrix}
= \mathcal{B}^{\dagger}_{\bm i}\,\psi_{\bm i}\,,
\qquad
\mathcal{S}_B \supset \sum_{ij} B^{\dagger}_{\bm i}\,u_{ij}\,B_{\bm j}\,,
\label{eq:fusion}
\end{equation}
where $\mathcal{B}_{\bm i}$ is an SU(2) matrix built from the spinless,
charge-$e$ chargon doublet
$B_{\bm i}=(B_{{\bm i},+},B_{{\bm i},-})^{\!\top}$,
\begin{equation}
\mathcal{B}_{\bm i}=\begin{pmatrix} B_{{\bm i},+} & -\bar B_{{\bm i},-}\\[2pt]
B_{{\bm i},-} & \phantom{-}\bar B_{{\bm i},+}\end{pmatrix}.
\label{eq:Bmatrix}
\end{equation}
The two components carry
\emph{opposite} emergent gauge charge but \emph{equal} physical electric
charge, and their hopping is governed by the same link field $u_{ij}$ that
defines the spinon ansatz [second relation in Eq.~\eqref{eq:fusion}].
Throughout, we hold $u_{ij}$ fixed at the link field of the undoped
state, that is, we assume that the doped carriers do not rearrange the
flux pattern of the parent ansatz.  Section~\ref{sec:strongcoupling}
estimates the doping range over which this assumption holds.  The
square- and triangular-lattice versions of this construction are given
in Refs.~\cite{Christos2023,Feuerpfeil2026}.

On kagome the sign of the chargon hopping relative to $u_{ij}$ is gauge
invariant, and it is fixed by the sign of the electron hopping.  The
sign amounts to a $\Z_2$ flux per triangle of the chargon hopping.  On
the bipartite square lattice this flux can be removed by a gauge
transformation, and on the triangular lattice it is undone by a
point-group operation (Sec.~\ref{sec:bands}).  With the nearest-neighbor
spinon correlator of the parent state,
$\avg{f^\dagger_{{\bm i}\sigma}f_{{\bm j}\sigma}}=c_1s_{ij}$ with
$c_1=0.221$ (Sec.~\ref{sec:bands}), and $u_{ij}=t\,s_{ij}\tau^3$ the
link field of Eq.~\eqref{eq:uij}, the electron hopping
$-t_c\sum_{\langle ij\rangle\sigma}(c^\dagger_{{\bm i}\sigma}c_{{\bm j}\sigma}+{\rm H.c.})$
decouples on the chargons as
\begin{align}
H_B&=-2c_1t_c\sum_{\langle ij\rangle}s_{ij}\bigl(\bar B_{{\bm i},+}B_{{\bm j},+}
-\bar B_{{\bm i},-}B_{{\bm j},-}\bigr)+{\rm H.c.}\nonumber\\
&=-\frac{2c_1t_c}{t}\sum_{ij}B^\dagger_{\bm i}u_{ij}B_{\bm j}\,,
\label{eq:decoupling}
\end{align}
so Eq.~\eqref{eq:fusion} as written, with $B_-$ the dispersive component
(Sec.~\ref{sec:bands}), is the case $t_c<0$.  The Cu--O--Cu hopping of
herbertsmithite has this sign, which places the flat band at the
bottom of the Cu manifold and the Dirac point at $n=4/3$
\cite{Mazin2014}.  In this decoupling the chargon hopping is
$t_B=2c_1|t_c|\simeq0.44\,|t_c|$.

For $t_c>0$, which is the case in the hole-doped $t$--$J$ numerics of
Sec.~\ref{sec:strongcoupling}, the roles of $B_+$ and $B_-$ are
exchanged.  The dispersive sector is then $B_+$, with
$c_\sigma=\bar B_+f_\sigma$.  The pair channel becomes the second term of
Eq.~\eqref{eq:pairop}, the charge-$2e$ composite becomes $B_-\avg{B_+}$,
and the Gauss law gives $n_f=1-x$ for hole doping.  Every spectral,
symmetry and Landau-theory statement below is the same in the two
labelings.  In both cases the chargon hopping scale of
Eq.~\eqref{eq:fusion} is $t_B=2c_1|t_c|$, not $|t_c|$ itself.  This
estimate does not capture the strong-coupling renormalization of the
holon bandwidth (Sec.~\ref{sec:strongcoupling}).  Throughout, $t$ in the
chargon sector means $t_B$.

The physical electron pair operator follows from Eq.~\eqref{eq:fusion} by
inspection.  We replace the spinon bilinears by their mean-field
amplitudes.  These are the spinon hopping amplitudes
$\chi_{ij}=\avg{f^\dagger_{{\bm i}\uparrow}f_{{\bm j}\uparrow}}$ and
$\chi'_{ij}=\avg{f^\dagger_{{\bm i}\downarrow}f_{{\bm j}\downarrow}}$ in the two spin
sectors, both equal to $c_1s_{ij}$ in the SU(2)-symmetric parent state,
and the singlet spinon pairing $\Delta^{\rm sp}_{ij}$, which vanishes in
the U(1) parent state and is nonzero in its $\Z_2$ descendants
(Sec.~\ref{sec:z2}).  One then finds
\begin{align}
\Delta^{\rm el}_{ij}&=\avg{c_{{\bm i}\uparrow}c_{{\bm j}\downarrow}}
=+\bar B_{{\bm i},+}\bar B_{{\bm j},-}\,\chi_{ji}
+\bar B_{{\bm i},+}\bar B_{{\bm j},+}\,\Delta^{\rm sp}_{ij}\nonumber\\
&\quad-\bar B_{{\bm i},-}\bar B_{{\bm j},-}\,(\Delta^{\rm sp}_{ij})^{*}
+\bar B_{{\bm i},-}\bar B_{{\bm j},+}\,\chi'_{ij}\,.
\label{eq:pairop}
\end{align}
With only $B_-$ condensed (Sec.~\ref{sec:bands}), the symmetry
classification below concerns the third term,
$-\bar B_{{\bm i},-}\bar B_{{\bm j},-}(\Delta^{\rm sp}_{ij})^{*}$.  In
this channel the electron pair amplitude is the condensate pair composite
$\bar B_{{\bm i},-}\bar B_{{\bm j},-}$ dressed by the spinon pair form
factor.  It is gauge invariant because the chargon and spinon factors
carry opposite gauge charge.  Because the spinon pairing of all four $\Z_2$
descendants is invariant under the projective translations (zero-momentum
spinon pairing), the electron pair momenta are those of the chargon pair
composites of Table~\ref{tab:ph}, and the pairing symmetries listed in
Table~\ref{tab:pairing} are the quantum numbers of $\Delta^{\rm el}$
itself.

In the strict rotor limit $\mathcal{B}_{\bm i}$ is unimodular,
$|B_{{\bm i},+}|^2+|B_{{\bm i},-}|^2=1$.  The chargon density
$\rho_r=B_r^\dagger B_r$ on site $r$ is then the same on every site, and an
on-site chargon repulsion $U\sum_r\rho_r^2$ would be a constant.  As in
Refs.~\cite{Chatterjee2016,Christos2023,Feuerpfeil2026} we relax this
constraint in the standard way and treat $B_{\bm i}$ as a soft complex
doublet whose amplitude is set by the condensation energy.  All of the
Landau theory from Sec.~\ref{sec:psg} onward refers to this soft field,
and the unimodular normalization is not used again.

In the fully condensed limit of Sec.~\ref{sec:gp} the soft field
carries the doped charge,
$\sum_rB_r^\dagger B_r=xN$, and the valley amplitudes of the Landau
theory are its projection onto the four condensation valleys,
$\sum_\eta n_\eta=xN$.  The valley Bloch states $\phi_\eta$ of
Eq.~\eqref{eq:expansion} are normalized to unity on the lattice, so that
$\sum_r\rho_r=\sum_\eta n_\eta$ identically.

The emergent Gauss law ties the spinon gauge charge
$\psi^\dagger_{\bm i}\tau^3\psi_{\bm i}=n_{f,{\bm i}}-1$ to that of the
chargons.  With only $B_-$ condensed, which is the case for the DSL and
its $\Z_2$ descendants (Sec.~\ref{sec:bands}), Eq.~\eqref{eq:fusion}
gives
$c_{{\bm i}\uparrow}=\bar B_{{\bm i},-}f^\dagger_{{\bm i}\downarrow}$ and
$c_{{\bm i}\downarrow}=-\bar B_{{\bm i},-}f^\dagger_{{\bm i}\uparrow}$, so
the electron is the particle--hole conjugate of the spinon.  In the
two-valley chiral regime of Sec.~\ref{sec:csl} it is $B_+$ that
condenses, and the mirrored bookkeeping is given there.  The Gauss law reads
$(n_f-1)+n_+-n_-=0$ on every site, with $n_\pm$ the number densities of
the two components of the soft doublet.  In the undoped parent $n_f=1$.
A $B_-$ condensate of density $x$ (Sec.~\ref{sec:gp}) shifts the spinon
filling to $n_f=1+x$ for hole doping and to $1-x$ for electron doping.
For electron doping it is the conjugate quanta that condense.  The
electron-like quasiparticles $f^\dagger$ then have density
$2-n_f=1\mp x$, which is the doped electron density.  Every
spinon-sector statement below (the spinon spectrum of
Sec.~\ref{sec:ansatz} and the pairing descendants of Sec.~\ref{sec:z2})
is made at $n_f=1$, the filling of the undoped parent and of its
descendants.  The doped Higgs phase is treated at $n_f=1\pm x$ in
Sec.~\ref{sec:electronspectrum}.

When the chargons are gapped the insulating spin liquid is stable, and
doping then produces a fractionalized Fermi liquid
\cite{SenthilSachdevVojta2003,Chatterjee2016,Bonetti2026}.  When the
effective chargon mass is tuned through zero, by pressure, doping, or
proximity to a metallic phase, the chargons condense at the minima of
their band structure.  The pattern of the condensate determines which
conventional orders (charge density waves, loop currents,
superconductivity) emerge from the spin liquid.
The projective symmetry group (PSG) of the ansatz therefore fixes,
lattice by lattice, the full set of proximate orders, in the sense of
the parent-state framing of Refs.~\cite{HermeleSenthilFisher2005,Kaul2008}.
In this paper we carry out that program for the kagome lattice.

\subsection{The $[0,\pi]$ ansatz and the chargon Hamiltonian}
\label{sec:ansatz}

\begin{figure}[b]
\includegraphics[width=\linewidth]{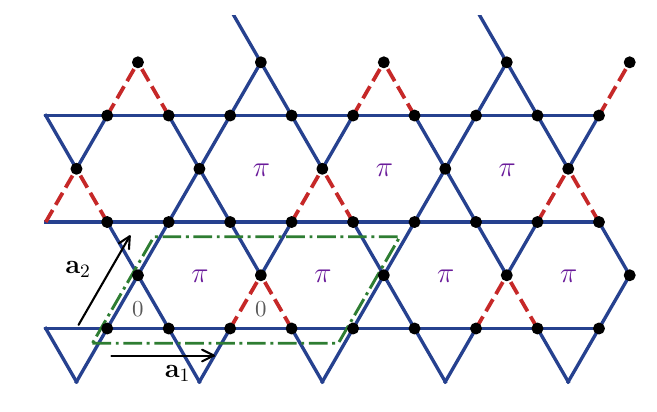}
\caption{Gauge used for the kagome $[0,\pi]$ DSL ansatz,
$u_{ij}=t\,s_{ij}\tau^3$ with $s_{ij}=+1$ on solid and $-1$ on dashed bonds.
The flux is $0$ through every triangle and $\pi$ through every hexagon.  The
$\pi$ flux per crystallographic cell doubles the cell.  The six-site
magnetic unit cell spanned by
$\mathbf{A}_1=2\mathbf{a}_1$, $\mathbf{A}_2=\mathbf{a}_2$ is shown as a
dash-dotted parallelogram.}
\label{fig:ansatz}
\end{figure}

The parent state is the $[0,\pi]$ Dirac spin liquid of Hastings and of Ran
\emph{et al.}\ \cite{Hastings2000,Ran2007,Hermele2008}, the energetically
best variational spin liquid for the nearest-neighbor kagome Heisenberg
antiferromagnet
\cite{Iqbal2011,IqbalZ22011,IqbalVBCNJP2012,Iqbal2013,IqbalGap2014,%
IqbalBreathing2018}.  The bracket lists the gauge flux through the
triangles and through the hexagons, in the convention of
Refs.~\cite{Ran2007,Lu2011}.  The ansatz is real nearest-neighbor hopping
with flux $0$ through every triangle and flux $\pi$ through every
hexagon,
\begin{equation}
u_{ij}=t\,s_{ij}\,\tau^3,\;\; s_{ij}=\pm1,\;\;
{\textstyle\prod_{\triangle}}s_{ij}=+1,\;\;
{\textstyle\prod_{\hexagon}}s_{ij}=-1.
\label{eq:uij}
\end{equation}
The flux through the crystallographic unit cell (two triangles plus one
hexagon) is $\pi$, so any gauge realizing Eq.~\eqref{eq:uij} doubles the unit
cell.  We use the six-site magnetic cell spanned by
$\mathbf{A}_1=2\mathbf{a}_1$, $\mathbf{A}_2=\mathbf{a}_2$ and the
sign pattern shown in Fig.~\ref{fig:ansatz}.  With $M(\kk)$ the $6\times6$ Bloch
matrix of the sign pattern $s_{ij}$, the two chargon components see opposite
overall signs of the hopping,
$H_{\pm}=\pm t\,M(\kk)$, while the spinons at half filling reproduce the
known features of the DSL: two Dirac cones at the Fermi level
$\varepsilon_F=(\sqrt3-1)t$ and a flat spinon pair at $+2t$
\cite{Ran2007,Hermele2008}.

The two sectors share the sign pattern $s_{ij}$, and hence the projective
symmetry group, but not the magnitude of the hopping.  The chargon
hopping is set by the electron hopping $t_c$ (Sec.~\ref{sec:parton}),
whereas the spinon hopping of the Mott insulator is set by the exchange.
Up to a decoupling constant of order one it is of order $c_1J$, with
$J=4t_c^2/U$ and $c_1=0.221$ the nearest-neighbor spinon correlator of
Sec.~\ref{sec:bands}.  This is a few percent of $t_c$ at the
parameters of Sec.~\ref{sec:experiment}.  The symbol $t$ in
Eq.~\eqref{eq:uij} and in the spinon spectrum denotes this spinon scale.
The chargon spectrum of Sec.~\ref{sec:bands} is written in units of the
chargon hopping, and the two scales enter together only in the
condensate-renormalized spinon problem of
Sec.~\ref{sec:electronspectrum}.

All statements below about spectra, symmetry representations, and
invariants have been verified to within numerical roundoff by exact
diagonalization of the single-particle Hamiltonian
on $12\times12$ and $24\times24$ tori with explicitly
constructed $\Z_2$ gauge transformations (Appendix~\ref{app:conventions}).
Equation~\eqref{eq:uij} is a nearest-neighbor truncation.  The one
further hopping that the same PSG admits at second-neighbor range, and
its effect on the chargon spectrum, are given at the end of
Sec.~\ref{sec:bands}.

\subsection{Exact chargon band structure}
\label{sec:bands}

\emph{The $\tau^3$ degeneracy is lifted.}  On the square and triangular
lattices the two chargon gauge components are exactly degenerate, and
the chargon bands come in doublets \cite{Christos2023,Feuerpfeil2026}.
Degeneracy of the doublet requires a transformation that exchanges
$B_+\!\leftrightarrow\!B_-$ while preserving the ansatz.  The natural
candidate, the global SU(2) gauge rotation $W=i\tau^1$, does exchange the
components, but it simultaneously maps $u_{ij}\to-u_{ij}$, shifting the
flux of every \emph{odd}-length loop by $\pi$
[$(\text{triangle},\text{hexagon})=(0,\pi)\to(\pi,\pi)$].  The two signs
of the ansatz thus describe the same spin liquid, since their
Gutzwiller-projected spinon wave functions are identical, as emphasized
in Appendix~A of Ref.~\cite{Hermele2008}.  The splitting discussed here
lives entirely in the charge sector.

The doublet is degenerate if and only if each component's spectrum is
symmetric under $E\to-E$, which holds whenever $-u$ is diagonal-gauge- or
symmetry-equivalent to $u$.  Odd loops are necessary for a splitting but
not sufficient, and the doublet splits if and only if no lattice
symmetry maps the shifted flux pattern back to the original.  On the square
lattice all loops are even, so the sign reversal preserves every flux and
the equivalence holds.  The triangular-lattice DSL has odd loops, but
the sign reversal sends its $(0,\pi)$ flux pattern on (up, down)
triangles to $(\pi,0)$, and a point-group operation that exchanges the
two triangle orientations maps this back.  The equivalence therefore
holds there as well, and the chargon spectrum is exactly particle--hole
symmetric \cite{Feuerpfeil2026}.  We have confirmed on a $12\times12$
torus that the spectrum is symmetric under $E\to-E$ to within numerical
roundoff.

On kagome the two triangle orientations carry \emph{equal} flux.  On the
$[0,\pi]$ ansatz the sign reversal therefore sends the (up, down)
triangle fluxes from $(0,0)$ to $(\pi,\pi)$ while leaving the even
hexagon at $\pi$, so every triangle now carries the gauge-invariant flux
$\pi$ instead of $0$, and no space-group operation can undo the change.
Accordingly $\mathrm{spec}\,[M(\kk)]$ is not particle--hole symmetric
(see below), and the chargon doublet splits into two spectrally
inequivalent sectors.  Much of the physics below follows from this
splitting.

\begin{figure*}
\includegraphics[width=\textwidth]{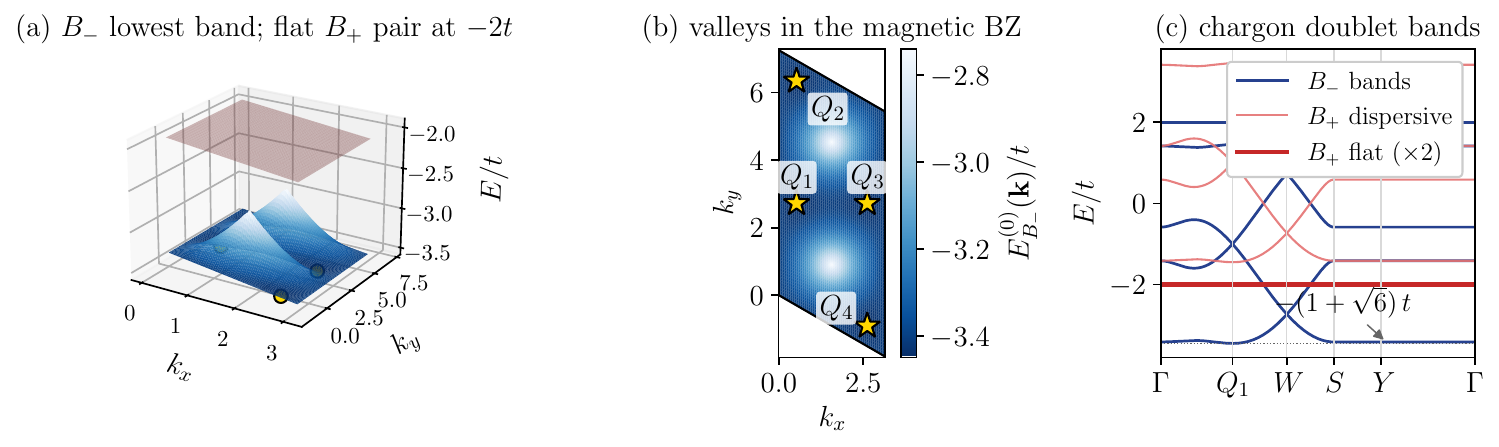}
\caption{Chargon band structure of the kagome $[0,\pi]$ DSL.
(a)~Lowest band of the dispersive $B_-$ sector over the magnetic Brillouin
zone, with the four condensation valleys marked by stars.  The translucent
plane marks the exactly flat, twofold $B_+$ multiplet at $-2t$.
(b)~The four valleys $\QQ_{1\dots4}$ of Eq.~\eqref{eq:valleys}.
(c)~All twelve chargon bands along $\Gamma\to Q_1\to W\to S\to Y\to\Gamma$,
with $\Gamma=(0,0)$, $Q_1=(\tfrac16,\tfrac5{12})$ the valley of
Eq.~\eqref{eq:valleys}, $W=(\tfrac12,\tfrac14)$ the maximum of the lowest
$B_-$ band, $S=(\tfrac12,\tfrac12)$ and $Y=(0,\tfrac12)$ in
fractional coordinates of the magnetic reciprocal vectors
$(\mathbf{G}_1,\mathbf{G}_2)$.
Blue lines show the $B_-$ sector and red lines the $B_+$ sector, with the
flat pair at $-2t$ drawn thick.  The highest pair of the $B_-$ sector is
the same flat multiplet, at $+2t$.  The global bottom, $-(1+\sqrt6)t$, is
reached at $Q_1$, and the flat $B_+$ multiplet lies $(\sqrt6-1)t$ above
it.  The lowest $B_-$ band rises from $-(1+\sqrt6)t$ at $Q_1$ to its
maximum $-(1+\sqrt3)t$ at $W$, so the path spans the full width
$(\sqrt6-\sqrt3)t$ of the lowest band.  Along $S\to Y\to\Gamma$ that band
is exactly flat at $-(2+\sqrt2)t$.}
\label{fig:bands}
\end{figure*}

\emph{The dispersive sector.}  Kagome with fluxes $(0,\pi)$ is the line
graph \cite{Mielke1991,Mielke1992,Bergman2008} of the $\pi$-flux honeycomb
model, with $M=\tilde{B}^{\dagger}\tilde{B}-2$ and $\tilde{B}$ the
$4\times6$ incidence matrix.  Consequently
$\mathrm{spec}\,[M(\kk)]=\{-2,-2\}\cup\{1+\epsilon_a(\kk)\}$, where
$\epsilon_a(\kk)$ ($a=1,\dots,4$) are the $\pi$-flux honeycomb bands and
$\sum_a\epsilon_a^2(\kk)=12$ identically.  The $B_-$ sector,
$\mathrm{spec}\,[-tM]$, thus attains its minimum
$E_{\min}=-(1+\sqrt6)\,t$ where the honeycomb spectrum reaches
$\epsilon=\sqrt6$.  This happens at four symmetry-related momenta
[Eq.~\eqref{eq:valleys} and Fig.~\ref{fig:bands}(a,b)],
\begin{equation}
\QQ_{1}=\QQ=\Big(\tfrac{\pi}{6},\tfrac{\sqrt3\pi}{2}\Big),\;\;
\QQ_2=\QQ+\tfrac{\mathbf{G}_2}{2},\;\;
\QQ_{3,4}=-\QQ_{1,2},
\label{eq:valleys}
\end{equation}
i.e., $(\tfrac16,\tfrac5{12})$, $(\tfrac16,\tfrac{11}{12})$,
$(\tfrac56,\tfrac7{12})$, $(\tfrac56,\tfrac1{12})$ in fractional coordinates
of $(\mathbf{G}_1,\mathbf{G}_2)$.

Each valley is nondegenerate, with an exactly isotropic curvature
$\partial^2E/\partial k^2=t/(2\sqrt6)$, and the next band at the
valley lies $\sqrt6\,t$ higher.  The lowest band is narrow, with full
width $(\sqrt6-\sqrt3)\,t\simeq0.72\,t$ and maxima at
$-(1+\sqrt3)t$.  The condensing fields are \emph{four}
nondegenerate complex bosons $B_{1,\dots,4}$, half the field content of
the triangular case \cite{Feuerpfeil2026}, where every valley is doubly
degenerate.  The dispersive sector coincides with the holon band of Ko,
Lee, and Wen \cite{Ko2009}, with the same four minima
(Sec.~\ref{sec:strongcoupling}).

\emph{The flat sector.}  The $B_+$ sector, $\mathrm{spec}\,[+tM]$, carries the
line-graph zero modes as its two \emph{lowest} bands, exactly flat at $-2t$
and separated by a gap $(3-\sqrt6)\,t\simeq0.55\,t$ from its dispersive bands.
The flat multiplet sits $\Delta_{\rm flat}=(\sqrt6-1)\,t\simeq1.45\,t$ above
the global band bottom.  At the Higgs transition the dispersive $B_-$
sector condenses.  The macroscopically degenerate flat multiplet remains
as the lowest gauge-charged excitation branch, with \emph{opposite}
gauge charge.  Neither the square nor the triangular lattice has a
counterpart of this branch.

Because $B_+$ and $B_-$ carry opposite emergent gauge charge and equal
physical charge, the composite $B_+\avg{B_-}$ is gauge neutral and
carries physical charge $2e$.  It has $Q_{\rm EM}-Q_g=2$ under the diagonal charge that stays
unbroken in the $B_-$ condensate (Sec.~\ref{sec:gp}).  Here we normalize
the emergent gauge charge $Q_g$ so that the condensed component $B_-$
carries $+1$ and, by Eq.~\eqref{eq:fusion}, the spinon carries $-1$.
Thus $Q_g$ is minus the $\tau^3$ generator of Sec.~\ref{sec:parton}.  In
the same particle convention, $Q_{\rm EM}(B_\pm)=+1$ for the quanta
created by $\bar B_\pm$, which are holes.  Once $B_-$ condenses, the flat
$B_+$ multiplet is a dispersionless branch of charge-$2e$ excitations
lying $\Delta_{\rm flat}=(\sqrt6-1)t$ above the condensate at the
transition.  The density modulation of the condensed crystals narrows
this gap, as quantified below.

At mean field the pair operator is present, but superconductivity is
suppressed because its band is pushed up and flat.  The line-graph zero
modes pin the $B_+$ bottom at $-2t$ and so fix in closed form the window
$(\sqrt6-1)t$ over which charge order occurs without superconductivity
at mean field.

In the absence of spinon pairing the physical pair operator requires both
gauge components, so condensing one alone gives no superconductivity, as
established in the SU(2) slave-boson theory
\cite{WenLee1996,LeeNagaosaNgWen1998,LeeNagaosaWen2006,Chatterjee2016}.
The degenerate chargon doublets of the square and triangular lattices
superconduct by condensing both components together
\cite{Christos2023,Feuerpfeil2026}.  On kagome the two components are
split by $\Delta_{\rm flat}$, and at mean field $\avg{B_+}=0$ is
therefore enforced over a finite window.  Whether that window survives
beyond mean field is formulated in Sec.~\ref{sec:gp}.

Within that window $B_+$ is gapped and, at mean field, has no source.
No term of the local potential allowed by gauge and charge conservation
is linear in it.  The electron hopping does contain one term that is
linear in $B_+$ once $B_-$ condenses,
$-t_c\sum_{\langle ij\rangle}\big[B_{{\bm i},+}\bar B_{{\bm j},-}
+B_{{\bm j},+}\bar B_{{\bm i},-}\big]\Delta^\dagger_{ij}+{\rm H.c.}$ with
$\Delta^\dagger_{ij}=f^\dagger_{{\bm i}\uparrow}f^\dagger_{{\bm j}\downarrow}
-f^\dagger_{{\bm i}\downarrow}f^\dagger_{{\bm j}\uparrow}$.  This term
follows from Eq.~\eqref{eq:fusion} and is gauge and charge neutral, but
it sources $B_+$ only through the spinon pair amplitude
$\avg{\Delta_{ij}}$.  That amplitude vanishes in the U(1) state and is
the origin of the $\Z_2$ channel of Sec.~\ref{sec:z2spectra}.  The
second-order effect of the term in the U(1) state is discussed in
Sec.~\ref{sec:gp}.  The $\bar B_+\bar B_-$ terms, which are the only
terms of Eq.~\eqref{eq:pairop} that survive in the U(1) state, where
$\Delta^{\rm sp}=0$, acquire no expectation value.

In gauge-invariant terms, under the diagonal charge $Q_{\rm EM}-Q_g$
introduced above, $B_-$ is neutral and $B_+$ carries $2$, so the $B_-$
condensate leaves the physical U(1) symmetry unbroken.  Superconductivity
requires a condensate of diagonal charge $2$, and in the U(1) state the
only such object in Eq.~\eqref{eq:pairop} is the gauge-neutral
composite
$\avg{B_+B_-}$, which vanishes as long as $B_+$ stays gapped.  Chargon
condensation produces density-wave order \emph{without}
superconductivity (Sec.~\ref{sec:phases}).  Superconductivity switches on
only through the $\Z_2$ pairing perturbations discussed in
Sec.~\ref{sec:z2}, which activate the
$\bar B_-\bar B_-(\Delta^{\rm sp})^{*}$ channel of Eq.~\eqref{eq:pairop}.
The diagonal charge $2$ is then carried by the spinon pair.

The selection rule is exhaustive for the local potential.  Since $B_+$
and $B_-$ carry the same electric charge and opposite gauge charge, an
on-site monomial $B_+^{a}B_+^{\dagger b}B_-^{c}B_-^{\dagger d}$ requires
$a+c=b+d$ (electric) and $a-c=b-d$ (gauge), i.e.\ $a=b$ and $c=d$.  Every
allowed potential term is then a polynomial in
$\rho_\pm=B_\pm^\dagger B_\pm$ alone, and no $B_+^\dagger B_-$ of any
power survives.  The leading coupling between the two components is
therefore the quartic $2U\rho_+\rho_-$ contained in $U(\rho_++\rho_-)^2$,
and its coefficient is positive.  At mean field a $B_-$
condensate of density $\bar\rho$ thus raises the $B_+$ mass by
$2U\bar\rho$ instead of lowering it.

Explicit lattice minimization at the Hartree--Gross--Pitaevskii level
confirms this (Sec.~\ref{sec:gp}).  For the lowest state found in the
scan of Sec.~\ref{sec:gp}, the stability gap of the unconstrained
minimizer against infinitesimal $B_+$ admixture falls from the bare value
$(\sqrt6-1)t\simeq1.449\,t$ to about $1.14\,t$ at the strongest coupling
($Ux=16/3$ in the normalization used there, $Ux=8/3$ in that of
$U\sum_r\rho_r^2$).  The $K/2$ crystal, a higher-lying local minimum at
that coupling, gives $1.154\,t$.  The decrease comes entirely from the
density modulation.  For a uniform condensate density the two sectors
shift rigidly by the same Hartree energy, and the gap stays at
$(\sqrt6-1)t$.  The gap also depends on the state.  For a condensate
constrained to the phase-I stripe it is $0.30\,t$ at the same coupling,
which is still positive (Sec.~\ref{sec:gp}).

Equation~\eqref{eq:uij} keeps only nearest-neighbor hopping, but the same
PSG admits exactly one further real hopping at second-neighbor range,
$\chi_2\,\nu_{ij}\tau^3$, with $\nu_{ij}=\pm1$ the unique symmetric
pattern of Appendix~\ref{app:z2psg}.  In closed form
$\nu_{ij}=-s_{ik}s_{kj}$, where $k$ is the common nearest neighbor of $i$
and $j$.  Every mixed triangle (two nearest-neighbor bonds and one
second-neighbor chord) then carries flux $\pi$ for $\chi_2>0$ and $0$ for
$\chi_2<0$, so this sign is gauge invariant.  The
term preserves time reversal and every triangle and hexagon flux, and it
leaves the spinon Dirac nodes intact, so it describes the same U(1)
$[0,\pi]$ spin liquid.

Although the term does not commute with $M(\kk)$, its effect on the
chargon band bottoms can be computed exactly.  The valley Bloch states
are simultaneous eigenvectors of the nearest- and second-neighbor form
factors, with eigenvalues $1+\sqrt6$ and $-2$, so
$\min E_-=-(1+\sqrt6)\,t+2\chi_2$, with the four valleys pinned at
$\QQ_{1\dots4}$ for $\chi_2<0.443\,t$.  For $\chi_2>0$ the bottom of the
$B_+$ sector is a flat-band eigenvector at $W=(\tfrac12,\tfrac14)$, with
$\min E_+=-2t-2\chi_2$.  Hence
$\Delta_{\rm flat}=(\sqrt6-1)\,t-4\chi_2$ for
$0\le\chi_2\le0.443\,t$, and the two sectors cross at
$\chi_2=(\sqrt6-1)\,t/4=0.362\,t$.  To leading order the flat pair
acquires a bandwidth $(12+4\sqrt2)\,|\chi_2|/7=2.52\,|\chi_2|$, which is
$0.50\,t$ at $\chi_2=0.2\,t$, and it merges with the dispersive $B_+$
bands above $\chi_2\simeq0.22\,t$.

For $\chi_2<0$ the gap widens, though more slowly,
$\Delta_{\rm flat}=(\sqrt6-1)\,t+1.48\,|\chi_2|+O(\chi_2^2)$ ($1.744\,t$ at
$\chi_2=-0.2\,t$).  For
$\chi_2<-(\sqrt6-\sqrt3)\,t/(3+\sqrt3)=-0.152\,t$ the half-filled spinon
spectrum acquires compensated pockets [the band minimum at $\QQ$,
$(\sqrt6-1)\,t+2\chi_2$, falls below the node energy
$(\sqrt3-1)\,t-(1+\sqrt3)\chi_2$], and the parent is no longer a Dirac
spin liquid.  The gap-opening branch can add at most $0.22\,t$
to $\Delta_{\rm flat}$.

Symmetry does not fix the sign of $\chi_2$.  In the mean-field
decoupling of Sec.~\ref{sec:parton}, Eq.~\eqref{eq:decoupling}, the
chargon second-neighbor amplitude is $2c_2t_2\,\nu_{ij}$.  It is
generated by the electron's second-neighbor hopping $t_2$ through the
second-neighbor correlator of the Dirac spin liquid itself,
$\avg{f^\dagger_if_j}=c_2\nu_{ij}$, with $c_2=0.080$ compared with
$c_1=0.221$ at nearest-neighbor range.  This gives
$\chi_2/t_B=(c_2/c_1)(t_2/t_c)=0.36\,t_2/t_c$, which can have either
sign.  It vanishes for the nearest-neighbor Hubbard model ($t_2=0$), and
for $t_2^2/t_c^2=J_2/J_1=0.1$ its magnitude is $0.11\,t_B$, far from the
crossing at $0.362\,t_B$.

If instead the chargons are taken to hop with the full spinon link field,
as Eq.~\eqref{eq:fusion} postulates, $\chi_2$ is inherited from the
spinon ansatz and is nonzero only if that ansatz carries a
second-neighbor amplitude.  For the nearest-neighbor Heisenberg model,
for which the $[0,\pi]$ state is the optimized ansatz, the
self-consistent $\chi_2$ vanishes identically at mean field, because
there is no second-neighbor decoupling channel.  For a $J_1$--$J_2$
model the same correlator sets the sign of this spinon-sector amplitude
at mean field.  An antiferromagnetic $J_2$ generates
$\chi_2=0.36\,(J_2/J_1)\,t$ at first iteration and $\chi_2=0.038\,t$
($0.058\,t$) self-consistently at $J_2/J_1=0.1$ ($0.15$).  These values
have the sign that narrows the gap, to $\Delta_{\rm flat}=1.30\,t$
($1.22\,t$).
The crossing would require $J_2/J_1\sim0.8$.  That is far outside the
range in which the Dirac spin liquid is the variational ground state of
the $J_1$--$J_2$ model, since a $q=0$ ordered state overtakes the spin
liquids for $J_2/J_1\gtrsim0.3$
\cite{IqbalJ1J22015,PhysRevB.104.144406}.  All statements below therefore
hold within the nearest-neighbor ansatz and, quantitatively, for
$\chi_2<(\sqrt6-1)\,t/4$.

\section{Projective symmetry action and low-energy theory}
\label{sec:psg}

\subsection{Valley representation}
\label{sec:valleys}

Near the condensation transition the chargon field is dominated by the four
band minima of Eq.~\eqref{eq:valleys},
\begin{equation}
B_{{\bm i},-}\;=\;\sum_{\eta=1}^{4} B_\eta\,\phi_\eta({\bm i})\,,
\label{eq:expansion}
\end{equation}
with $\phi_\eta$ the Bloch states at $\QQ_\eta$, normalized to
$\sum_{\bm i}|\phi_\eta({\bm i})|^2=1$, and $B_{1\dots4}$ four slowly
varying complex fields.  Because the chargons hop in the $[0,\pi]$ gauge, the
physical symmetries act on $B_\eta$ \emph{projectively}.  The chargon
sees the same emergent flux as the spinon, because the electron
$c\sim\mathcal{B}^\dagger\psi$ must transform linearly \cite{Bonetti2026}.
Table~\ref{tab:psg} summarizes the resulting representation.  We
constructed it from the explicit $\Z_2$ gauge transformations and
verified it as a set of operator identities, to within numerical
roundoff, on $12\times12$ and $24\times24$ tori
(Appendix~\ref{app:conventions}).  Translations act as
$T_1\!:B_1\leftrightarrow e^{-i\pi/6}B_2,\ B_3\leftrightarrow e^{+i\pi/6}B_4$
and $T_2=\mathrm{diag}\,(e^{-i5\pi/6},e^{+i\pi/6},e^{+i5\pi/6},e^{-i\pi/6})$
($=e^{-i\QQ_\eta\cdot\mathbf{a}_2}$), so that
\begin{equation}
T_1T_2=-T_2T_1\,,
\label{eq:magalg}
\end{equation}
which is the magnetic algebra of the $\pi$ flux per crystallographic
cell.  The sixfold rotation and the mirror mix the two translation
pairs with $1/\sqrt2$ weights,
\begin{equation}
C_6:\
\begin{pmatrix}B_1\\ B_2\\ B_3\\ B_4\end{pmatrix}
\mapsto\frac{1}{\sqrt2}
\begin{pmatrix}
e^{3i\pi/4}\,(B_3+B_4)\\
e^{i\pi/4}\,(B_3-B_4)\\
e^{-3i\pi/4}\,(B_1+B_2)\\
e^{-i\pi/4}\,(B_1-B_2)
\end{pmatrix},
\label{eq:C6act}
\end{equation}
\begin{equation}
\sigma:\
\begin{pmatrix}B_1\\ B_2\\ B_3\\ B_4\end{pmatrix}
\mapsto\frac{1}{\sqrt2}
\begin{pmatrix}
i\,(B_3+iB_4)\\
i\,(iB_3+B_4)\\
-i\,(B_1-iB_2)\\
-(B_1+iB_2)
\end{pmatrix},
\label{eq:sigact}
\end{equation}
and time reversal acts antiunitarily as $\Theta\!:B_\eta\mapsto-B_{\eta+2}$
(indices mod 4).  The projective algebra closes as
\begin{gather}
C_6^{\,6}=-1,\quad C_6^{\,3}=-C_2,\quad \sigma^2=+1,\quad(\sigma C_6)^2=+1,
\nonumber\\
C_6T_1C_6^{-1}=-T_2,\qquad C_6T_2C_6^{-1}=T_2T_1^{-1},
\nonumber\\
\sigma T_1\sigma^{-1}=T_1^{-1},\qquad \sigma T_2\sigma^{-1}=-T_2T_1^{-1},
\nonumber\\
\Theta^2=+1,\qquad [\Theta,\,g]=0\ \ \forall g.
\label{eq:psgalg}
\end{gather}
\begin{table}[t]
\caption{Projective symmetry action on the valley fields $B_{1\dots4}$.
The action is exact and has been verified as operator identities on
$12\times12$ and $24\times24$ tori.  Entries refer to the fixed gauge of
Appendix~\ref{app:conventions}.  Individual phases depend on the gauge,
but the composite quantum numbers do not.  The signs in
Eq.~\eqref{eq:psgalg} refer to the generators as printed.  The
spanning-tree construction of Appendix~\ref{app:conventions} leaves the
sign of each generator free, and multiplying a single generator by $-1$
changes individual signs (e.g.\ $T_2\to-T_2$ turns
$C_6T_1C_6^{-1}=-T_2$ into $+T_2$).  It changes neither the class of the
projective algebra nor any composite quantum number.}
\label{tab:psg}
\begin{ruledtabular}
\begin{tabular}{ll}
$T_1$ & $B_1 \leftrightarrow e^{-i\pi/6}B_2$, \quad $B_3 \leftrightarrow e^{+i\pi/6}B_4$ \\
$T_2$ & $\mathrm{diag}\,(e^{-i5\pi/6},e^{+i\pi/6},e^{+i5\pi/6},e^{-i\pi/6})$ \\
$C_6$ & $(1,2)\!\leftrightarrow\!(3,4)$ blocks, $1/\sqrt2$ mixing [Eq.~\eqref{eq:C6act}] \\
$\sigma$ & $(1,2)\!\leftrightarrow\!(3,4)$ blocks, $1/\sqrt2$ mixing [Eq.~\eqref{eq:sigact}] \\
$\Theta$ (antiunitary) & $B_\eta \to -B_{\eta+2}$ \ (indices mod 4) \\
\end{tabular}
\end{ruledtabular}
\end{table}

This action involves four nondegenerate valley fields, half the eight of
its triangular-lattice counterpart, Table~II of
Ref.~\cite{Feuerpfeil2026}, because the $\tau^3$ degeneracy is lifted
(Sec.~\ref{sec:bands}).

Projective symmetry groups of bosonic fields on the kagome lattice, and
the orders produced by their condensation, were worked out earlier for
Schwinger-boson spinons \cite{Sachdev1992,WangVishwanath2006} and for
visons \cite{HuhPunkSachdev2011}.  Table~\ref{tab:psg} is the chargon
counterpart.  The translation sector of this representation was
anticipated by Ko, Lee, and Wen \cite{Ko2009}, who obtained the
anticommuting action and the phase set $\{\pm\pi/6,\pm5\pi/6\}$ from a
Chern--Simons analysis of holon quantum numbers [their Eq.~(30)].  To our
knowledge, the projective point-group and time-reversal actions in
Eqs.~\eqref{eq:C6act}--\eqref{eq:psgalg} have not been presented before.
The valley \emph{positions} and the individual phases in
Table~\ref{tab:psg} depend on the gauge \cite{Ran2007}.  All physical
statements below are made through the gauge-invariant
composites of Table~\ref{tab:ph}, whose momenta include the projective
offsets.

\subsection{Landau--Higgs functional}
\label{sec:landau}

The projective action of Table~\ref{tab:psg} strongly constrains the
potential for the valley fields.  With
$n_\eta=|B_\eta|^2$, we find \emph{exactly one} quadratic invariant,
$\sum_\eta n_\eta$, so the projective action enforces equal valley
masses.  There are \emph{exactly three} quartic invariants, and in a
closed basis the potential reads
\begin{equation}
V = r\sum_\eta n_\eta
+ u_0\Big(\sum_\eta n_\eta\Big)^{2}
+ u_P\,\mathcal{P}
+ u_X\,\mathcal{X}\,,
\label{eq:potential}
\end{equation}
\begin{align}
\mathcal{P}&=(n_1+n_2)^2+(n_3+n_4)^2\,,
\label{eq:Pinv}\\
\mathcal{X}&=(n_1-n_2)(n_3-n_4)+4\,\mathrm{Re}\big(\bar B_1\bar B_3B_2B_4\big)\,.
\label{eq:Xinv}
\end{align}
Three couplings is a small number for four condensing fields.  The
unconstrained space of Hermitian quartics is 100-dimensional, and the
triangular-lattice analog of this expansion has eight independent
couplings \cite{Feuerpfeil2026}.  Part of the difference is the field
count, since the nondegenerate valleys give four fields instead of
eight.  Within the four-field space, the magnetic translation algebra
[Eq.~\eqref{eq:magalg}], which acts on the valleys through
twelfth-root-of-unity phases and an intra-pair valley swap, reduces the
100 Hermitian quartics to ten.  These are six density--density
terms, the two inter-pair pair-hoppings
$\mathrm{Re}(\bar B_1\bar B_3B_2B_4)$ and
$\mathrm{Re}(\bar B_1\bar B_4B_2B_3)$, and the two intra-pair
pair-hoppings $\mathrm{Re}[(\bar B_1B_2)^2]$ and
$\mathrm{Re}[(\bar B_3B_4)^2]$.  Translations cannot forbid the last two,
because $\bar B_1B_2$ carries the $M$-point momentum
$\QQ_2-\QQ_1=\mathbf{G}_2/2$ [Eq.~\eqref{eq:valleys}] and its square a
reciprocal-lattice vector.

The point-group action of
Eqs.~\eqref{eq:C6act}--\eqref{eq:sigact}, with its $1/\sqrt2$ mixing,
reduces these ten to three.  The threefold
rotation $C_3=C_6^{\,2}$, which preserves each pair, removes every
intra-pair anisotropy and ties the cross-pair terms into
$(n_1+n_2)(n_3+n_4)$ and the single combination $\mathcal{X}$, leaving
four invariants.  On a pair sphere the translation-invariant quartics
still span the two $\ell=2$ harmonics $n_z^2$ and $n_x^2-n_y^2$ of the
Bloch vector of Sec.~\ref{sec:moduli}, the latter being
$4\,\mathrm{Re}[(\bar B_1B_2)^2]/(n_1+n_2)^2$.  The pair-exchanging
elements ($C_6$, $C_2$ or $\sigma$) then identify $(n_1+n_2)^2$ with
$(n_3+n_4)^2$ in $\mathcal{P}$, leaving three.

This count comes from an explicit computation.  The
unitary part of the projective symmetry group has 288 elements as
$4\times4$ matrices.  The scalar $-\mathbb 1$ acts trivially on every
composite, so modulo the overall sign there are 144.  The full
288-element group of Appendix~\ref{app:theorem}, which contains the
unitary and antiunitary elements together, is counted modulo the same
sign.  The Reynolds operator, the average of the induced action over
these 144 elements, has trace $1$ on the space of Hermitian quadratics
and trace $3$, with rank $3$, on the 100-dimensional space of Hermitian
quartics.  The same average over the translation subgroup alone (24 of
the 288 matrices) has trace $10$, over
$\langle T_1,T_2,C_3\rangle$ (72 matrices) trace $4$, and over
$\langle T_1,T_2,C_6\rangle$ (144 matrices, the index-two subgroup
without the mirror) already trace $3$.  Time reversal is not needed for
this count, because the three unitary invariants are already
$\Theta$-even [item (iii) below].

In terms of these invariants the quartic potential has the following
properties.
\begin{enumerate}[label=(\roman*),leftmargin=*,itemsep=0pt,topsep=2pt,parsep=0pt,partopsep=0pt]
\item \emph{Intra-pair lock.}  Within each translation pair the
densities enter only through $(n_1+n_2)^2+(n_3+n_4)^2$, and no symmetric
quartic can distinguish points on a pair sphere $\{(B_1,B_2,0,0)\}$.
This is a consequence of the point group, since translations alone
would leave the anisotropies $n_1n_2$ and $\mathrm{Re}[(\bar B_1B_2)^2]$
(Sec.~\ref{sec:moduli}).
\item \emph{Umklapp lock.}  The inter-pair Umklapp term is not
separately invariant.  Symmetry ties it to the product of pair
polarizations in the single combination $\mathcal{X}$, and neither
$(n_1{-}n_2)(n_3{-}n_4)$ nor the Umklapp term alone is invariant.  We
have verified both statements to within numerical roundoff.
\item \emph{No chiral quartic.}  All three invariants are even under
$\Theta$, so time reversal can be broken only spontaneously, by the
condensate itself (Sec.~\ref{sec:phases}).
\end{enumerate}

Since $u_0$ multiplies a constant on the condensate sphere, the
mean-field phase diagram depends on a single angle $\theta$ in the plane
of the two nontrivial couplings [Fig.~\ref{fig:wheel}].  We use an
orthonormalized frame $(J_1,J_2)$ made of two fixed linear combinations
of $(J_0,\mathcal{P},\mathcal{X})$ that are orthogonal to $J_0$.  Here
$J_0=(\sum_\eta n_\eta)^2$, and the frame is given in
Appendix~\ref{app:theorem}, Eq.~\eqref{eq:dictionary}.
Phase~I [$\theta\in(-\arctan\sqrt5,+\arctan\sqrt5)$, i.e.\
$|\theta|<0.3661\pi$] condenses a single translation pair, with
$(\mathcal{P},\mathcal{X})=(1,0)$.  The complementary arc is phase~II, in
which the condensate spreads over both translation pairs.  It consists of
two Umklapp branches, II$^\mp$ with
$(\mathcal{P},\mathcal{X})=(\tfrac12,\mp\tfrac14)$, separated by a
boundary at $\theta=\pi$.  On each branch the quartic minimum is
the entire level set of $(\mathcal{P},\mathcal{X})$, which forms a
continuous manifold of degenerate states (Sec.~\ref{sec:phasediagram}).

The quartic is positive definite on the sphere, so that the condensate
amplitude is finite at quartic order, if and only if
$u_0>-\min\big(u_P,\ \tfrac12u_P-\tfrac14u_X,\ \tfrac12u_P+\tfrac14u_X\big)$.
The three arguments are the values of $u_P\mathcal{P}+u_X\mathcal{X}$
at the phase-I, II$^-$ and II$^+$ vertices of the closed triangle
$\tfrac12\le\mathcal{P}\le1$, $|\mathcal{X}|\le(1-\mathcal{P})/2$ that the
two invariants sweep on the unit sphere.  At equality the quartic is
still bounded below but flat along a vertex direction, and stability is
then set by the sextic.  Both microscopic representatives located below
satisfy the condition with a finite margin,
$u_0+\min=\tfrac43\,U/N$ at the projected on-site point and
$\tfrac73\,V/N$ at the projected nearest-neighbor point.  For the
mixture the margin is $(\tfrac43U+\tfrac73V)/N>0$.

We now locate microscopic interactions in this plane.  The projected
on-site Hubbard repulsion sits \emph{exactly on the I/II$^-$ phase
boundary}, at $\theta_U=\arctan\sqrt5$ in closed form
[Eq.~\eqref{eq:hubbardquartic}].  Nearest-neighbor repulsion, projected
in the same way onto the valley Bloch states, gives
\begin{equation}
V\!\sum_{\langle rr'\rangle}\!\rho_r\rho_{r'}=\frac{V}{N}\Big[
\tfrac{9+2\sqrt6}{4}\Big(\sum_\eta n_\eta\Big)^{2}
-\tfrac{6\sqrt6-1}{12}\,\mathcal{P}
+\tfrac{5+2\sqrt6}{6}\,\mathcal{X}\Big],
\label{eq:vquartic}
\end{equation}
i.e.\ $u_X/u_P=-2(5+2\sqrt6)/(6\sqrt6-1)=-1.4454$ and
$\theta_V=0.32362\pi$, which lies inside phase~I.  For the mixture
$U\sum_r\rho_r^2+V\sum_{\langle rr'\rangle}\rho_r\rho_{r'}$ with $U,V>0$
the angle is $\tan\theta=\tfrac{\sqrt5}{2}\,u_X/(-u_P)$ in the frame of
Eq.~\eqref{eq:dictionary}.  It decreases monotonically from
$\theta_U=\arctan\sqrt5$ at $V=0$ to $\theta_V$ as $V/U\to\infty$, so any
$V>0$ places the quartic theory strictly inside phase~I, whereas the
on-site direction alone selects neither phase.  The choice
between the $M$-star and $K/2$-star orders is made beyond quartic order
(Sec.~\ref{sec:phasediagram}).

The gauge-invariant composites and their quantum numbers are collected in
Table~\ref{tab:ph}.  Particle--hole bilinears realize momenta
$\Gamma\,(\times2)$, the full $M$ star, the full $K/2$ star, and $\pm K$;
symmetric pairs realize $\Gamma\,(\times1)$, the $M$ star, and the $K/2$
star.

In the extended-point-group language of the kagome metals
\cite{Venderbos2016,RMP2026}, the $\mathcal{T}$-even charge/bond
$M$-triplet transforms as $F_3$ ($C_2$-odd) and the $\mathcal{T}$-odd
current $M$-triplet as $F_2'$, the loop-current representation.
Throughout, a prime on an irrep label denotes the $\mathcal{T}$-odd copy
of the corresponding $C_{6v}'''$ irrep, with the spatial transformation
of the unprimed label.  The cubic invariant $\Delta_1\Delta_2\Delta_3$,
which exists only for $F_1$ \cite{RMP2026} and drives the first-order
trihexagonal/Star-of-David selection in the kagome metals, is
therefore forbidden here, and with it the symmetry-mandated first-order
mechanism of that CDW.  In fact \emph{no} invariant cubic exists
anywhere in the $(F_3\oplus F_2')$ $M$-space.
The $2\times2$ charge order of the chargon mechanism is therefore
$C_2$-odd at leading order in the condensate, because the site density
is bilinear in $B$ and its $M$-point content is pure $F_3$ for every
condensate.  The $C_2$-even $F_1$ CDW of the kagome metals has the
opposite parity.  Beyond leading order the
triple-$M$ state, in which all three arms of the triplet are nonzero,
admits a $C_2$-even $F_1$ admixture through the composite
$\Delta_i\Delta_j$ (Sec.~\ref{sec:catalog}).  Its relative amplitude is
of order $Ux/t$, set by the higher-band admixture of
Sec.~\ref{sec:phasediagram}.  The uni-$M$ stripe has a single arm and
hence no cross term, so it is $C_2$-odd at every order.

Whether the transition is continuous is determined by fluctuations of
the valley fields and of the emergent $U(1)$ gauge field.  This
fixed-point question is beyond the present work, but
Appendix~\ref{app:betas} records what the two-loop beta functions
printed there imply for it.  On the Hubbard ray, the direction of the
projected on-site repulsion (Sec.~\ref{sec:hubbardray}), the scalar
sector of the continuum theory of the transition is, within the scope
stated there, the gauged $O(4)\times U(1)$ model.

The channels found above are new to the doped problem.  The
order-parameter catalog of the \emph{undoped} DSL below scaling
dimension two consists of all sixteen fermion bilinears and all six
monopoles.  It carries only $\Gamma$- and $M$-point momenta, and its
spin-singlet $M$-triplets are all $C_2$-even and $\mathcal{T}$-even
($F_1$ and $F_2$), while its spin-triplet members are $\mathcal{T}$-odd
by construction \cite{Hermele2008,SongPRX2020,SongNC2019}.  The $F_3$
triplets, the $K$ star, and the entire $K/2$ star lie in
symmetry channels that no slow mode of the undoped spin liquid
occupies.

The $K$-star particle--hole composite, the Hermitian combination of the
$\pm K$ bilinears, is $\mathcal{T}$-odd (Table~\ref{tab:ph}).  Its
site-density form factor and its gauge-invariant bond charge out to
third-neighbor range vanish identically at $K$ for every condensate, so
in the present construction the $K$ star is a $\sqrt3\times\sqrt3$
loop-current crystal and carries no local charge-density or
bond-charge order.  On the triangular lattice a $K$-point charge order requires the
$\Z_2$ descendant \cite{Feuerpfeil2026}, whereas on kagome the $K$ star
appears already in the U(1) state, as a loop-current channel.

\subsection{The moduli-space theorem}
\label{sec:moduli}

At quartic order the Landau functional of the four valley fields is
exactly constant on each translation-pair sphere
$\{(B_1,B_2,0,0)\}\simeq\mathbb{CP}^1$, and it remains constant at
sextic order.  The first symmetry-allowed anisotropy appears at eighth
order.  As we now show, this follows from the projective symmetry
action alone.  The square-lattice precedent is the pair of
$\pi$-flux boson fields whose residual quartic degeneracy is lifted only
at eighth order \cite{Lannert2001,SachdevPark2002,Balents2005}.  The
kagome case differs in the size of the in-pair group
(Appendix~\ref{app:theorem}).

Restricting attention to the pair $(B_1,B_2)$, write $B=(B_1,B_2)^{\!\top}$ and
introduce the Bloch vector $\hat n=B^\dagger\bm\sigma B/B^\dagger B$.  Of the
288 elements of the projective symmetry group, 144 (72 unitary and 72
antiunitary) preserve the pair subspace $\{(B_1,B_2,0,0)\}$, and they induce
48 distinct $O(3)$ elements on the Bloch sphere.  These are closed under
multiplication and comprise 24 proper and 24 improper elements,
including the inversion (Appendix~\ref{app:theorem}).  The proper part
is generated by two orthogonal $\pi$ rotations ($T_2$ about $\hat z$,
$T_1$ about $\hat x$), a $2\pi/3$ rotation about a body diagonal of the
frame they define ($C_3$), and a $\pi$ rotation about a $(110)$-type
axis of that frame induced by $\sigma C_6$, which also preserves the
pair.  Its multiset of rotation angles (in
degrees, with multiplicities) $\{0\!:\!1,\ 90\!:\!6,\ 120\!:\!8,\
180\!:\!9\}$ identifies it as the octahedral rotation group $\mathrm{O}$
(order 24).  The tetrahedral group $\mathrm{T}$ generated by $T_{1,2}$
and $C_3$ alone is an index-two subgroup.  The elements of
$\mathrm{O}$ that $\mathrm{T}$ lacks are the $90^\circ$ rotations about
the coordinate axes and the $\pi$ rotations about $(110)$-type axes, and
all of them involve the mirror.  All 24 proper elements come
from the unitary elements of the projective symmetry group alone.  In
addition, the antiunitary element $\mathcal{A}=\Theta C_2$ maps the pair to
itself, and a direct computation shows that its induced action on the Bloch
sphere is the \emph{inversion} $\hat n\mapsto-\hat n$.  The full in-pair
symmetry action on $\mathbb{CP}^1$ is the full octahedral group
$\mathrm{O}_h=\mathrm{O}\times\{1,-1\}$ (order 48).

A gauge-invariant potential term of order $2n$ in $B$ restricts, at fixed
total density, to a polynomial of degree $n$ in $\hat n$, i.e.\ to spherical
harmonics with $\ell\le n$.  The octahedral character sums
\begin{equation}
n_\ell^{\rm inv}=\frac{1}{24}\Big[(2\ell{+}1)+6\,\chi_\ell(\pi/2)
+8\,\chi_\ell(2\pi/3)+9\,\chi_\ell(\pi)\Big],
\label{eq:charsum}
\end{equation}
with $\chi_\ell(\theta)=\sin\big[(\ell+\tfrac12)\theta\big]/\sin(\theta/2)$,
give $n_\ell^{\rm inv}=(1,0,0,0,1)$ for $\ell=0,\dots,4$, so the harmonics
with $\ell=1$, $2$ and $3$ are removed by the unitary rotations alone.
Under the tetrahedral subgroup $\mathrm{T}$ the $\ell=3$ harmonic $xyz$
would survive, with $n_3=1$, and would then be removed by the inversion
induced by $\mathcal{A}$.  In $\mathrm{O}$, however, it is already
removed by a $90^\circ$ rotation about a coordinate axis, so time
reversal is not needed for the sextic-order degeneracy.  The inversion
independently removes all odd $\ell$, and the first surviving invariant is the
$\ell=4$ cubic harmonic.

Hence \emph{any} symmetric quartic ($\ell\le2$) or sextic ($\ell\le3$)
potential is constant on the pair sphere, and the moduli space of
degenerate density-wave states, interpolating between the
unidirectional-$M$ stripe and the triple-$M$ crystal, is lifted only at
octic order.  The on-site octic moment of the two orbits differs by
exactly $1/57=1.75\%$, Eq.~\eqref{eq:octicmoment}, and the octic that
on-site repulsion generates through the non-condensing states selects
the stripe orbit (Sec.~\ref{sec:phases}).  Through sextic order the
unitary subgroup $\mathrm{O}$ alone suffices, so the argument applies
verbatim to all four $\Z_2$ descendants.  The on-site moments make the
degeneracy explicit.  With $\sum_r\rho_r=N$, the moments
$N^{-1}\sum_r\rho_r^2=\tfrac43$ and $N^{-1}\sum_r\rho_r^3=2$ are
constant on the pair sphere, while the octic moment is the single
$\ell=4$ cubic harmonic
\begin{equation}
\frac1N\sum_r\rho_r^4=\frac{13}{4}-\frac{1}{12}\big(n_x^4+n_y^4+n_z^4\big)
\label{eq:octicmoment}
\end{equation}
in the Bloch frame of the pair.  It equals $19/6$ on the uni-$M$ stripe
orbit (its minimum, at the six coordinate-axis points) and $29/9$ on the
triple-$M$ orbit (its maximum, at the eight body diagonals), so the two
orbits are split by $(29/9-19/6)/(19/6)=1/57$.  Every
symmetry-allowed sextic invariant is likewise constant on the sphere to
within numerical roundoff.

On the square lattice, Christos \emph{et al.}\ \cite{Christos2023}
relate $d$-wave superconductivity and period-2 charge stripes through a
possible emergent $\mathrm{SO}(5)$ symmetry of the Higgs sector at a
proposed deconfined critical point, and they argue that the terms
selecting among these orders are likely irrelevant there.  The
degeneracy is then a property of the critical theory, and its
resolution rests on couplings that are irrelevant at the critical point.

The degeneracy of the competing charge crystals found here has a
different, representation-theoretic origin.  Any function on the
pair sphere that is invariant under the in-pair group has no harmonic
content at $\ell=1$, $2$ or $3$, at tree level and at every loop order,
for every term that respects the projective symmetry group.  The first
symmetry-allowed angular dependence is the $\ell=4$ cubic harmonic,
whose coefficient is dynamical and may receive tree-level, loop or
nonanalytic contributions.  The microscopically generated potential
does contain this anisotropy, the octic term of Sec.~\ref{sec:phaseI}.

At quartic order the degeneracy is stronger still.  For arbitrary
couplings $(u_0,u_P,u_X)$ the spectrum of the full Hessian of
Eq.~\eqref{eq:potential} is constant along the pair sphere.  Write
$\rho=B^\dagger B$ for the density of the condensed pair.  The block
$\partial^2V/\partial B_a\partial\bar B_b$ ($a,b=3,4$) in the
uncondensed pair is $r+2u_0\rho+u_X\rho\,\hat n\cdot\bm\sigma$, with
eigenvalues $r+(2u_0\pm u_X)\rho$.  At the minimum on the sphere, where
$r=-2(u_0+u_P)\rho$, these are $(-2u_P\pm u_X)\rho$, both positive
if and only if $|u_X|<-2u_P$, which is the phase-I condition.  In
phase~I the pair sphere is, at quartic order, a manifold of
degenerate minima with identical energies and identical fluctuation
spectra.

The theorem describes the interior of phase~I, where a single
translation pair condenses.  It is protected by the projective
$\mathrm{O}_h$ action on the pair sphere (no emergent continuous
symmetry is involved), and its octic lifting is parametrically small.  It
also suggests a mechanism for the empirical softness of kagome charge
orders, seen in the uni- versus tri-directional CDW competition and its
extreme strain sensitivity \cite{RMP2026}.  We return to this point in
Sec.~\ref{sec:experiment}.

In the chiral descendant the counting changes qualitatively.  Time
reversal and the mirror survive only in the combination $\sigma\Theta$,
so the mirror-derived elements that enlarge $\mathrm{T}$ to $\mathrm{O}$
are absent and the unitary in-pair action is the tetrahedral group
$\mathrm{T}$ (Sec.~\ref{sec:csllandau}).  The antiunitary in-pair
element ($\sigma\Theta$) \emph{preserves} the harmonic $xyz$, and the
two-valley moduli space is lifted already at sextic order, by
intersite loop terms alone (Sec.~\ref{sec:csl}).

\subsection{All-orders protection of the Hubbard ray}
\label{sec:hubbardray}

The projected on-site repulsion sits exactly on the I/II$^-$ boundary,
at $\theta_U=\arctan\sqrt5$ [Eq.~\eqref{eq:hubbardquartic}].  This
bare-coupling coincidence survives fluctuations, including those of the
emergent gauge field, to all orders in the quartic and gauge couplings,
because it reflects an enhanced symmetry of the on-site quartic.  With
$\rho=\sum_\eta n_\eta$,
\begin{equation}
\begin{aligned}
\mathcal{P}-2\mathcal{X}&=\rho^2-4|Q|^2,&\quad Q&=B_1B_3+B_2B_4,\\
\mathcal{P}+2\mathcal{X}&=\rho^2-4|\tilde Q|^2,&\quad
\tilde Q&=\bar B_1B_4-\bar B_2B_3,
\end{aligned}
\label{eq:rayforms}
\end{equation}
so the Hubbard ray $u_X=-2u_P$ is built from the uniform pair amplitude
of Appendix~\ref{app:composites} (there normalized as $Q/\sqrt2$), which
carries gauge charge two.  Its mirror image under $u_X\to-u_X$ is the
ray $u_X=+2u_P$, the I/II$^+$ boundary, which we call the antipodal ray
below.  It is built from a gauge-neutral particle--hole amplitude.  In
the basis
\begin{equation}
\begin{aligned}
C_1&=\tfrac{1}{\sqrt2}(B_1+B_3),&\quad C_2&=-\tfrac{i}{\sqrt2}(B_1-B_3),\\
C_3&=\tfrac{1}{\sqrt2}(B_2+B_4),&\quad C_4&=-\tfrac{i}{\sqrt2}(B_2-B_4),
\end{aligned}
\label{eq:Cbasis}
\end{equation}
one has $\rho=C^\dagger C$ and $Q=\tfrac12\sum_aC_a^2$, so that
on the Hubbard ray the quartic reads
\begin{equation}
V=(u_0+u_P)\,(C^\dagger C)^2-u_P\,|C^{\!\top}C|^2,
\label{eq:o4quartic}
\end{equation}
which is the quartic of the $O(4)\times U(1)$ abelian Higgs model,
invariant under $C\to e^{i\alpha}OC$ with $O$ real orthogonal.  The
factor $-i$ on the antisymmetric combinations is what makes
$C^{\!\top}C$ the $O(4)$ invariant.  Every unitary element of the
projective symmetry group preserves $Q$ up to a phase.  The unitary part
of the group (144 elements modulo the overall sign) lies
inside the enhanced group $O(4)_Q\times U(1)$.

The renormalization-group flow preserves every symmetry shared by the
action and the regularization, and a $U(4)$-symmetric regularization
exists for this compact internal symmetry.  A plane of quartic couplings
is therefore invariant under the flow, to all orders, whenever the
quartics invariant under its stabilizer span only that plane.

Once the gauge field is included, the relevant symmetries must commute
with the photon, which couples to all four valleys alike through
$B^\dagger B$, so they lie in $U(4)$.  The kinetic term is $U(4)$
symmetric because the four valleys are degenerate with the same
isotropic curvature $t/(2\sqrt6)$ (Sec.~\ref{sec:bands}).  The Dirac
spinons of the parent state, which enter the gauged scalar sector only
through the photon, carry no chargon valley index and are singlets under
$O(4)_Q$.  Their direct couplings to the valley fields break $O(4)_Q$
and are among the terms listed at the end of this subsection.

The stabilizer of the Hubbard quartic inside $U(4)$ has Lie algebra
$\mathfrak{so}(4)\oplus\mathfrak u(1)$, of dimension seven.  Within
the three-dimensional space of PSG-invariant quartics spanned by
$\{\rho^2,\mathcal{P},\mathcal{X}\}$, the quartics invariant under it
span exactly the two-dimensional plane of $\rho^2$ and $|Q|^2$.  The
ray $u_X=-2u_P$ is therefore invariant under the gauged
renormalization group, to all orders in the quartic and gauge
couplings.  At two loops, with all gauge terms included,
$\beta_X+2\beta_P=0$ holds identically on the ray, where $\beta_P$ and
$\beta_X$ are the beta functions of $u_P$ and $u_X$
(Appendix~\ref{app:betas}).  The antipodal ray is invariant only in the
ungauged theory.

The third mean-field boundary, the II$^-$/II$^+$ boundary $u_X=0$ of
Sec.~\ref{sec:landau}, is likewise protected, gauged or not.  The
stabilizer of $\mathcal{P}$ inside $U(4)$ has Lie algebra
$\mathfrak u(2)\oplus\mathfrak u(2)$, of dimension eight.  Its invariant
quartics in the same space are $\rho^2$ and $\mathcal{P}$ alone, so
$\beta_X$ is proportional to $u_X$ to all orders, as
Eq.~\eqref{eq:betauX} shows at two loops.

In the ungauged theory the map
$g:(B_1,B_2,B_3,B_4)\mapsto(\bar B_1,-\bar B_2,B_4,B_3)$ is an element
of $SO(8)$ that fixes $\rho^2$ and $\mathcal{P}$ and reverses
$\mathcal{X}$.  It exchanges the two rays and makes the ungauged flow
covariant under $u_X\to-u_X$, with $\beta_{u_0}$ and $\beta_{u_P}$ even
and $\beta_{u_X}$ odd in $u_X$.  At one loop, where no gauge term
distinguishes the rays, each is preserved, with
$\beta_X=-2\beta_P$ on $u_X=-2u_P$ and $\beta_X=+2\beta_P$ on
$u_X=+2u_P$.  However, $g$ reverses the gauge charge of $B_1,B_2$
relative to $B_3,B_4$.  It acts as a partial charge conjugation and
therefore lies outside $U(4)$.

The $U(4)$ stabilizer of the antipodal quartic is the five-dimensional
$\mathfrak{su}(2)\oplus\mathfrak u(1)^2$.  Its invariant quartics in the
same space are all three of $\rho^2$, $\mathcal{P}$ and $\mathcal{X}$,
so once the theory is gauged the antipodal ray is not protected.  At two
loops $\beta_X-2\beta_P=-384\,f\,u_P^2$ on $u_X=+2u_P$.  The asymmetry
between the two rays has a single source, the term $-48fu_X^2$
of $\beta_{u_X}$, which is the only monomial in the whole system that
violates the $u_X$-parity of $g$.  It contributes $-192fu_P^2$ on both
rays and is canceled on the Hubbard ray alone by the covariant remainder
(Appendix~\ref{app:betas}).

The enhanced symmetry accounts for the degeneracy noted after
Eq.~\eqref{eq:hubbardquartic}.  In the basis of Eq.~\eqref{eq:Cbasis}
the minimum manifold $Q=0$ of the Hubbard quartic on the unit sphere is
the set $C=(x+iy)/\sqrt2$ with $(x,y)$ an orthonormal pair of vectors in
$\mathbb{R}^4$, which is a single orbit of $O(4)$.  The single-valley
states, the whole pair sphere, and the phase-locked four-valley states
are degenerate because they lie on this one orbit of the enhanced
symmetry.

In the gauged scalar sector the Hubbard ray is an invariant submanifold
of the renormalization-group flow for every coupling that preserves
$O(4)_Q\times U(1)$, which is the class the flow generates once it
starts on the ray.  We refer to this statement as the ray theorem.  A
flow that starts on the Hubbard ray, with higher couplings that respect
the enhanced symmetry, stays on it, so the location of the on-site
quartic on the phase boundary is structural.  The coincidence
$\theta_U=\arctan\sqrt5$ is a property of the projected on-site quartic,
which has more symmetry than the projective symmetry group requires, and
not an artifact of the $(J_1,J_2)$ frame of Eq.~\eqref{eq:dictionary}.
The theorem extends beyond quartic order, because the
$O(4)_Q\times U(1)$-invariant gauge-neutral sextics again span only two
dimensions, $\rho^3$ and $\rho|Q|^2$.

A local sextic of the form $u_6\sum_r\rho_r^3$, whose moment ranks the
II$^-$ level set in the same order as the derived sextic
(Sec.~\ref{sec:phaseII}), is not of this class.  With $\sum_r\rho_r=1$
its moment $N^2\sum_r\rho_r^3$ equals $2$ on every phase-I state and
spans $[2-1/(4\sqrt3),\,2+1/(4\sqrt3)]$ on the II$^-$ level set,
although all of these states lie on the single $O(4)_Q$ orbit $Q=0$.
It therefore breaks the enhanced symmetry explicitly, at one power of
the condensate density above the quartic, and this breaking lets it
select among the minima that the quartic leaves degenerate.

The theorem does not cover the complete lattice critical theory.
Couplings that the projective symmetry group allows but that break
$O(4)_Q$ fall outside it.  Any $V>0$ moves the quartic angle off the
ray, by up to $\theta_U-\theta_V=0.0425\pi$, which is $5.8\%$ of the
$0.7323\pi$ width of the phase-I arc, and inside phase~I the enhanced
symmetry is explicitly broken.  Further examples are the sextic just
discussed, a
mixed chargon--spinon term, a monopole--chargon dressing of the kind
discussed in Sec.~\ref{sec:conclusions}, and higher-derivative terms.
The $\mathcal{T}$-odd couplings of the $M$-sector composites are of this
class, as are two couplings generated by the electron hopping itself.
The first is the linear coupling of the chargon $M$-bilinears to the
parent's conserved $M$-point currents (Sec.~\ref{sec:catalog}).  In the
relativistic theory its tree-level scaling dimension is $1+2=3=d$, so it
is marginal at tree level.  Its one-loop fate is not computed here.  The
second is the mixed chargon--spinon quartic
$\bar B_\eta B_{\eta'}\,\psi^\dagger\Gamma\psi$, which the
condensate-induced spinon hopping of Sec.~\ref{sec:electronspectrum}
generates at second order in $t_c$.  It is gauge invariant and marginal
in $d=3$.

Within these limits, the scalar sector of the continuum theory of the
transition on the Hubbard ray is the gauged $O(4)\times U(1)$ model, not
a generic four-boson theory, and at the order the theorem controls the
Dirac spinons enter only through the photon.
In the basis of Eq.~\eqref{eq:Cbasis} with $C=(x+iy)/\sqrt2$,
$x,y\in\mathbb{R}^4$, the ray quartic is
$\tfrac14u_0(x^2+y^2)^2+u_P\,[x^2y^2-(x\cdot y)^2]$.  The ungauged
theory on the Hubbard ray is the $N=4$ member of the
$O(N)\times O(2)$ Landau--Ginzburg--Wilson model of noncollinear magnets
\cite{Kawamura1988,PelissettoRossiVicari2001,Calabrese2003,Delamotte2004},
in its chiral sector, since $u_P<0$ for the projected on-site repulsion
[Eq.~\eqref{eq:hubbardquartic}].  Its fate in $d=3$ is unsettled.
Six-loop resummations find a stable fixed point for $N<5.7(3)$
\cite{Calabrese2003}, whereas the nonperturbative renormalization group
finds none below $N_c\simeq5.1$ \cite{Delamotte2004}.  The gauged
theory, with four charged bosons and $N_f=4$ Dirac fermions coupled to a
single $U(1)$ gauge field, is of the type studied at large $N$ by Kaul
and Sachdev \cite{KaulSachdev2008}.

\section{Phases of the condensed chargon}
\label{sec:phases}

\subsection{Mean-field phase diagram}
\label{sec:phasediagram}

\begin{figure}[t]
\includegraphics[width=0.94\columnwidth]{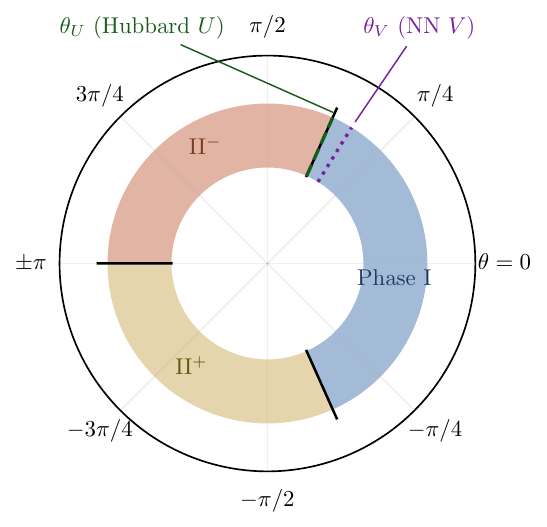}
\caption{Mean-field phase diagram of the quartic Landau theory,
$V_\theta=\cos\theta\,J_1+\sin\theta\,J_2$, on the condensate sphere
(where $J_0=(\sum_\eta n_\eta)^2$ is constant), in the canonically fixed
frame of Appendix~\ref{app:theorem}.  In phase~I
[$|\theta|<\arctan\sqrt5=0.3661\pi$] the condensate occupies a $T_1$
pair, and its minima form the exactly degenerate moduli sphere of
$M$-point charge orders.  The complementary arc splits at $\theta=\pi$
into two Umklapp branches II$^\mp$
[$(\mathcal{P},\mathcal{X})=(\tfrac12,\mp\tfrac14)$].  Each branch is a
continuous manifold of $K/2$ crystals that contains both four-valley and
cross-pair two-valley representatives.  The projected on-site Hubbard
functional sits \emph{exactly} on the I/II$^-$ boundary
($\theta_U=\arctan\sqrt5$), and nearest-neighbor repulsion lies inside
phase~I ($\theta_V=0.3236\pi$).}
\label{fig:wheel}
\end{figure}

Minimizing Eq.~\eqref{eq:potential} on the condensate sphere yields the
one-angle phase diagram of Fig.~\ref{fig:wheel}.

In \emph{phase~I} the condensate occupies a single translation pair,
$(B_1,B_2,0,0)$ or $(B_3,B_4,0,0)$, and the intra-pair lock makes the
entire pair sphere degenerate (Sec.~\ref{sec:moduli}).  In
\emph{phase~II} the condensate spreads over both translation pairs.  On
each Umklapp branch the quartic minimum is the full continuous level set
of $(\mathcal{P},\mathcal{X})=(\tfrac12,\mp\tfrac14)$.  This level set
interpolates between four-valley states, whose relative phases are
locked by the Umklapp part of $\mathcal{X}$ [Eq.~\eqref{eq:Xinv}], and
\emph{cross-pair} two-valley states of the type $(B_1,0,0,B_4)$, on
which the Umklapp term vanishes identically.

Projecting the microscopic interactions onto the condensation valleys
locates them on this diagram.  Throughout, we take $U\sum_r\rho_r^2$
with $\rho_r=|B_r|^2$ as a representative on-site chargon repulsion
acting on the soft condensate field of Sec.~\ref{sec:parton}.  Its
coefficient $U$ is not derived from the microscopic electron
interaction, and the projection below fixes only its direction on the
phase wheel.  Its natural scale is not small.  If the soft-field quartic
is identified with the charging energy of the rotor, which is of the
order of the Hubbard $U$ of the parent insulator, then $U/t$ is of order
thirty and $Ux/t$ is of order one to ten over the doping range of
Sec.~\ref{sec:experiment}.  Every selection below that rests on an
expansion in $Ux/t$ (the sextic and octic selections of
Sec.~\ref{sec:phases} and the crossover of Eq.~\eqref{eq:Vcrossover}) is
therefore a weak-coupling mean-field statement.  It is controlled for
$Ux/t\ll1$ and extrapolated beyond it, and Sec.~\ref{sec:gp} gives the
controlled range.  At the physical charging energy, both the choice
between the $M$ and $K/2$ stars and the choice between the stripe and
triple-$M$ orbits remain open.

Inserting Eq.~\eqref{eq:expansion}, $B_r=\sum_\eta B_\eta\varphi_\eta(r)$,
with $\varphi_\eta$ the exact valley Bloch states, into the on-site
repulsion $U\sum_r\rho_r^2$ gives a quartic form in the valley
amplitudes.  The form is PSG invariant, so it is a combination of the
three invariants of Eq.~\eqref{eq:potential}, and the valley overlap sums
$\sum_r\bar\varphi_\eta\varphi_{\eta'}\bar\varphi_\kappa\varphi_{\kappa'}$
evaluate exactly to
\begin{equation}
U\sum_r\rho_r^2=\frac{U}{N}\Big[2\Big(\sum_\eta n_\eta\Big)^{2}
-\tfrac23\,(\mathcal{P}-2\mathcal{X})\Big],
\label{eq:hubbardquartic}
\end{equation}
i.e.\ $u_X=-2u_P$ with $u_P<0$, which is the I/II$^-$ boundary direction
$\theta_U=\arctan\sqrt5$ of Appendix~\ref{app:theorem}.  Since
$\mathcal{P}-2\mathcal{X}=(\sum_\eta n_\eta)^2-4|B_1B_3+B_2B_4|^2$, the
Hubbard functional is minimized on the quadric $B_1B_3+B_2B_4=0$.  This
quadric contains the single-valley states, the $T_1$-pair sphere, and
the phase-locked four-valley states, and the functional takes the same
value $F_U/U=4/(3N)$ on all of it (on the unit sphere).  A phase-unlocked
equal-modulus four-valley state is in general not degenerate with these
states.  As the relative phases vary, $\mathcal{P}-2\mathcal{X}$ sweeps
the full interval $[0,1]$ and $F_U/U$ ranges over $[4/(3N),\,2/N]$.  The
random-phase value $\mathcal{P}-2\mathcal{X}=\tfrac12$ gives
$F_U/U=5/(3N)$, and the maximum $2/N$ is reached at the II$^+$ vertex.
Nearest-neighbor repulsion, projected in the same way, gives
Eq.~\eqref{eq:vquartic}, with $\theta_V=0.3236\pi$ inside phase~I.  The
combined direction $\theta(V/U)$ decreases monotonically from
$\theta_U=\arctan\sqrt5$ at $V=0$ to $\theta_V$ as $V/U\to\infty$
(Sec.~\ref{sec:landau}).  Any $V>0$ places the projected
quartic theory strictly inside phase~I, because the projected on-site
term contributes no quartic energy difference between the two stars.
Once the two quartics named in Sec.~\ref{sec:hubbardray} are included,
the boundary moves in $V$ by an amount that depends on the charging
energy through $u=U/|t_c|$.  The spinon-polarization quartic alone gives
$V^*\simeq0.19\,u\,t_c^2/t_B$.  Above $u=0.563$ the two-boson ladder
overturns it, drives $V^*$ negative, and so selects phase~I already at
$V=0$.

Nearest-neighbor repulsion favors phase~I by
$\mathcal{V}_{\rm I}-\mathcal{V}_{\rm II^-}=\tfrac{3-2\sqrt6}{12}\,\tfrac{V}{N}=-0.158\,V/N$
per unit $(\sum_\eta n_\eta)^2$, a value we verified directly on the
lattice by comparing the stripe with the $K/2$ crystal of
Eq.~\eqref{eq:phase2min}.

\begin{figure*}[t]
\includegraphics[width=0.92\textwidth]{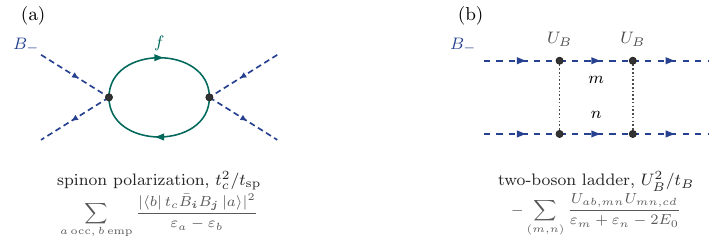}
\caption{The two quartics missed by the projection onto the four valley
Bloch states.  Both are generated at second order, and both lie off the
Hubbard ray.  Solid teal lines are spinons, dashed indigo lines the
condensing chargon $B_-$, filled circles the second-order insertions,
and the dotted rungs the chargon contact repulsion $U_B$.
(a)~Spinon-polarization quartic.  The condensate dresses the spinon link
field and the filled Dirac sea responds at second order, so the energy
denominators are spinon particle--hole energies and the scale is
$t_c^2/t_{\rm sp}$.  (b)~Two-boson ladder, the second-order part of the
chargon two-body $T$ matrix.  Its intermediate pair $(m,n)$ is excluded
from the fourfold band bottom, so the denominators are chargon
two-particle energies and the scale is $U_B^2/t_B$.  The different
topologies of the two diagrams give them different scales.  Both are
evaluated in Appendix~\ref{app:offray}.}
\label{fig:quartics}
\end{figure*}

The pure projected on-site direction lies on the I/II$^-$ boundary,
where the projected quartic gives the $M$-star orders of phase~I and the
$K/2$-star crystals of phase~II$^-$ the same energy, so the choice
between them is made beyond the projected quartic.  Two further
quartics, absent from the projection, enter one power of $x$ below the
sextic (Fig.~\ref{fig:quartics} and Appendix~\ref{app:offray}):
\begin{enumerate}[label=(\roman*),leftmargin=*,itemsep=0pt,topsep=2pt,parsep=0pt,partopsep=0pt]
\item the spinon-polarization quartic generated at second order by the
condensate-induced spinon hopping, with
$N(u_0,u_P,u_X)=(-0.1374,\,+0.0635,\,+0.1096)\,t_c^2/t_{\rm sp}$.  It
lies far off both rays ($\theta=0.652\pi$, inside phase~II$^-$), is
constant on the pair sphere and on the II$^-$ level set, and
contributes
$E_{\rm I}-E_{\rm II^-}=+0.059\,x^2t_c^2/t_{\rm sp}$ per site.
\item the off-ray part of the two-boson (ladder) correction to the
projected on-site quartic.  Its coefficients grow logarithmically with
system size, but only along the Hubbard ray, so the off-ray part
converges.  It gives $E_{\rm I}-E_{\rm II^-}=-0.061$, $-0.054$ and
$-0.053\,U^2x^2/t_B$ per site at $L=12$, $24$ and $36$, which favors
phase~I.
\end{enumerate}

The two terms carry different energy denominators (spinon particle--hole
energies in the first, chargon two-particle energies in the second) and
also different interactions.  The spinon scale descends from the
electronic Hubbard parameter through $t_{\rm sp}=c_1J$ with
$J=4t_c^2/U_H$.  The ladder is built from the soft-chargon coupling
$U_B$ of Eq.~\eqref{eq:gpfunctional}, which, as noted at the start of
this section, is not derived from $U_H$.  If we identify the two,
$U_B=U_H\equiv U$, as the charging-energy estimate there does, the
comparison reduces to one ratio.  With $t_B=2c_1|t_c|$ we get
$t_{\rm sp}/t_B=2|t_c|/U$, so the single parameter $u=U/|t_c|$ controls
both terms.  In the common units $x^2t_c^2/t_B$ the two contributions
are then $+0.0296\,u$ and $-0.0525\,u^2$.  They cross at $u=0.563$.
This is far below the strong-coupling regime in which $J=4t_c^2/U_H$ was
obtained, so the crossing does not mark an accessible transition of this
theory.  Above the crossing the ladder dominates.  At the physical
$u=28$ to $36$ (Sec.~\ref{sec:experiment}) it exceeds the spinon term by
one to two orders of magnitude, so the two terms do not nearly cancel
and together favor phase~I.

Neither term, however, is a controlled correction there.  The expansion
parameter of the ladder is $U/t_B=63$ to $81$, so the correction is
larger than the quartic it corrects.  For the spinon term the parameter
is $t_cx/t_{\rm sp}=0.9$ to $6.8$ across the doping window.  Their net
preference for phase~I at the physical $u$ is therefore not a controlled
selection.  They do show that the placement of the \emph{projected}
on-site coupling on the I/II$^-$ boundary is a property of the
projection, and that the physical coupling acquires off-ray components
under lattice matching and need not sit on that boundary at all.
Compared with the sextic, $+1.469\,U^2x^3/t_B$, the ladder is
larger below $x\simeq0.036$, independently of $u$ because both scale as
$U^2/t_B$.  The spinon quartic exceeds the sextic only below
$x\simeq0.001$ at $u=22$.

At the sextic level, on-site repulsion generates, by second-order
elimination of the non-condensing $B_-$ states, a sextic that is
negative definite (Sec.~\ref{sec:phaseII}).  The $B_+$ sector is
not sourced at mean field, since no allowed term of the local potential
is linear in $B_+$ (Sec.~\ref{sec:bands}).  The magnitude of the sextic
is set by the resolvent matrix element $v^\dagger Gv$, with $v$ the
projection of the Hartree source $|B_r|^2B_r$ onto the eliminated states
and $G$ their resolvent.  For $\sum_r\rho_r=N$ this element equals
$(12-\sqrt6)N/108$ ($38.20$ on the $432$-site torus) everywhere in
phase~I, as the moduli theorem requires.  At the selected $K/2$ crystal
of the II$^-$ quadric it is $\simeq0.456N$ ($\simeq197$).  In closed
form the phase-I sextic is
$e_6=-\tfrac{12-\sqrt6}{27}\,U^2\bar\rho^3/t$ per site at mean chargon
density $\bar\rho$ [Eq.~\eqref{eq:octicderived}].  The pure projected
on-site representative therefore selects the $K/2$ crystal, and for the
same reason the unconstrained lattice minimization of
Sec.~\ref{sec:gp}, which contains only on-site $U$, lands in
phase~II$^-$.

Nearest-neighbor repulsion, by contrast, lowers the quartic energy of
phase~I linearly in $V$, at one power of the condensate density below
the sextic gain of II$^-$.  At mean chargon density $\bar\rho=x$ the
quartic gain of Eq.~\eqref{eq:vquartic},
$\tfrac{2\sqrt6-3}{12}\,Vx^2N$, competes with the derived sextic gain of
the $K/2$ crystal over phase~I,
$4\big[(v^\dagger Gv)_{K/2}-(v^\dagger Gv)_{\rm I}\big]\,U^2x^3/t
=4\,(0.4556-0.0884)\,U^2x^3N/t=1.47\,U^2x^3N/t$.  The sextic energy per
site is $e_6=-4\,(v^\dagger Gv/N)\,U^2\bar\rho^3/t$, whose phase-I
value is the $u^2$ term of Eq.~\eqref{eq:octicderived}.  At mean field
the phase-I $M$ state is selected for
\begin{equation}
\begin{aligned}
V>V_c&=\frac{48}{2\sqrt6-3}\,
\frac{(v^\dagger Gv)_{K/2}-(v^\dagger Gv)_{\rm I}}{N}\,
\frac{U^2x}{t}\\
&=9.3\,\frac{U^2x}{t},
\end{aligned}
\label{eq:Vcrossover}
\end{equation}
and the $K/2$ crystal below it.  Phase~I thus requires
$V/U>9.3\,(U/t)\,x$.  For example, $V_c=0.26\,t$ at $U=t$, $x=1/36$,
but $V_c=3.1\,t$ at $U=2t$, $x=1/12$.  With increasing coupling and
doping, only an intersite repulsion comparable to $U$ itself overturns
the on-site sextic.

Equation~\eqref{eq:Vcrossover} compares the two vertices at fixed total
density and is perturbative in $Ux/t$.  It neglects the deformation of
the minimizer away from the vertices at finite $V$, as well as the
sextic terms generated by $V$ itself, so it locates the phase boundary
to within an order of magnitude.  The derived sextic is also not of the
local form $u_6\sum_r\rho_r^3$.  At equal on-site moment
$N^2\sum_r\rho_r^3=2$, its strength on the pair sphere and on the
cross-pair state of the II$^-$ level set differs by the factor
$0.2415/0.0884=2.7$, so no single $u_6$ reproduces both vertices.  On
the II$^-$ level set itself, however, it is an affine function of the
moment (Sec.~\ref{sec:phaseII}).  The $K/2$ crystal and the $M$-star
orders therefore both remain candidates, and which star is selected
depends on how the intersite repulsion compares with
$V_c\simeq9.3\,U^2x/t$, a comparison that lies beyond the quartic
theory.

On the triangular lattice the eight-coupling quartic theory selects
discrete condensates only after couplings are chosen by hand
\cite{Feuerpfeil2026}.  On kagome, the three-coupling structure reduces
the diagram to one angle, and the on-site interaction lands on its
multicritical direction.  The Landau theory also tests the earliest
treatment of doped kagome holons directly.  For the flux-free holon
state, Ko, Lee, and Wen argued that by Bose statistics the holons
condense at each of the quadratic band bottoms \cite{Ko2009}, which
amounts to a balanced occupation of the four valleys.  In the present
theory $U$ selects nothing at $\theta_U$, $V$ tilts
the condensate to a single translation pair, and the microscopic sextic
tilts it to an unbalanced four-valley $K/2$ crystal with weights
$\tfrac14(1\pm1/\sqrt3)$.  The balanced occupation is selected in
neither limit.

\subsection{Phase I: loop-current charge stripes and octic selection}
\label{sec:phaseI}

\begin{figure*}[t]
\includegraphics[width=\textwidth]{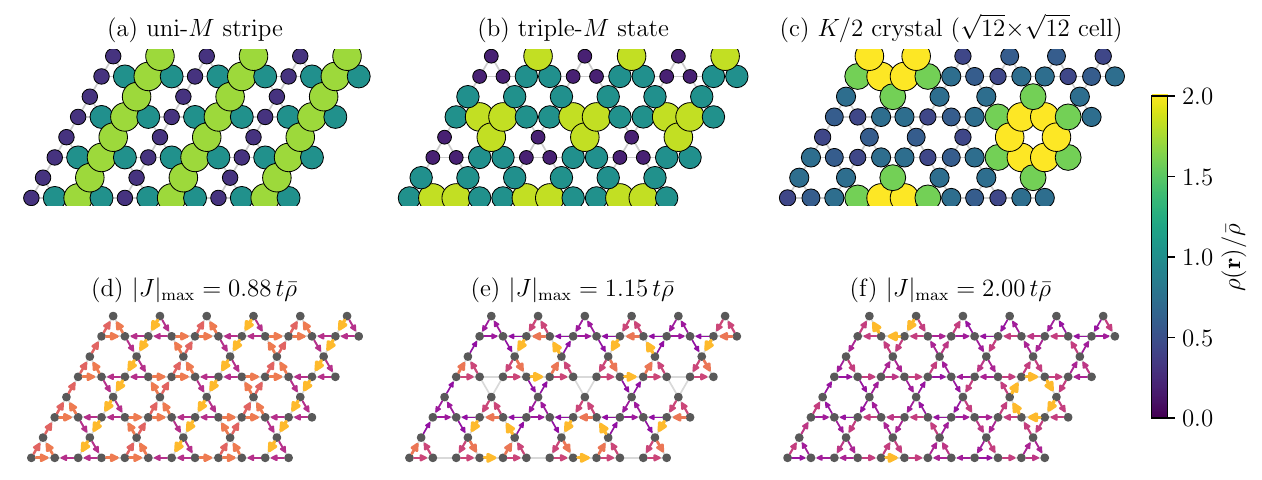}
\caption{Real-space structure of the three key condensates.  Top row:
site densities $\rho(\mathbf{r})/\bar\rho$, with $\bar\rho$ the mean
site density, for (a) the uni-$M$ stripe (doubled cell, three levels),
(b) the triple-$M$ state at the body-diagonal point
$\hat n=(1,1,1)/\sqrt3$ of the pair sphere ($2\times2$ cell, three
levels), and (c) the
sextic-selected $K/2$ crystal ($\sqrt{12}\times\sqrt{12}$, $R30^\circ$,
thirty-six-site cell, five levels).  Bottom row (d)--(f): the
corresponding bond currents $J_{ij}=2\,\mathrm{Im}[\bar B_i s_{ij}B_j]$,
shown by arrow direction and width.  Every condensate is a magnetic
charge crystal, with staggered loop-current textures accompanying the
density order.  The panels show the valley-projected condensates of the
Landau theory.  On the weak-coupling branch $U\bar\rho/t\le1/3$, where
the lattice minimization of Sec.~\ref{sec:gp} returns the $K/2$ crystal,
the full minimizers have the same symmetry and the same stars but
smaller amplitudes, because the weight outside the four valleys screens
the modulation (Sec.~\ref{sec:phaseII}).  Above that branch, the lattice
minimizers found there have unbalanced valley content and no residual
translation symmetry on the torus (Sec.~\ref{sec:gp}).}
\label{fig:condensates}
\end{figure*}

The pair sphere interpolates between two crystal classes
[Fig.~\ref{fig:condensates}(a,b)].  A single-valley condensate produces
the \emph{unidirectional-$M$ stripe}, as one condensed valley yields one
stripe on the square lattice \cite{Christos2023}.  The stripe is a
two-cell charge order that breaks $C_3$, with three site densities,
$\rho/\bar\rho=1-1/\sqrt2$, $1$, and $1+1/\sqrt2$ ($0.2929$, $1$,
$1.7071$), each on one third of the sites.  The three levels give
$\langle(\rho/\bar\rho)^2\rangle=4/3$, as
Eq.~\eqref{eq:hubbardquartic} requires.

The three $M$ amplitudes of a pair condensate are proportional to the
three components of its Bloch vector $\hat n$
(Appendix~\ref{app:composites}), so a pair activates one, two, or three
arms of the $M$ star.  The axis, angle and determinant of the action
induced on $\hat n$ by the in-pair elements $T_1$, $T_2$, $C_3$,
$\sigma C_6$, $C_6\sigma$ and $\Theta C_2$ are tabulated in
Table~\ref{tab:inpair} of Appendix~\ref{app:theorem}.  The one-arm
states are the six coordinate-axis points, and the body-diagonal points
$\hat n=(\pm1,\pm1,\pm1)/\sqrt3$ give the \emph{triple-$M$} $2\times2$
state (four cells, twelve sites), with the three levels
$\rho/\bar\rho=1-\sqrt{2/3}$, $1$, and $1+\sqrt{2/3}$ ($0.1835$, $1$,
$1.8165$) in the ratio $1\!:\!2\!:\!1$.

All of these are exactly degenerate through sextic order.  The on-site
moments are constant on the pair sphere,
$\langle(\rho/\bar\rho)^2\rangle=\tfrac43$ and
$\langle(\rho/\bar\rho)^3\rangle=2$, and the first splitting arises at
octic order.  By the $\mathrm{O}_h$ invariance of Sec.~\ref{sec:moduli},
the octic moment $\langle(\rho/\bar\rho)^4\rangle$ is a constant plus
the single $\ell=4$ cubic harmonic,
$\langle(\rho/\bar\rho)^4\rangle=\tfrac{13}{4}-\tfrac{1}{12}\,(n_x^4+n_y^4+n_z^4)$
in the Bloch frame of the pair.  It equals $19/6$ on the uni-$M$ stripe
orbit (its minimum, the six coordinate-axis points) and $29/9$ on the
triple-$M$ orbit (its maximum, the eight body diagonals), so the two
orbits differ by the relative amount
$(29/9-19/6)/(19/6)=1/57=1.75\%$.  A representative octic proportional
to this moment with a positive coefficient selects the stripe orbit,
with the triple-$M$ state as its maximum.

We do not need such a representative, because the octic that on-site
repulsion generates can be derived directly.  Eliminating the
non-condensing $B_-$ states at fixed total chargon density $\bar\rho$,
to third order in $U$, gives on the pair sphere the energy per site
\begin{equation}
\begin{aligned}
\frac{e}{\bar\rho\,t}&=-(1+\sqrt6)+\tfrac43\,u-\tfrac{12-\sqrt6}{27}\,u^2\\
&\quad+\big(w_0-w_4\,K_4\big)u^3+O(u^4),\qquad u=\frac{U\bar\rho}{t},\\
w_0&=\tfrac{225-64\sqrt6}{81}=0.8424,\qquad
w_4=\tfrac{73-24\sqrt6}{27}=0.5264,
\end{aligned}
\label{eq:octicderived}
\end{equation}
with $K_4=n_x^4+n_y^4+n_z^4$ in the Bloch frame of the pair.  The
quartic and sextic terms are constant on the sphere, as the moduli
theorem requires.  The $\ell=4$ harmonic is the first anisotropy, and
$w_4>0$.  The uni-$M$ stripe orbit ($K_4=1$) is the minimum
of the microscopic octic and the triple-$M$ orbit ($K_4=\tfrac13$) its
maximum.  The stripe selection follows from the polynomial potential of
the chargon sector and is not put in by hand.

The filled spinon sea adds an anisotropy of the opposite sign, favoring
the triple-$M$ state.  The two depend on different hoppings.  The spinon
sea depends on $|t_c|$ and $x$ only through
$gx=(|t_c|/t_{\rm sp})x$, and the octic depends on
$\epsilon_B=U\bar\rho/t_B$, with $|t_c|/t_{\rm sp}=u/4c_1$ and
$U/t_B=u/2c_1$ for $u=U/|t_c|=28$ to $36$ (Sec.~\ref{sec:experiment}).
Setting $|t_c|=t_{\rm sp}$ and $U=t_B/2$ brings the two to comparable
size, but that is a different Hamiltonian and not a different choice of
units.  With the scales restored, the spinon anisotropy saturates, and
its local power of $gx$ falls from $3.8$ below $gx=0.2$ to about unity
by $gx=3$.  The octic grows as $u^3x^4$, and the ratio of the two is
$2.3\times10^{-3}$ at $x=1/36$ and $2\times10^{-4}$ by $x=1/12$.  This
ratio is a formal extrapolation.  Both terms are evaluated at
$U/t_B=63$ to $81$, where $\epsilon_B$ is $1.8$ to $2.3$ at $x=1/36$ and
$5.3$ to $6.8$ at $x=1/12$.  The octic splitting quoted below, a
fraction $0.263\,\epsilon_B^2$ of the quartic energy, is then comparable
to the quartic at the lower doping and several times larger at the
upper one, so the octic lies far outside its own expansion and the
small ratio places no bound on the physical anisotropy.  As stated in
Sec.~\ref{sec:phasediagram}, the choice between the stripe and
triple-$M$ orbits at the physical charging energy therefore remains
open.

The anisotropy is in any case not a single power of $x$.  Within weak
coupling, the two orbits are split by $\tfrac23w_4\,\epsilon_B^3\bar\rho
t=\tfrac{146-48\sqrt6}{81}\,\epsilon_B^3\bar\rho t
=0.351\,\epsilon_B^3\bar\rho t$ per site, both at fixed density and at
fixed chemical potential.  This is a fraction $0.263\,\epsilon_B^2$ of
the quartic energy.  The splitting is small only because of the
$\epsilon_B^3$ suppression.  The octic energies of the two orbits
themselves, $\tfrac{6+8\sqrt6}{81}\,\epsilon_B^3\bar\rho t$ and
$\tfrac{152-40\sqrt6}{81}\,\epsilon_B^3\bar\rho t$ per site, differ
by $111\%$ of the stripe value.  Fixing the total density instead of
the condensate amplitude shifts only the sphere-constant part of the
octic, from $\tfrac{131-40\sqrt6}{27}$ to $w_0$.  The reason is that the
depletion feedback of the eliminated states is the product of the
quartic invariant and a sextic form in the condensate, both of which are
constant on the sphere.

All coefficients are obtained in closed form for the infinite lattice.
The Hartree source $|\psi|^2\psi$ of a pair condensate carries only the
momenta $\QQ_1$ and $\QQ_2$.  At these momenta it couples to the twofold
$B_-$ level at $-t$ and to the $B_-$ flat multiplet at $+2t$, and the
resolvent is diagonal in momentum.  A $24\times24$ torus reproduces
every coefficient to within numerical roundoff.  A nonperturbative
minimization over all non-condensing components at fixed density
reproduces the octic on both orbits to better than one part in $10^7$
and places the stripe below the triple-$M$ state at every coupling
studied, $u=U\bar\rho/t\le0.04$ (eleven values from $10^{-3}$, with
condensate depletion below $2.2\times10^{-4}$).

The octic selection is a mean-field result derived for on-site
repulsion alone.  It applies only inside phase~I, and for the on-site
representative phase~I requires an intersite repulsion above the
crossover $V_c\simeq9.3\,U^2x/t$ of Eq.~\eqref{eq:Vcrossover}, whose
octic terms we have not computed.  Because the uni- versus
multi-directional competition is parametrically soft, geometry, strain,
doping commensuration, or disorder can flip the winner.  A comparable
sensitivity is observed in the doped kagome numerics, where stripe and
two-dimensional holon crystals swap with cylinder geometry and doping
\cite{Jiang2017} (Sec.~\ref{sec:doped}).  It is also seen in the kagome metals, in the
unresolved debate over uni- versus tri-directional $2\times2$ CDW order
and its extreme strain sensitivity \cite{RMP2026}
(Sec.~\ref{sec:experiment}).

Every phase-I condensate is a \emph{magnetic} charge-density wave, and
its current pattern has a uniform and a modulated part.  The
$\Gamma$-point valley polarization $\Delta N$ between the two
translation pairs is saturated ($|\Delta N|=\sum_\eta n_\eta$ everywhere
on the pair sphere), so every phase-I state carries the
$\mathcal{T}$-odd $B_2'$ staggered flux of Table~\ref{tab:ph}.  In the
convention $J_{ij}=2\,\mathrm{Im}[\bar B_is_{ij}B_j]$ of
Fig.~\ref{fig:condensates}, this flux is a uniform, triangle-averaged
circulation of
$\tfrac32(\sqrt3-\sqrt2)\,t\bar\rho=0.4768\,t\bar\rho$, with one sense
on the up-triangles and the opposite sense on the down-triangles.  The
$M$-modulated part adds to it.  At each activated $M$ point, the
$\mathcal{T}$-even ($F_3$) charge/bond composite and the
$\mathcal{T}$-odd ($F_2'$) current composite are activated with
\emph{equal weight}.

For the uni-$M$ stripe, which activates one arm, this adds an
$M$-modulated circulation
$\pm\tfrac12(3-\sqrt6)\,t\bar\rho=\pm0.2753\,t\bar\rho$, smaller than
the uniform part by a factor $\sqrt3$.  The stripe's triangle
circulations are therefore
$\tfrac32(\sqrt3-\sqrt2)\mp\tfrac12(3-\sqrt6)=\{0.2015,\,0.7520\}\,t\bar\rho$
on up-triangles and the reverse on down-triangles.  The multi-arm states
distribute the same $M$ weight over two or three arms
[Fig.~\ref{fig:condensates}(e)].

Time reversal is spontaneously broken, and phase~I is an orbital
loop-current crystal in the family of orbital antiferromagnets.  Its
primary order parameters, the $\Gamma$-point $B_2'$ flux of
Table~\ref{tab:ph} and the $M$-point $F_2'$ current triplet, are both
staggered, and neither couples linearly to $\sigma_{xy}$, which
transforms as $A_2'$.  The magnetic stabilizer of each condensate,
which we find by explicit group search in the full 288-element
projective symmetry group, determines whether a net orbital moment is
nevertheless symmetry-allowed, and with it a polar Kerr or zero-field
anomalous Hall response.

The octic-selected uni-$M$ stripe retains unitary glide mirrors, which
force $\sigma_{xy}=0$ and a vanishing net moment.  Its antiunitary
element $C_6^2\sigma\Theta$ is a $\Theta$-mirror, which alone would not
force either.  The stripe is Kerr- and Hall-blind by symmetry, and
so are the two-arm points of the pair sphere (where one component of
$\hat n$ vanishes).  Those points retain an antiunitary element,
$\Theta C_2$ combined with an in-pair rotation, whose spatial part is a
proper rotation.  Such an element forces $\sigma_{xy}=0$ as effectively
as a unitary mirror.

The triple-$M$ state retains only $C_3$ and antiunitary mirrors.  Its
magnetic point group is $3m'$, a ferromagnetic class.  Correspondingly,
it carries the $A_2'$ composite $\Delta_1'\Delta_2'\Delta_3'$ of its
three current arms, where $\Delta_i'$ are the components of the $F_2'$
current triplet.  Because the arms are proportional to the components
of $\hat n$ (Appendix~\ref{app:composites}), this composite is
$\propto n_xn_yn_z$ on the pair sphere, and it couples linearly to
$\sigma_{xy}$.  The same holds for the $K/2$ crystal of
Sec.~\ref{sec:phaseII} ($6m'm'$, three equal current arms).  Generic
points of the pair sphere retain no antiunitary element at all.  The
composite is cubic in the current bilinears, hence sixth order in the
chargon amplitude and third order in the condensate density.  Since no
$\Gamma$-point $A_2'$ bilinear exists (Table~\ref{tab:ph}), any such
moment enters at order $x^3$, with a size that we leave undetermined.

Local probes ($\mu$SR, NMR line broadening, and polarized neutron
diffraction at the nuclear Bragg positions) couple directly to the
staggered $B_2'$ flux of every phase-I state.  The uniform responses
distinguish the uni-$M$ stripe, which is blind to them, from the
triple-$M$ state and the $K/2$ crystal, where they are allowed
(Sec.~\ref{sec:experiment}).  The accompanying bond-density textures
are fixed by the composite momenta.  The stripe modulates a single arm
of the $M$ bond star (doubled cell, four distinct bond strengths), and
the triple-$M$ state modulates all three ($2\times2$ cell, five distinct
bond moduli $|\bar B_is_{ij}B_j|$ and six distinct bond charges
$\mathrm{Re}[\bar B_is_{ij}B_j]$).

\subsection{Phase II and the sextic-selected $K/2$ crystals}
\label{sec:phaseII}

The quartic phase-II manifold is continuous.  An explicit group search
over random points of the II$^-$ level set shows that its generic points
are fully chiral in the sense used in this section, meaning that they
retain no antiunitary element of the projective symmetry group.
Antiunitary mirrors survive only on particular representatives, among
them the sextic-selected $K/2$ crystals ($6m'm'$, below) and the
cross-pair states.

Sextic terms lift the manifold, and their sign is not free.  The sextic
that on-site repulsion generates by second-order elimination of the
non-condensing $B_-$ states is negative definite,
$e_6=-4\,(v^\dagger Gv/N)\,U^2\bar\rho^3/t$ per site
(Sec.~\ref{sec:phasediagram}), so the physical state maximizes the
resolvent form $v^\dagger Gv$.  On the II$^-$ level set that form is an
affine function of the on-site moment with positive slope,
\begin{equation}
\frac{v^\dagger Gv}{N}=\frac{48+23\sqrt6}{432}
+\frac{2+\sqrt6}{3}\Big(N^2\sum_r\rho_r^3-2\Big),
\label{eq:affine}
\end{equation}
an identity that we verified to $10^{-14}$ on random level-set points
and by direct maximization of $v^\dagger Gv$ over the level set.  On the
II$^-$ level set the physical state is the \emph{maximum} of
the on-site moment $\sum_r\rho_r^3$.  Off the level set the two
functions differ (Sec.~\ref{sec:phasediagram}).  A global maximization
of $\sum_r\rho_r^3$ over the quartic level set of the II$^-$ branch
selects an \emph{unbalanced four-valley} state with the closed-form
valley weights
\begin{equation}
n_\eta=\tfrac14\Big(1\pm\tfrac{1}{\sqrt3}\Big),
\qquad
\textstyle\sum_r\rho_r^3
=\big(2+\tfrac{1}{4\sqrt3}\big)\big/N^2,
\label{eq:phase2min}
\end{equation}
with heavy and light valleys alternating within each translation pair
[Fig.~\ref{fig:condensates}(c)].  To derive
Eq.~\eqref{eq:phase2min} in closed form, parametrize the II$^-$ level
set as
$B=\big(\cos\tfrac a2,\ \sin\tfrac a2\,e^{i\alpha},\ \sin\tfrac a2\,e^{i(\alpha+\gamma-\pi)},\ \cos\tfrac a2\,e^{i\gamma}\big)/\sqrt2$,
with $\sum_r\rho_r=1$ throughout this paragraph.  The quartic moment is
constant on it, $N\sum_r\rho_r^2=\tfrac43$, and the sextic moment is
$N^2\sum_r\rho_r^3=2-\tfrac38\,\sin^2\!a\,\cos a\,\sin2\alpha\,\sin3\gamma$.
Its global extrema are $2\pm1/(4\sqrt3)$, attained for example at
$\cos a=\pm1/\sqrt3$ with $\alpha=7\pi/4$ and $\gamma=3\pi/2$, while
the two-valley endpoints $a=0,\pi$ and the balanced four-valley states
sit at $2$.  The negative-definite sextic that on-site repulsion
generates selects the maximum, $2+1/(4\sqrt3)=2.1443$ at
$\cos a=+1/\sqrt3$, which is the state of Eq.~\eqref{eq:phase2min}.

That argument maximizes only on the II$^-$ level set.  At the Hubbard
quartic, however, the whole degenerate manifold is flat, including
states of intermediate pair population that the level set does not
contain.  The selection is nevertheless global.  To see this, consider
the wider family
\begin{equation}
B=\big(\sqrt s\,c,\ \sqrt s\,d\,e^{i\alpha},\
-\sqrt{1-s}\,d\,e^{i(\alpha+\gamma)},\ \sqrt{1-s}\,c\,e^{i\gamma}\big),
\label{eq:quadric}
\end{equation}
with $c=\cos\tfrac a2$, $d=\sin\tfrac a2$ and $0\le s\le1$.  Its
endpoints $s=0,1$ are the phase-I pair sphere, its midpoint
$s=\tfrac12$ is the level set above, and the projected quartic moment is
$N\sum_r\rho_r^2=\tfrac43$ on all of it.  On this family
\begin{equation}
N^2\sum_r\rho_r^3=2-3\,[s(1-s)]^{3/2}\,\sin^2\!a\,\cos a\,
\sin2\alpha\,\sin3\gamma,
\label{eq:m6global}
\end{equation}
and the resolvent element that the sextic ranks is
\begin{equation}
\frac{v^\dagger Gv}{N}=r_{\rm I}+4s(1-s)\,(r_0-r_{\rm I})
+\frac{2+\sqrt6}{3}\Big(N^2\sum_r\rho_r^3-2\Big),
\label{eq:resglobal}
\end{equation}
with $r_{\rm I}=(12-\sqrt6)/108$ the phase-I value and
$r_0=(48+23\sqrt6)/432$.  Both $s$-dependent terms increase with
$s(1-s)$, since $r_0>r_{\rm I}$ and $|\sin^2\!a\,\cos a|\le2/(3\sqrt3)$,
so the maximum over $\alpha$ and $\gamma$ at fixed $s$ is monotone in
$s(1-s)$ and is attained at $s=\tfrac12$.  The angular maximum there
lies at $|\cos a|=1/\sqrt3$ and gives
$v^\dagger Gv/N=r_0+(2+\sqrt6)/(12\sqrt3)=0.4555999$, which is the state
of Eq.~\eqref{eq:phase2min}.  Direct maximization of $v^\dagger Gv$
over all four parameters on the $432$-site torus returns the same
maximizer (to $4\times10^{-8}$ in $s$ and $\cos a$) and the same value.
Equations~\eqref{eq:m6global} and \eqref{eq:resglobal} reproduce the
lattice element to $3\times10^{-15}$ over random points of the
manifold.  The minimum, $2-1/(4\sqrt3)$ at $\cos a=-1/\sqrt3$, is what
a positive representative sextic $+\sum_r\rho_r^3$ would select.  It is
the density conjugate $\rho\to2\bar\rho-\rho$ of the maximum up to a
lattice translation, with the same valley weights (heavy and light
valleys interchanged), the same cell, and the same stabilizer.  The
sextic fixes the sign of the $K/2$ modulation but not the crystal
itself.

In its valley projection, the unconstrained lattice minimization of
Sec.~\ref{sec:gp}, which contains only on-site $U$, returns the valley
weights $0.394$ and $0.106$, i.e.\ $\tfrac14(1\pm1/\sqrt3)$, and the
projected moment $N^2\sum_r\rho_r^3=2.1443=2+1/(4\sqrt3)$.  The moment
of the lattice density itself is lower, because the weight outside the
four valleys screens the modulation (Sec.~\ref{sec:gp}).  That
minimization realizes the state of Eq.~\eqref{eq:phase2min},
not its conjugate.  Its real-space density is a
\emph{thirty-six-site} crystal (twelve primitive cells forming the
$\sqrt{12}\times\sqrt{12}$ $R30^\circ$ supercell) with five density
levels per cell, $\rho/\bar\rho=1-1/\sqrt3$, $1-1/\sqrt2+1/(2\sqrt3)$,
$1-1/(2\sqrt3)$, $1+1/\sqrt3$, and $1+1/\sqrt2+1/(2\sqrt3)$ ($0.4226$,
$0.5816$, $0.7113$, $1.5774$, and $1.9958$), with multiplicities
$(6,6,12,6,6)$.  Its Bragg content separates cleanly.  The charge order
lives purely on the $K/2$ star, because the $M$-point charge components
of the two pairs cancel identically, while the loop currents modulate
on the $M$ and $K$ stars.

A full group closure within the full 288-element projective symmetry
group finds a stabilizer generated by the supercell translations
(twelve on the $12\times12$ torus, the lattice of the thirty-six-site
cell) together with a magnetic point group of order twelve.  That point
group consists of the six unitary rotations $C_6^k$ about a hexagon
center and the six mirrors, each present only in combination with
$\Theta$.  No unitary mirror and no translation beyond the supercell
lattice survives, so the stabilizer is the magnetic point group
$6m'm'$ modulo the supercell translations.

Like the phase-I states, the selected state is a magnetic
($\mathcal{T}$-breaking, current-carrying) crystal, and its phase
degrees of freedom are locked already at sextic order.  Unlike the
uni-$M$ stripe, however, it is not protected against a net orbital
moment or a polar Kerr response (Sec.~\ref{sec:phaseI}).  Its point
group $6m'm'$ is a ferromagnetic class, and with three equal current
arms the crystal carries the $A_2'$ composite
$\Delta_1'\Delta_2'\Delta_3'$.

The competing representatives are cleanly ranked by the on-site moment
$N^2\sum_r\rho_r^3$ with $\sum_r\rho_r=1$, i.e.\
$\langle\rho^3\rangle/\langle\rho\rangle^3$.  The selected state has
$2+1/(4\sqrt3)=2.1443$, while the cross-pair two-valley states and the
balanced four-valley states are degenerate at $2$.  The generic members
of the latter are fully chiral, with no antiunitary element in their
stabilizer.  The sextic \emph{moment} spreads by
$1/(8\sqrt3)=7.2\%$ of the competitors' value [$1/(8\sqrt3+1)=6.7\%$ of
$2.1443$].  The derived sextic separates the states more strongly,
because on the level set $v^\dagger Gv$ is an affine, not a
proportional, function of the moment [Eq.~\eqref{eq:affine}].  The
selected state has $v^\dagger Gv/N=0.4556$ against $0.2415$ for both
competitors.  This gives a sextic-energy gain of
$4\times0.214\,U^2x^3/t=0.86\,U^2x^3/t$ per site, or $89\%$ of the
competitors' sextic energy.  Symmetry fixes which anisotropies are
allowed, whereas the sign and the competition of the higher-order
coefficients are set microscopically, and a sextic of the opposite sign
would select the density-conjugate point of the manifold.

The $K/2$ star has appeared before in the kagome problem.  The
thirty-six-site valence-bond crystal \cite{MarstonZeng,SinghHuse2007},
the classic competitor of the Dirac spin liquid
\cite{YanHuseWhite2011,Iqbal2011}, is a
$\sqrt{12}\times\sqrt{12}$ ($R30^\circ$)
crystal whose superlattice Bragg vectors have length $|K|/2$
along the $K$ directions \cite{Iqbal2011}.  The sextic-selected chargon
crystal of Eq.~\eqref{eq:phase2min} occupies \emph{the same
thirty-six-site cell}, which no order of the flux sector has.  The
singlet monopoles of the DSL carry the three $M$ momenta
\cite{SongPRX2020}, which generate only the $2\times2$ twelve-site
superlattice, and every nonzero vector of that superlattice is longer
than $|K|/2$.  Monopole condensation alone therefore cannot produce a
thirty-six-site cell, whereas chargon condensation produces it directly.

The two states are nonetheless sharply different.  The VBC preserves
time reversal and, in the pinwheel form of
Refs.~\cite{MarstonZeng,SinghHuse2007}, the sixfold rotations but not
the mirrors (point group $C_6$).  Only the variant of
Ref.~\cite{Iqbal2011} has the full $C_{6v}$ \cite{IqbalVBCNJP2012}.
The chargon crystal breaks $\mathcal{T}$ and retains mirrors only in
combination with time reversal, so its symmetry is a magnetic space
group on the lattice of the VBC.

That the projected Hubbard functional lies on the I/II boundary may be
read as a Landau-theoretic counterpart of the long-standing competition
between $M$-type and $K/2$-type orders in kagome numerics.  The
microscopic interaction is poised between the two stars, and terms
beyond the quartic level select the winner, through spin-Peierls
energetics for the undoped VBC \cite{Iqbal2011} and through sextic
selection and doping commensuration for the chargon crystals
(Sec.~\ref{sec:doped}).

\subsection{Order-parameter catalog}
\label{sec:catalog}

\begin{table}[t]
\caption{Particle--hole composites (identical for the $U(1)$ state and
all four $\Z_2$ descendants).  The last column gives the irrep of the
extended point group $C_{6v}'''$ (space group modulo the $2\times2$
translations, Appendix~\ref{app:composites}) for the $\Gamma$- and
$M$-point composites.  The $K$ and $K/2$ stars lie outside that quotient
and are labeled by their $C_3$ or mirror eigenvalue and their $\Theta$
parity.  $\Theta$ is antiunitary and maps $K$ to $K'$, so the Hermitian
$K$-star composite is built from the bilinears at both points,
$\bar B_1B_4$ and $\bar B_2B_3$ at $K$ and $\bar B_3B_2$ and
$\bar B_4B_1$ at $K'$.  This composite is $\mathcal{T}$-odd, so the $K$
star carries loop current and no charge.  Under the little group
$C_{3v}$ of $K$ it is $C_3$-even and odd under the three mirrors
$\sigma_K$ that fix $K$ (the mirrors through $\Gamma$ and $K$, namely
$\sigma C_6$ and its conjugates).  It therefore transforms as $A_2$ of
$C_{3v}$, the pseudoscalar of the little group, as a loop-current order
must, and for this row the irrep column gives the $C_{3v}$ label.  The
six $K/2$ bilinears form three $(+Q,-Q)$ mates, and the six symmetric
$K/2$ pair composites $B_aB_b$ form the same three mates, all nonzero
on the phase-II$^-$ crystal.}
\label{tab:ph}
\begin{ruledtabular}
\begin{tabular}{@{}l@{\hspace{4pt}}c@{\hspace{4pt}}c@{\hspace{4pt}}c@{\hspace{4pt}}c@{\hspace{4pt}}c@{}}
composite & momentum & $C_2$ & $\sigma_v$ & $\mathcal T$ & irrep \\
\colrule
$n_{\rm tot}$ & $\Gamma$ & $+$ & $+$ & $+$ & $A_1$ \\
$\Delta N$ (flux order) & $\Gamma$ & $-$ & $-$ & $-$ & $B_2'$ \\
charge/bond wave & $M$ & $-$ & $+$ & $+$ & $F_3$ \\
current wave & $M$ & $+$ & $-$ & $-$ & $F_2'$ \\
$\sqrt3{\times}\sqrt3$ loop current & $K$, $K'$ & \multicolumn{2}{c}{$C_3$ $+$, $\sigma_K$ $-$} & $-$ & $A_2$ \\
$K/2$ charge crystal & $K/2$, 3 mates & \multicolumn{2}{c}{mirror: $+$} & $+$ & --- \\
\end{tabular}
\end{ruledtabular}
\end{table}

Table~\ref{tab:ph} lists the gauge-invariant particle--hole composites
and their quantum numbers.  The list is the same for the U(1) state and
all four $\Z_2$ descendants.  The momentum-star decomposition
of the sixteen bilinears and ten pair composites, and the calibration
of the $C_{6v}'''$ labels, are given in Appendix~\ref{app:composites}.

Beyond the $M$-triplets discussed above, the $K$-point composite
realizes a $\sqrt3\times\sqrt3$ \emph{loop-current} crystal that is
\emph{$C_3$-symmetric} ($C_3$ eigenvalue $+1$) and $\mathcal{T}$-odd.
It is not a charge order.  The site-density form factor at $K$ vanishes
identically for every condensate, and so does the gauge-invariant bond
charge at $K$ out to third-neighbor range, so within the constructions
tested the $K$ star is not a local charge-density or bond-charge
channel.  It can carry loop current, which makes it a target for probes
of orbital magnetism and not for x-ray or electron diffraction
(Sec.~\ref{sec:experiment}).

The chargon mechanism produces the enlarged $\sqrt3\times\sqrt3$
cell without rotational symmetry breaking, but in the current channel.
The $K$-point CDW of ScV$_6$Sn$_6$, whose three CDW peaks show no $C_3$
anisotropy (unlike the $2\times2$ kagome-metal CDWs \cite{RMP2026}),
shares this cell and its point symmetry, but not the channel or the
(phononic) mechanism.

The $\Gamma$-point valley polarization $\Delta N$ is a staggered
(Haldane-like) flux order, $\mathcal{T}$-odd, $C_6$-odd, $\sigma_v$-odd
and $\sigma_d$-even ($B_2'$).  The same valley-polarization bilinear is
a \emph{charge} stripe on the square lattice \cite{Christos2023} and an
opposite-flux order on the triangular lattice \cite{Feuerpfeil2026}.
Kagome behaves like the triangular case, in that valley imbalance comes
with time-reversal breaking.

The $F_3$ triplet also supports a vestigial composite.  Because every
element of $C_{6v}'''$ acts on the $F_3$ triplet
$(\Delta_1,\Delta_2,\Delta_3)$ by a signed permutation of the three $M$
points, the translation-invariant squares
$(\Delta_1^2,\Delta_2^2,\Delta_3^2)$ carry the permutation
representation $A_1\oplus E_2$ (the cross terms $\Delta_i\Delta_j$,
$i\neq j$, form a $\mathcal{T}$-even $F_1$ $M$-triplet).  The $E_2$
doublet
$\mathcal{N}=\big(2\Delta_1^2-\Delta_2^2-\Delta_3^2,\ \sqrt3(\Delta_2^2-\Delta_3^2)\big)$
is a $\mathcal{T}$-even, $C_2$-even, translation-invariant $\Z_3$
nematic that rotates by $120^\circ$ under $C_3$.  Because it is
quadratic in the triplet, symmetry allows it to order where
$\langle\Delta_i\rangle=0$, so a vestigial nematic phase above the
$M$-triplet transition is permitted.  Whether such a phase occurs, and
at what temperature, is a question about fluctuations.

To evaluate the nematic on the mean-field condensates, take
$\sum_\eta|B_\eta|^2=1$ and the one-arm normalization
$|\Delta|=\tfrac12$ of Appendix~\ref{app:composites}.  The uni-$M$
stripe, with one arm activated, then has
$|\mathcal{N}|=\tfrac12\,\bar\rho^{\,2}$, and the three one-arm states
give the three nematic directions.  The triple-$M$ state at the body
diagonal has $\mathcal{N}=0$ (three equal arms,
$|\Delta_i|=1/(2\sqrt3)$).  The $K/2$ crystal of
Eq.~\eqref{eq:phase2min} has $\mathcal{N}=0$ because its $F_3$ triplet
vanishes identically.  Its stabilizer contains $C_2$ about a hexagon
center, and $F_3$ is $C_2$-odd.  Its particle--hole $M$ weight resides
entirely in the $\mathcal{T}$-odd current triplet $F_2'$, with three
equal arms.  Only the zeros and the ratios of these values are
independent of the normalization.  On the pair sphere $\Delta_i=n_i/2$
in an orthonormal frame, so $\mathcal{N}$ is the $\ell=2$ harmonic of
the Bloch vector.  The in-pair group $\mathrm{O}_h$
(Sec.~\ref{sec:moduli}), whose first anisotropy is $\ell=4$, excludes
this harmonic from the condensate energy, yet it is present as an
$E_2$ order parameter.

The charge and current $M$-triplets are locked to each other by symmetry
as long as the $\Gamma$-point flux is ordered.  At mean field the
locking is automatic, because at each activated $M$ point the $F_3$ and
$F_2'$ composites are the two Hermitian parts of a single bilinear of
the condensate, activated with equal weight (Sec.~\ref{sec:phaseI}).
Beyond mean field the gauge-invariant composites are the physical order
parameters, and one may ask whether fluctuations can split the single
Higgs transition into separate charge-order and loop-current
transitions.  Averaging over the full 288-element projective symmetry
group shows that the space of invariant trilinears built from one
component of $\Delta N$, one of the $F_3$ triplet and one of the $F_2'$
triplet is exactly one-dimensional,
\begin{equation}
\mathcal{L}_3=\Delta N\sum_{i=1}^{3}\epsilon_i\,\Delta_i\Delta_i',
\qquad \epsilon=(+,-,+),
\label{eq:lock}
\end{equation}
in the component convention of Appendix~\ref{app:composites}.  No
invariant of the forms $F_3F_2'F_2'$, $F_3F_3F_2'$, $\Delta N\,F_3F_3$
or $\Delta N\,F_2'F_2'$ exists, and the first coupling that does not
involve the flux is quartic (four invariants of the form
$F_3F_3F_2'F_2'$).

Wherever the $B_2'$ flux is ordered, Eq.~\eqref{eq:lock} mixes the two
triplets linearly, so neither can order without inducing the other, and
the $M$-point charge order and the loop-current order set in at a
single transition.  This holds on every phase-I condensate and
throughout the upper stage of the two-stage finite-temperature
structure of Sec.~\ref{sec:experiment}.  Only when the flux is absent
and time reversal is restored are the two irreps coupled at quartic
order alone, and only then can they order at separate transitions.  On
every phase-I state $\mathcal{L}_3$ takes the same value,
$-\tfrac14\bar\rho^{\,3}$ in the normalization used for the vestigial
nematic above, as the equal-weight activation of Sec.~\ref{sec:phaseI}
implies.  On the
$K/2$ crystal the trilinear vanishes, since there $\Delta N=0$ and the
$F_3$ triplet vanishes identically.

In the \emph{undoped} spin liquid, the operator content of the DSL
below scaling dimension two (sixteen fermion bilinears and six
monopoles) carries only $\Gamma$- and $M$-point momenta, and all its $M$-triplets
are $C_2$-even.  The VBS channels (Hastings' David-star pattern among
them) are $F_1$, and the three spin-singlet monopoles form
$F_1\oplus F_2$, two $\mathcal{T}$-even triplets distinguished by their
mirror eigenvalue \cite{Hermele2008,SongPRX2020,SongNC2019}.

Each monopole triplet carries a single time-reversal parity.  This
follows from $\mathcal{T}^2=1$.  For an antiunitary $\mathcal{T}$ with
$\mathcal{T}\Phi^\dagger\mathcal{T}^{-1}=e^{i\beta}\Phi$ one has
$\beta\in\{0,\pi\}$, so the two Hermitian components of a monopole share
one parity instead of splitting into an even and an odd part.  The
argument does not fix which parity is realized, since $\beta=0$ makes
both components even and $\beta=\pi$ makes both odd.  We take $\beta=0$
for the singlet monopoles as an input from the explicit monopole
computation of Refs.~\cite{SongNC2019,SongPRX2020}, which identifies the
spin-singlet monopoles with $\mathcal{T}$-even valence-bond orders and
the spin-triplet monopoles with $\mathcal{T}$-odd magnetic orders.

The chargon current triplet $F_2'$ is the $\mathcal{T}$-odd
partner of the monopole triplet $F_2$ in the same spatial channel
($C_2$-even, both mirrors odd).  The two are distinct order parameters,
and at the level of masses and monopoles the Higgs (chargon) sector of
the doped system and the flux (monopole) sector of the undoped magnet
share no channel at all.  $F_2'$ and $F_3$ do reappear in the parent,
among its conserved SU(4) currents at the $M$ points, whose scaling
dimension is exactly two.  The chargon composites couple linearly to
these currents, while the parent's masses and monopoles do not enter.
The $F_2'$ channel is also the one selected by the weak-coupling
loop-current theory of the kagome metals \cite{FuNSR2025,ZhanLCO2026}.
The $M$-point current order of Ref.~\cite{FuNSR2025} transforms as
$A_2$ of the little group $C_{2v}$ (see its supplementary
classification), which is $F_2'$ in the present labeling.  In both works
the Landau trilinear vanishes by time reversal, just as every cubic
vanishes here.

Among masses and monopoles the $F_3$ channel is populated by the
chargon sector alone.  It is also the representation of the classical
kagome ground states at the $M$ points, whose symmetry, as Hermele
\emph{et al.}\ noted, is absent from every enhanced ASL observable
\cite{Hermele2008}.  The ``missing'' classical $M$-point representation
thus reappears in the charge sector of the doped system, as a spin
singlet and not as a magnetic order.

Because the $M$-point fluctuations of the undoped DSL are numerically
weak in projected wave functions \cite{Hermele2008}, the orders
cataloged here are best regarded as doping induced and new in the
symmetry channels they occupy, and not as amplified intrinsic
fluctuations of the parent magnet.

\section{$\Z_2$ descendants}
\label{sec:z2}

\subsection{The four neighbors of the Dirac spin liquid}
\label{sec:fourZ2}

Lu, Ran, and Lee \cite{Lu2011} classified the $\Z_2$ spin liquids
continuously connected to the $[0,\pi]$ DSL and found four, which we
label $\alpha,\beta,\gamma,\delta$.  All four share trivial translation
dressings and $g_{\mathcal T}=i\tau^1$,
$\eta_{\mathcal T}=\eta_{12}=\eta_{C_6T_1}=-1$.  They differ only by
uniform $i\tau^3$ dressings of $C_6$ and/or $\sigma$, that is, by the
sign triple $(\eta_\sigma,\eta_{\sigma C_6},\eta_{C_6})$.  Here
$\eta_{\sigma C_6}$ is the phase of the projective relation
$\sigma^{-1}C_6\sigma C_6=e$,
$g_\sigma^\dagger(C_6(i))\,g_{C_6}(C_6(i))\,g_\sigma(i)\,g_{C_6}(\sigma(i))
=\eta_{\sigma C_6}\tau^0$ [Eq.~(A14) of Ref.~\cite{Lu2011}].  Defining
the middle entry through $(g_\sigma g_{C_6})^2$ instead would multiply
it by $\eta_\sigma$ and reverse it for $\gamma$ and $\delta$.  The four
triples are
\begin{equation}
\begin{aligned}
\alpha\ (\text{No.~6}):&\ (+,-,+),\qquad &\beta\ (\text{No.~2}):&\ (+,+,-),\\
\gamma\ (\text{No.~14}):&\ (-,-,+),\qquad &\delta\ (\text{No.~16}):&\ (-,+,-),
\end{aligned}
\label{eq:etatriples}
\end{equation}
in the numbering of Ref.~\cite{Lu2011}.  We have constructed all four
explicitly and validated their symmetric pairing content, range by range,
against Eqs.~(C16)--(C19) of Ref.~\cite{Lu2011} (twenty independent
counts; Table~\ref{tab:z2content}).  We quote the real-space form
factors as the ratio of the pairing amplitude to the hopping amplitude
on the same bond, $\Delta_{ij}/\chi_{ij}$, which
is invariant under the residual $\Z_2$ gauge transformations of the
descendants, whereas the sign of $\Delta_{ij}$ on any single bond is
not.  A uniform U(1) rotation of the parent's invariant gauge group
multiplies the ratio by a global phase, so its overall sign is
conventional and only its bond-to-bond pattern is physical.

Symmetry fixes the pattern of this ratio but not its magnitude, which
is a free variational parameter.  The state $\alpha$ carries
nearest-neighbor pairing with
$\mathrm{sgn}(\Delta_{ij}/\chi_{ij})=+1$ on up-triangles and $-1$ on
down-triangles, which is an $f$-wave sign structure.  The state
$\gamma$ carries the analogous alternation on the second-neighbor
inscribed triangles, and the second-neighbor pairing of $\beta$
follows the 2NN hopping sign pattern.  Each of these patterns is
fixed uniquely by symmetry (Appendix~\ref{app:z2psg}).

Of the four, only $\beta$ gaps the spinon Dirac cones \cite{Lu2011}.
For this reason $\beta$ was originally proposed as the candidate for
the fully gapped spin liquid supported by density-matrix renormalization
and tensor-network studies
\cite{YanHuseWhite2011,Depenbrock2012,Mei2017}.  It is also the kagome
channel whose Higgs criticality is analyzed in
Ref.~\cite{FeuerpfeilDepleted2026} (Appendix~\ref{app:z2psg}).
Variational calculations found the Dirac state stable against opening
the $\beta$ gap \cite{IqbalZ22011}, and Lanczos-improved wave functions
placed the gapless DSL marginally below it with a vanishing spin gap
\cite{Iqbal2013,IqbalGap2014}.  Subsequent DMRG work shifted the
balance toward the DSL, finding a much smaller spin gap together with
Dirac-cone signatures matching the $[0,\pi]$ ansatz
\cite{HeZaletel2017}, and tensor-network and entanglement studies
concur \cite{Liao2017,ZhuEnt2018}.  The $\alpha$ state has numerical
support of its own.  A pseudofermion functional-renormalization study
of the kagome Heisenberg model, combined with a self-consistent spinon
mean field, identified a gapless $\Z_2$ spin liquid of this type
\cite{Hering2019}.  For our purposes the four states enter on an equal
footing, as the four symmetry-distinct pairing channels through which
the doped DSL can superconduct, whatever the verdict on the undoped
system.  The classification does not select among them energetically,
and we make no claim about which one is realized.

The classification assumes translation-invariant spinon pairing.  A
recent machine-learning variational study proposes instead that the
kagome ground state is a spinon pair-density wave descending from the
same $[0,\pi]$ DSL, with spinon Cooper pairs whose center-of-mass
momentum lies at the $M$ point \cite{DuricPDW2025}, a finite-momentum
$\Z_2$ class outside Eq.~\eqref{eq:etatriples}.  Its variational
energies have been challenged as an artifact of non-ergodic Monte Carlo
sampling, and ergodic sampling is reported to give energies above the
DMRG benchmarks \cite{KamalComment2026}.  Independently of these
energetics, translations by one lattice constant act as pure gauge
transformations because the pair momentum is $M$, and the induced
composites sit at $2M\equiv0$.  The projective symmetry analysis that
underlies the composite catalog of Sec.~\ref{sec:catalog} then shows
that the state breaks neither time reversal nor any lattice symmetry.

\begin{table}[b]
\caption{Number of symmetric pairing patterns of the four $\Z_2$
descendants at each range, checked against Appendix~C of
Ref.~\cite{Lu2011}.}
\label{tab:z2content}
\begin{ruledtabular}
\begin{tabular}{lccccc}
state & on-site & NN & 2NN & 3NN$_{\rm chain}$ & 3NN$_{\rm hex}$ \\
\colrule
$\beta$  & 1 & 1 & 1 & 1 & 0 \\
$\alpha$ & 0 & 1 & 0 & 0 & 1 \\
$\gamma$ & 0 & 0 & 1 & 0 & 0 \\
$\delta$ & 0 & 0 & 0 & 1 & 0 \\
\end{tabular}
\end{ruledtabular}
\end{table}

\subsection{Chargon spectra with pairing}
\label{sec:z2spectra}

In the chargon sector the $\Z_2$ perturbation is strongly constrained.
For the $\beta$ state, the only symmetric Higgs channel at on-site and
nearest-neighbor range is the uniform on-site term
$\lambda\,B^\dagger\tau^1 B$.  The single symmetric NN pattern of
$\beta$ is an invariant-gauge-group rotation of the hopping and is
physically trivial, as in the triangular case \cite{Feuerpfeil2026}.
The NN pattern of $\alpha$, unlike that of $\beta$, has zero overlap
with the hopping and is the genuine $f$-wave pairing of
Table~\ref{tab:pairing}, which, as shown below, preserves the flat
bands.  Hereafter $\Delta_2$
denotes the second-neighbor pair amplitude of $\beta$, the pairing
partner of the second-neighbor hopping $\chi_2$ of Sec.~\ref{sec:bands}
(Appendix~\ref{app:z2psg}), and no longer the $M$-triplet component of
Sec.~\ref{sec:catalog}.  The on-site term alone is the case
$\Delta_2=0$.

With only the on-site term switched on, the $\beta$ chargon doublet
spectrum takes the closed form
\begin{equation}
E_\pm(\kk)=\pm\sqrt{t^2 m(\kk)^2+\lambda^2},
\label{eq:z2spectrum}
\end{equation}
where $m(\kk)$ are the eigenvalues of the $6\times6$ Bloch hopping
matrix $M(\kk)$ of Sec.~\ref{sec:ansatz} (unrelated to the $M$ point
of the Brillouin zone).  The corresponding $12\times12$ Bloch
problem is
$\begin{pmatrix}H_+&\lambda\Lambda\\ \lambda\Lambda&H_-\end{pmatrix}$,
where $H_\pm$ are the Bloch hopping matrices of the two chargon
components, $H_+=-H_-$, and $\Lambda(\kk)$ is that of the pairing
pattern.  It has the Bogoliubov--de~Gennes form in the $(B_+,B_-)$
isospin space, which produces the $E\to-E$ symmetry.  It nevertheless
conserves chargon number and contains no anomalous term, so all twelve
bands are physical chargon states.  We have verified
Eq.~\eqref{eq:z2spectrum} on the torus for $\lambda/t=0$--$4$ at
$\Delta_2=0$.  The four valleys remain pinned at $\QQ_{1\dots4}$ and
nondegenerate for all $\lambda$, and the flat pair stays exactly flat
at $-\sqrt{4t^2+\lambda^2}$.  Because the second-neighbor pair form
factor does not commute with $M(\kk)$, the $\beta$ spectrum has no
closed form once $\Delta_2$ is switched on.  At the valleys the
effective pair field combines the on-site and second-neighbor amplitudes as
\begin{equation}
\lambda_{\rm eff}=\lambda-2\Delta_2\,,
\label{eq:lambdaeff}
\end{equation}
because the 2NN form factor has valley eigenvalue exactly $-2$ in the
flux-canonical sign convention $\nu_{ij}=-s_{ik}s_{kj}$ of
Sec.~\ref{sec:bands} (reversing the overall sign of the form factor
requires $\Delta_2\to-\Delta_2$).  The corresponding combination at the
\emph{spinon} Dirac points, a different momentum in a different sector,
has different weights [$(-1:\sqrt3{+}1:-(\sqrt3{-}1))$ for (on-site :
2NN : NN), Eq.~(C24) of Ref.~\cite{Lu2011}].

\begin{figure*}[t]
\includegraphics[width=0.92\textwidth]{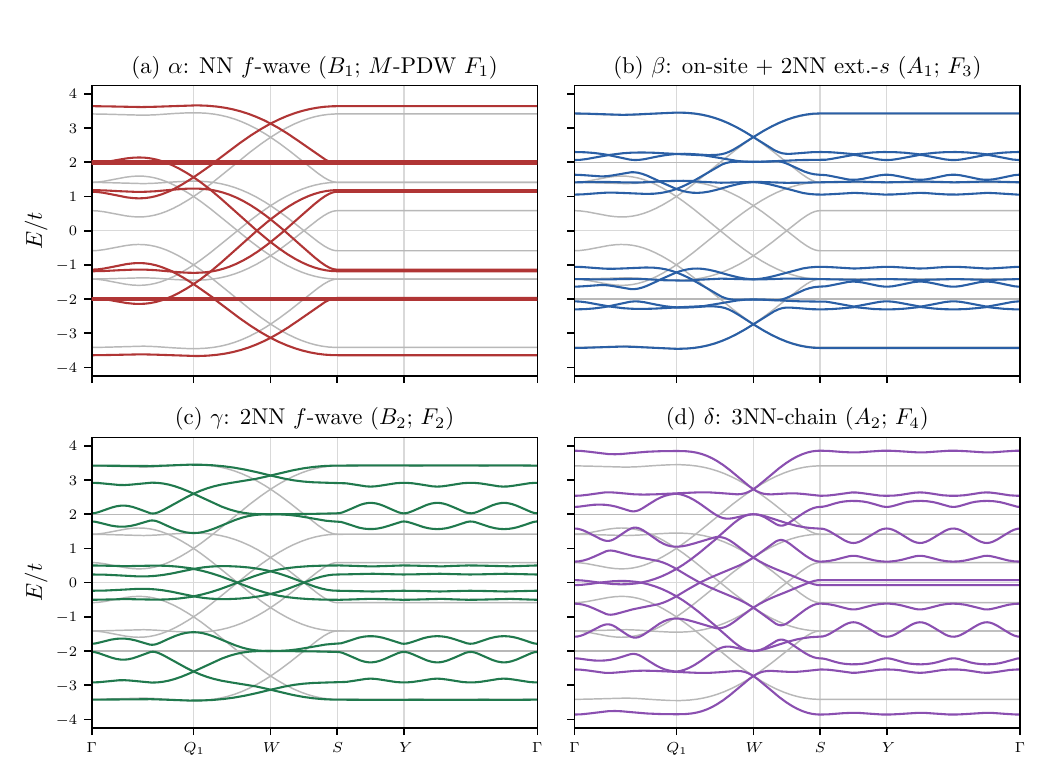}
\caption{Chargon doublet spectra of the four $\Z_2$ descendants along
$\Gamma\to Q_1\to W\to S\to Y\to\Gamma$, the path of
Fig.~\ref{fig:bands}(c).  Gray lines show $\lambda=0$.  Colored lines
show each descendant's symmetry-fixed pairing pattern at amplitude
$0.4\,t$ for $\alpha$, $\gamma$, $\delta$ and the on-site term of
$\beta$, and $0.3\,t$ for the second-neighbor term of $\beta$.  Every
panel agrees with exact $12\times12$-torus diagonalization to within
numerical roundoff ($3\times10^{-14}$).  The NN $f$-wave pairing of
$\alpha$ leaves the chargon flat bands \emph{exactly} flat at $\pm2t$
(thick lines).  The $\beta$ state obeys the closed form
$E=\pm\sqrt{t^2m(\kk)^2+\lambda^2}$ of Eq.~\eqref{eq:z2spectrum},
with $m(\kk)$ the eigenvalues of the Bloch matrix $M(\kk)$, only at
$\Delta_2=0$.  For nonzero $\Delta_2$ the second-neighbor pair form
factor does not commute with $M(\kk)$, the $\beta$ spectrum has no
closed form, and the valley pair field is
$\lambda_{\rm eff}=\lambda-2\Delta_2$ [Eq.~\eqref{eq:lambdaeff}].  The
segment $S\to Y$ lies on a dispersionless line of the DSL chargon
spectrum.}
\label{fig:descendants}
\end{figure*}

Figure~\ref{fig:descendants} shows the four doublet band structures.
The NN $f$-wave pairing of $\alpha$ leaves the chargon flat bands
\emph{exactly} flat at $\pm2t$ for all pairing amplitudes.  The
line-graph kernel persists in the pairing channel, so the kagome
flat-band structure survives both the $[0,\pi]$ flux and the
$\alpha$-type Higgs field.  With $\lambda$ switched on, the condensed $B_-$
acquires a $B_+$ admixture.  For the on-site term of $\beta$ its weight
is $[\lambda/2(1{+}\sqrt6)t]^2+O(\lambda^4)$, which we have verified
numerically at the sub-percent level.  A physical pair amplitude
\emph{linear} in $\lambda$ also appears.  Its leading term, which
fixes its symmetry, is the composite of Eq.~\eqref{eq:elpair} below.
The admixture channel depends on the descendant and is determined by
the $6\times6$ Bloch block at the valley momentum, where the $B_+$
spectrum is $\{-2t\,(\times2),\ (1-\sqrt6)t,\ t\,(\times2),\
(1+\sqrt6)t\}$, so the admixture statements below hold exactly for the
infinite lattice.

Normalize each pattern so that its largest bond amplitude is $\lambda$.
Every symmetric pattern of $\beta$ then acts on the four valley states
as a scalar, with eigenvalues of modulus $1$, $2$ and $\sqrt6$ for its
on-site, second-neighbor and third-neighbor-chain patterns, and maps
them onto the top level of the $B_+$ spectrum at $(1+\sqrt6)t$.  The
nearest-neighbor pattern is the pure-gauge rotation of the hopping
noted above, and the second-neighbor eigenvalue is $-2$ in the
flux-canonical convention $\nu_{ij}=-s_{ik}s_{kj}$ of
Sec.~\ref{sec:bands}.  The admixture goes through the energy
denominator $2(1+\sqrt6)t=6.90\,t$, and the on-site weight is
$[\lambda/2(1+\sqrt6)t]^2=(7-2\sqrt6)\lambda^2/100t^2$.

The patterns of $\alpha$ (nearest-neighbor and across-hexagon
third-neighbor) and of $\delta$ (third-neighbor chain) instead map each
valley state entirely onto the fourfold $B_+$ level at $(1-\sqrt6)t$.
That level is the partner of the valley level.  The dispersive $B_-$
spectrum has the exact symmetry $E\to-2t-E$.  It follows from the
line-graph structure of Sec.~\ref{sec:bands}, because the honeycomb
bands $\epsilon_a(\kk)$ are symmetric under $\epsilon\to-\epsilon$.
The valley at $-(1+\sqrt6)t$ has a partner at $(\sqrt6-1)t$
at the same momentum, and the $B_+$ sector, whose spectrum is the
negative of the $B_-$ spectrum, carries this partner at $(1-\sqrt6)t$.
The $\alpha$ and $\delta$ admixtures go through an energy
denominator of $2t$, with
$B_+$ weights $3\lambda^2/4t^2$ ($\alpha$) and $3\lambda^2/2t^2$
($\delta$), larger than the on-site weight of $\beta$ by the factors
$3(1+\sqrt6)^2=35.7$ and $6(1+\sqrt6)^2=71.4$ (amplitude ratios $5.97$
and $8.45$).

For the nearest-neighbor pattern of $\alpha$ the mapping follows from
an algebraic identity.  The
pattern is $\tilde B^\dagger\Xi\tilde B$, with $\tilde B$ the signed
$4\times6$ incidence matrix of Sec.~\ref{sec:bands} and $\Xi=\pm1$ on
the two sublattices of the honeycomb of triangle centers.  It
anticommutes with $1-M$ and squares to $9-(M-1)^2$, which equals $3$ on
the valleys, so it maps each valley state, with squared norm $3$, onto
its partner.  The flat multiplet at $\Delta_{\rm flat}$ receives no
weight from any pattern.

Because the pairing maps each valley onto a single eigenvector and
back, the coupled problem closes on a two-level block per valley, and
the condensate energy has a closed form.  It is
$-\sqrt6\,t-\sqrt{t^2+3\lambda^2}$ for $\alpha$ and
$-\sqrt6\,t-\sqrt{t^2+6\lambda^2}$ for $\delta$, compared with
$-\sqrt{(1+\sqrt6)^2t^2+\lambda^2}$ for the on-site term of $\beta$
[Eq.~\eqref{eq:z2spectrum}].  These values reproduce the exact
diagonalization of the paired chargon Hamiltonian on the $12\times12$
torus to $10^{-14}$.  For the on-site term of $\beta$ the four valleys
remain the nondegenerate minimum of the paired one-body problem for
all $\lambda$.  For the nearest-neighbor pattern of $\alpha$ they
remain so for $\lambda<0.85\,t$ [in closed form
$\lambda_c=t\sqrt{(3+\sqrt{33})/12}$, beyond which the minimum moves
away from the valleys continuously], and for $\delta$ for
$\lambda<0.32\,t$.  Beyond that value the $\delta$ band minimum jumps
onto six degenerate points of the magnetic Brillouin zone, $\Gamma$,
$Y$, $S$, $(\tfrac12,0)$ and $(0,\pm\tfrac14)$.  These fall into three
classes under the projective translations [$\Gamma\sim Y$,
$(\tfrac12,0)\sim S$ and
$(0,\tfrac14)\sim(0,-\tfrac14)$] and all lie on the lines
$\Gamma$--$Y$--$S$ of Fig.~\ref{fig:bands}(c) and
$\Gamma$--$(\tfrac12,0)$, along which the lowest band is
dispersionless at $\lambda=0$.  The $\delta$ panel of
Fig.~\ref{fig:descendants}, drawn at $0.4\,t$, lies beyond this point,
and there the lowest band at $S$, $Y$ and $\Gamma$ lies below the
valley.

For the $\gamma$ state the unique symmetric second-neighbor pattern
annihilates the four valley states and their
partners at $(\sqrt6-1)t$, though not the rest of the lowest band.
The kernel of the pattern is a doubly degenerate flat zero band.  The
valley state therefore remains an exact eigenvector of the paired
one-body Hamiltonian at $-(1+\sqrt6)t$, and it is the band minimum for
$\lambda<0.56\,t$.  No $B_+$ admixture is generated at any order in
$\lambda$.  The annihilation does not remove the uniform pair channel
of $\gamma$.  On a condensate with a uniform pair component
(Sec.~\ref{sec:pairingmap}) the composite of Eq.~\eqref{eq:elpair}
retains a nonzero $q=0$ component transforming as $B_2$.  Only its
bond-summed ($A_1$) part vanishes, as it does for $\alpha$ ($B_1$) and
$\delta$ ($A_2$).  The second route to a physical pair amplitude,
through the $\bar B_+\bar B_-$ term of Eq.~\eqref{eq:pairop}, which the
other three descendants possess (Sec.~\ref{sec:pairingmap}), is absent
for $\gamma$.  This protection holds only for the pure
valley condensate.  On-site repulsion admixes non-condensing states
(Sec.~\ref{sec:phasediagram}), which the pattern does not annihilate,
so an admixture of order $U\lambda$ appears in a density-modulated
condensate.

Compare the square-lattice formula for the induced electron pairing
[Eq.~(3.7) of Ref.~\cite{Chatterjee2016}].  Its cross term is the
product of the two chargon components times the spinon hopping.  The
kagome theory realizes the limit in which this term vanishes on the
DSL condensate itself, because $\langle B_+\rangle=0$ at $\lambda=0$.
It returns only through the $\lambda$-induced $B_+$ admixture, via the
$\bar B_+\bar B_-$ term of Eq.~\eqref{eq:pairop}, which multiplies the
spinon hopping and is absent for $\gamma$ (Sec.~\ref{sec:pairingmap}).
Every channel is then linear in the spinon pairing, so
superconductivity here requires a paired $\Z_2$ parent and does not
follow automatically from confinement.

\subsection{Pairing-symmetry map and pair-density waves}
\label{sec:pairingmap}

Table~\ref{tab:pairing} lists the superconducting channels of the four
descendants in terms of the physical electron pair operator.
Equation~\eqref{eq:fusion} gives
$c_{{\bm i}\uparrow}c_{{\bm j}\downarrow}
=-\bar B_{{\bm i},-}\bar B_{{\bm j},-}
f^\dagger_{{\bm i}\downarrow}f^\dagger_{{\bm j}\uparrow}+\cdots$.  The
omitted terms are proportional either to $\bar B_+\bar B_+$ times the
spinon pair operator, where the product $\bar B_+\bar B_+$ is of
second order in $\lambda$, or to $\bar B_\pm\bar B_\mp$ times the
spinon hopping.  The latter is fed by the $B_+$ admixture of
Sec.~\ref{sec:z2spectra} at the same order in $\lambda$ as the
displayed term.  Its amplitude relative to the displayed term is of
the order of the admixture amplitude $|B_+|/|B_-|$ divided by the
pairing-to-hopping ratio $\lambda/t$ of the strongest pairing bond.
This quotient, the admixture amplitude per unit $\lambda/t$, is
$(\sqrt6-1)/10=0.145$ for the on-site channel of $\beta$,
$\sqrt3/2=0.87$ for $\alpha$, $\sqrt6/2=1.22$ for $\delta$ and zero
for $\gamma$.  The hopping term carries the same $\Gamma$-point
irreducible representation as the displayed term ($A_1$ for $\beta$
and $B_1$ for $\alpha$, while for $\delta$ its uniform part vanishes
identically because the $A_2$ irrep is forbidden on nearest-neighbor
bonds).  The channel assignments below are unaffected,
although the term is small only for $\beta$.  With only $B_-$
condensed, the electron pair amplitude that carries the classification
in a $\Z_2$ descendant is therefore
\begin{equation}
\Delta^{\rm el}_{ij}=-\bar B_{{\bm i},-}\bar B_{{\bm j},-}\,
\big(\Delta^{\rm sp}_{ij}\big)^{*},
\label{eq:elpair}
\end{equation}
the symmetric chargon pair composite times the conjugate spinon pair
amplitude $\Delta^{\rm sp}_{ij}=\avg{f_{{\bm i}\uparrow}f_{{\bm j}\downarrow}}$.
Because the spinon pairing of all four descendants is translation
invariant, the electron pair momenta are those of the chargon
composites.  The ten symmetric composites $B_\eta B_{\eta'}$ decompose
into $\Gamma$ ($\times1$), the $M$ star ($\times3$) and the $K/2$ star
($\times6$), and the uniform member is $(B_1B_3+B_2B_4)/\sqrt2$
(Appendix~\ref{app:composites}).  Equation~\eqref{eq:elpair} is gauge
invariant and transforms under the space group by pure lattice
permutation.  On the space spanned by the ten composites times the
descendant's projective-symmetry-invariant spinon form factor, its
action coincides with the chargon pair representation dressed by the
signs that the form factor acquires under $C_6$ and $\sigma$.  The
uniform ($\Gamma$) channel is at most one-dimensional, so
the uniform pair channel of each descendant is a one-dimensional
irrep.  The state $\beta$ pairs in $A_1$ (extended $s$), $\alpha$ in
$B_1$ ($f$-wave), $\gamma$ in $B_2$ (the second $f$-wave), and
$\delta$ in $A_2$, so no chiral $d\pm id$ channel appears at this
order.

We have verified this exclusion, which follows from symmetry and
involves no energetics, on the $12\times12$ torus for every
symmetry-allowed spinon form factor of the four descendants through
third neighbors.
The $\Gamma$ sector transforms as $A_1$ ($\beta$), $B_1$ ($\alpha$),
$A_2$ ($\delta$) and $B_2$ ($\gamma$), the $M$ triplet as $F_3$, $F_1$,
$F_4$ and $F_2$, and the $E_1$ and $E_2$ multiplicities at $\Gamma$
vanish in every case.  At mean field, and at the level of the
composite of Eq.~\eqref{eq:elpair}, which is bilinear in the valley
amplitudes, no descendant of the kagome DSL can therefore produce a
two-component ($E_2$) uniform pair order parameter, and so no $d+id$
state.

For the $M$-star condensates of phase~I the statement holds beyond
the bilinear level, at every order in the condensate.  A
translation pair carries the crystal momenta $\QQ_1$ and
$\QQ_1+\mathbf{G}_2/2$ [Eq.~\eqref{eq:valleys}], so every uniform
charge-$2e$ composite $\bar B^{\,p}B^{\,q}$ with $p-q=2$ built from it
carries momentum $-2\QQ_1$ modulo $\mathbf{G}_2/2$, whose first
fractional coordinate is $-1/3$.  Since this is never a
reciprocal-lattice vector, the composite vanishes identically.

On the II$^-$ branch the statement holds only at bilinear order.  The
next composites have bidegree
$(3,1)$, a pair times a density, and contain a uniform doublet whose
$\Gamma$ sector is six-dimensional.  It decomposes as
$A_1+B_2+E_1+E_2$ undressed ($A_2+B_1+E_1+E_2$ for the $\alpha$ and
$\delta$ dressings).  On the $K/2$ crystal of Eq.~\eqref{eq:phase2min}
the crystal's stabilizer, a sixfold rotation combined with a
translation, forces the induced uniform composite to be a $C_6$
eigenvector with chiral eigenvalue.  It transforms as $E_2$ for the
descendants $\alpha$ and $\gamma$, whose form factors are odd under
$C_6$, and as $E_1$ for $\beta$ and $\delta$.

The pair-times-density composite has the form of the standard coupling
of intertwined-order Landau theory, in which a pair-density wave times
a charge density wave induces a uniform pair component.  Here the
coupling arises through the feedback of the condensate on the spinon
pairing, and its coefficient is
dynamical and of relative order $x$.  On the $K/2$ branch a small
chiral $d\pm id$-like admixture is therefore allowed for $\alpha$ and
$\gamma$, and the exclusion there is a leading-order statement.  In
the $\alpha$ state the $B_1$ uniform amplitude is carried by the
nearest-neighbor form factor.  It vanishes identically on the
across-hexagon third-neighbor form factor, since $B_1$ is odd under
the $C_2$ that exchanges the two ends of such a bond.

The frequently proposed kagome $d+id$ state is thus ruled out for
every descendant of the DSL at leading order, and at any order on the
$M$-star condensates.  On the triangular lattice, by contrast, uniform
$d\pm id$ is allowed by symmetry and is not selected energetically
\cite{Feuerpfeil2026}.

Other approaches give the same channel structure near $1/3$ doping.
Weak-coupling functional renormalization at the Dirac filling finds a
$B_{1u}$ $f$-wave leader with a $B_{2u}$ alternative \cite{Mazin2014},
and the same $B_{1u}$/$B_{2u}$ pair leads in a
random-phase-approximation study of $A$V$_3$Sb$_5$ at its van Hove
filling \cite{WuPairing2021}.  Strong-coupling tensor-network
calculations find that the surviving pairing is the $Q=0$,
sign-alternating nearest-neighbor structure of our $\alpha$ channel,
and that $d+id$ is never the ground state \cite{XuGuYang2024}.  These
are dynamical calculations, whereas the present projective
classification is a symmetry constraint that forbids a uniform $d+id$
at the level of Eq.~\eqref{eq:elpair} without ranking the allowed
channels.  The fRG $f$-wave of Ref.~\cite{Mazin2014} is a spin
triplet and the parton channels here are classified as singlets, so
the agreement with it concerns the spatial form factor only.

A Landau theory of three-component pair-density waves on the kagome
lattice and of their charge-$6e$ vestigial phase is developed in
Ref.~\cite{LinSongZhang2025}.  The $M$-point pair triplets of the four
descendants realize \emph{all four} three-dimensional irreps of the
extended point group ($\alpha\to F_1$, $\gamma\to F_2$,
$\beta\to F_3$, $\delta\to F_4$), in one-to-one correspondence with
the dressing table~\eqref{eq:etatriples}.  The states $\beta$ and $\delta$
carry \emph{odd-parity} ($C_2$-odd) $M$-PDWs.  The $K/2$ pair sextet
carries a mirror parity that likewise distinguishes the states
(Table~\ref{tab:pairing}).  The uniform pair component of the selected
condensates is fixed branch by branch by the exact identity following
Eq.~\eqref{eq:hubbardquartic}, in which the uniform composite
$(B_1B_3+B_2B_4)/\sqrt2$ enters.  On the unit sphere
$|B_1B_3+B_2B_4|=\tfrac12\big[(\sum_\eta n_\eta)^2-(\mathcal{P}-2\mathcal{X})\big]^{1/2}$.
It vanishes identically on all of phase~I and branch~II$^-$, and it
reaches its maximum value $\tfrac12$ only at the II$^+$ vertex
$(\mathcal{P},\mathcal{X})=(\tfrac12,+\tfrac14)$, represented by the
translation-symmetric pair condensate $(c_1,0,c_3,0)$.  On the
phase-I and II$^-$ arcs this condensate is never selected, and at the
projected on-site and nearest-neighbor directions $\theta_U$ and
$\theta_V$ it is the global maximum of the potential.  The condensates
selected there carry \emph{zero uniform pair component}.
Phase-I condensates produce pure $K/2$ pair-density waves, and II$^-$
condensates produce $K/2$ plus $M$ PDWs.  On the II$^+$ arc,
$\theta/\pi\in(1,\,1.63386)$, which covers $31.7\%$ of the one-angle
phase diagram of Fig.~\ref{fig:wheel}, the same condensate
$(c_1,0,c_3,0)$ is instead a global minimum.  The selected states
there carry the full uniform pair component,
$|B_1B_3+B_2B_4|=\tfrac12$, in addition to their finite-momentum pair
content.

On phase~I and branch~II$^-$ the superconductor descending from the
kagome DSL is a pair-density wave in the intertwined-order
sense \cite{Agterberg2020,Fradkin2015}, with no uniform Josephson
component at leading order.  The same conclusion was reached for
kagome metals in a large-$V$ patch model at intermediate coupling, in
which sublattice interference renders the on-site repulsion irrelevant
\cite{WuThomaleRaghu2023}.  Here it holds exactly on the condensate
manifold of those two branches.  On the triangular lattice,
finite-momentum pairing is selected energetically while uniform
pairing remains allowed \cite{Feuerpfeil2026}.  On kagome the uniform
channel is orthogonal at the quartic level to the condensates selected
on phase~I and branch~II$^-$, and it appears only on the II$^+$ arc.

The lowest-order \emph{uniform} pair composite that the
finite-momentum content generates depends on the condensate and
follows from the momenta of Table~\ref{tab:ph}.  On the II$^-$ crystal
the three $M$ components close at second order ($2M\equiv0$) and the
six $K/2$ components form three $(+Q,-Q)$ mates, all nonzero, so
$\Delta_Q\Delta_{-Q}$ is a uniform charge-$4e$ composite at second
order.  On the phase-I sphere the $-Q$ arms vanish identically, so
only one member of each mate is nonzero and no charge-$4e$ composite is
generated at second order.  The triple-$M$ orbit carries a $C_3$
triple of $K/2$ components with $Q_1+Q_2+Q_3=0$, whose product is a
uniform charge-$6e$ composite at third order.  This channel coincides
with the vestigial order identified in Ref.~\cite{LinSongZhang2025}.
The octic-selected uni-$M$
stripe carries two $K/2$ components differing by an $M$ vector, which
close only at sixth order (charge $12e$).

Phase-sensitive tunneling on KV$_3$Sb$_5$, read through a
Ginzburg--Landau analysis of its doping dependence, supports a
sign-changing pair-density wave at the full $2\times2$ $M$ star that is
\emph{primary}, not induced \cite{YanPDW2026}.  Nonmagnetic
pair-breaking suppresses the pair modulation while the charge order
persists and the uniform component strengthens, which is the opposite
of an induced-PDW response.  The accompanying analysis classifies PDW
order parameters by irreps of the same extended point group
$C_{6v}'''$ used here, and Table~\ref{tab:pairing} gives the complete
$F_1$--$F_4$ assignment.

Because KV$_3$Sb$_5$ retains a large uniform pair component
\cite{YanPDW2026}, the chargon-descendant prediction of no uniform
charge-$2e$ component at the quartic level is not tested there, and
the data in fact argue against it.  The agreement concerns only the
irrep classification of the $M$-star PDW.

The vortex- and edge-spectroscopy signatures computed for candidate
kagome pairings \cite{DingVortex2022} indicate what to look for,
channel by channel.  The states $\alpha$ and $\gamma$ correspond to the
six-pointed-star vortex states and orientation-selective flat Andreev
edge bands of the two $f$-waves, and $\beta$ to the featureless
Caroli--de~Gennes--Matricon ladder of extended $s$.  The chiral-$d$
signatures (gapped vortex core, Chern number two) belong to a channel
that is absent from the $M$-star condensates at any order and induced
only at relative order $x$ on the $K/2$ branch, for $\alpha$
and $\gamma$ (Sec.~\ref{sec:pairingmap}).

The correspondence is between pairing channels.  The spectra
themselves do not carry over, because those calculations assume
uniform pairing and the condensates selected here are pair-density
waves.  Their Bogoliubov--de~Gennes problem couples $\kk$ to
$-\kk+\QQ$ in a folded zone, so their vortex and edge spectra depend on
the modulation components, the normal dispersion and the magnetic
symmetries, and we do not compute them here.

\begin{table}[t]
\caption{Superconducting channels of the four $\Z_2$ descendants.
Columns give the $\Gamma$ irrep, the $M$-triplet parities and irrep,
and the $K/2$ mirror parity of the physical electron pair amplitude of
Eq.~\eqref{eq:elpair}.  The $K/2$ mirror parity refers to the
hexagon-centered mirror whose axis lies along a lattice vector, which
is the only mirror family that fixes a $K/2$ momentum.}
\label{tab:pairing}
\begin{ruledtabular}
\begin{tabular}{lcccc}
state & $\Gamma$ channel & $M$-PDW $(C_2,\sigma_v)$ & irrep & $K/2$ mirror \\
\colrule
$\beta$  & $A_1$ (ext.\ $s$) & $(-,+)$ & $F_3$ & $+$ \\
$\alpha$ & $B_1$ ($f$)       & $(+,+)$ & $F_1$ & $-$ \\
$\delta$ & $A_2$             & $(-,-)$ & $F_4$ & $-$ \\
$\gamma$ & $B_2$ ($f$)       & $(+,-)$ & $F_2$ & $+$ \\
\end{tabular}
\end{ruledtabular}
\end{table}

\section{Chiral descendant}
\label{sec:csl}

\subsection{Ansatz and identification}
\label{sec:cslansatz}

The chiral neighbor of the DSL is the Kalmeyer--Laughlin chiral spin liquid
stabilized by further-neighbor or explicitly chiral couplings
\cite{KalmeyerLaughlin1987,Gong2014,Bauer2014,Zhu2015,Wietek2015,Hu2015}.  Its parton ansatz is
the oriented flux family of Hu \emph{et al.}\ \cite{Hu2015},
\begin{equation}
[\theta_\triangle,\theta_{\hexagon};\theta_{abc},\theta_{acd}]
=[3\varphi_1,\ \pi-6\varphi_1;\ \pi-2\varphi_1-\varphi_2,\ 3\varphi_2],
\label{eq:cslfluxes}
\end{equation}
with nearest- and second-neighbor hoppings $t_1e^{i\varphi_1}$,
$t_2e^{i\varphi_2}$ (the site labels $a,b,c,d$, the orientation
conventions of $\varphi_1$ and $\varphi_2$ and the second-neighbor sign
pattern $\nu_{ij}$ are defined in Appendix~\ref{app:csl}).  The U(1)
DSL is the point $[0,\pi;\pi,0]$.

We have verified that this state is the chiral spin liquid ``CSL~C'' of
the Bieri--Lhuillier--Messio classification [No.~10 of Table~VIII in
Ref.~\cite{Bieri2016}; No.~15 of Table~I in Ref.~\cite{Bieri2015}].  It
carries the Kalmeyer--Laughlin projective class
$(\tau_\sigma,\tau_R)=(1,0)$, the only class that admits a nonzero
spinon Chern number \cite{Bieri2016}.  Our torus implementation
reproduces all four gauge-invariant fluxes of Eq.~\eqref{eq:cslfluxes},
including the mixed-triangle flux $\theta_{abc}$, which fixes the
otherwise arbitrary global sign of the second-neighbor pattern
(Appendix~\ref{app:csl}).

Symmetry also admits a purely imaginary diagonal (intra-hexagon)
hopping.  The constraint relating a directed bond to its reverse is
\emph{antilinear}, and its solution locks the diagonal phase to
$\beta_d=\pi/2$, as tabulated in Ref.~\cite{Bieri2016}.  In the
time-reversal-symmetric DSL the same bond is forbidden outright, so
$\beta_d$ is a $\mathcal{T}$-breaking order-parameter direction and
cannot be tuned freely.  With nearest-neighbor hopping only, the family
reduces to the uniform-flux ansatz $[\theta,\pi-2\theta]$ of Marston
and Zeng \cite{MarstonZeng} and Hastings \cite{Hastings2000}.  In the
monopole language the CSL is the DSL with the $\Gamma$-point singlet
mass condensed, and the resulting Chern--Simons term suppresses
monopole proliferation \cite{SongNC2019}.

\subsection{Chargon sectors: the two-valley theory and its boundary}
\label{sec:cslsectors}

\begin{figure}[t]
\includegraphics[width=\columnwidth]{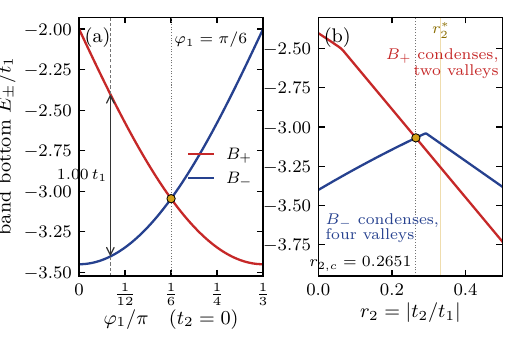}
\caption{Band bottoms of the two chargon sectors of the chiral ansatz,
in units of $t_1$.  The sector $B_+$ hops with Hu's amplitudes and
$B_-$ with minus their complex conjugate.  (a)~Dependence on
$\varphi_1$ at $t_2=0$.  The sectors are exactly degenerate at the
nearest-neighbor Kalmeyer--Laughlin point $\varphi_1=\pi/6$ (circle)
and split by $1.00\,t_1$ at Hu's $\varphi_1=0.0567\pi$ (dashed).
(b)~Dependence on $r_2=|t_2/t_1|$ at Hu's phases
$(0.0567\pi,-0.2807\pi)$.  The sectors cross at $r_{2,c}=0.2651$,
Eq.~\eqref{eq:crossing}.  The four-valley $B_-$ sector condenses below
the crossing and the two-valley $B_+$ sector above it.  The band marks
the spinon ratio $r_2^*\simeq0.33$ obtained by inverting the mean-field
gap of Ref.~\cite{Hu2015}.}
\label{fig:chiralsectors}
\end{figure}

The chiral chargon problem has a sector structure with no triangular
analog (Fig.~\ref{fig:chiralsectors}).  Consider first the
nearest-neighbor-only point $[\pi/2,0]$ ($\varphi_1=\pi/6$, $t_2=0$).
It was historically the first kagome CSL candidate, favored at mean
field but losing to the DSL after projection \cite{Ran2007}.  At this
point the two chargon sectors are \emph{exactly} degenerate.  The
$B_+$ and $B_-$ hoppings are the two
diagonal entries of $u_{ij}$ [Eq.~\eqref{eq:fusion}], $t_{ij}$ and
$-t_{ij}^{*}$ (Sec.~\ref{sec:parton}), so the $B_-$ hopping is minus the
complex conjugate of the $B_+$ hopping.  The sign reversal shifts the
flux of every odd loop by $\pi$ (Sec.~\ref{sec:bands}) and the
conjugation reverses every flux, so $B_-$ sees the fluxes
$[\pi-3\varphi_1,\,6\varphi_1-\pi]$, which is the same nearest-neighbor
family at $\varphi_1\to\pi/3-\varphi_1$.  At $\varphi_1=\pi/6$ these
coincide with the $B_+$ fluxes $[\pi/2,0]$.  The two sectors are then
U(1) gauge copies of one another (we construct the gauge
transformation explicitly on the torus, Appendix~\ref{app:csl}).  They
have identical spectra and band Chern numbers, band bottoms at
$-3.045474876\,t_1$, and four valleys each at $\QQ_{1\dots4}$.  The
NN-only Kalmeyer--Laughlin point is a $4{+}4$-valley theory
of a single chirality and has no two-valley description.  (At mean
field the $[\pi/2,0]$ ansatz itself has gapped Chern bands
\cite{Wietek2015}.)

Moving $\varphi_1$ away from $\pi/6$ splits the sectors, by
$1.00\,t_1$ at Hu's $\varphi_1=0.0567\pi$ with $t_2=0$
[Fig.~\ref{fig:chiralsectors}(a)].  Second-neighbor hopping then drives
them to cross.  At Hu's optimized phases
$(\varphi_1,\varphi_2)=(0.0567\pi,-0.2807\pi)$ the global chargon
minimum switches from the four-valley $B_-$ sector to the $B_+$ sector
at
\begin{equation}
r_{2,c}\equiv|t_2/t_1|_c=0.265082\ldots,
\label{eq:crossing}
\end{equation}
beyond which the $B_+$ bottom has \emph{two} nondegenerate valleys
[Fig.~\ref{fig:chiralsectors}(b)].  Their separation, a gauge-invariant
quantity, equals the spinon internode vector (an $M$ vector), and in
the gauge of Appendix~\ref{app:conventions} they sit at the spinon
Dirac momenta.  (On the square lattice chargon minima and spinon nodes
coincide automatically \cite{Bonetti2026}.  On the kagome DSL they do
not, and the coincidence reappears only in the chiral regime.)

With $B_+$ condensed instead of $B_-$, the bookkeeping of
Sec.~\ref{sec:parton} is mirrored.  Equation~\eqref{eq:fusion} gives
$c_{{\bm i}\sigma}=\bar B_{{\bm i},+}f_{{\bm i}\sigma}$, so the electron
is the spinon itself, not its particle--hole conjugate.  The unbroken
diagonal charge is $Q_{\rm EM}+Q_g$, under which $B_+$ is neutral and
$B_-$ carries charge $2$.  The charge-$2e$ object is again the
gauge-neutral composite $\avg{B_+B_-}$, which now reduces to
$B_-\avg{B_+}$ and vanishes as long as $B_-$ stays gapped.  In the
emergent Gauss law $(n_f-1)+n_+-n_-=0$ (Sec.~\ref{sec:parton}) the
condensate enters with the opposite sign.  A $B_+$ condensate of
density $x$ gives the spinon filling $n_f=1-x$ for hole doping and
$1+x$ for electron doping, which is the electron density itself, since
the electron is now the spinon.  The no-superconductivity argument of
Sec.~\ref{sec:bands} carries over with the roles of the two sectors
exchanged.

Hu \emph{et al.}\ provide optimized \emph{phases} at the coupling point
$(J_2,J_3)=(0.5,0.6)$ of their $J_1$--$J_2$--$J_3$ model (in units of
$J_1$), and their quoted mean-field gap can be inverted against our
bands.  That gap is $\simeq0.6$ at the optimum in their normalization
$\mathrm{Re}\,t_1=1$, or $0.6\cos\varphi_1=0.591$ in units of
$|t_1|$.  It is a $16\times16\times3$-cluster value, which we take as
the bulk gap (grid-resolved readings give larger values,
Appendix~\ref{app:csl}).  Inverting it yields a \emph{spinon} ratio
$r_2^*\simeq0.334$, which lies beyond $r_{2,c}$.

Using this value as the chargon ratio assumes that the chargon
second-neighbor amplitude is proportional to the spinon variational
one, although the two are distinct.  The chargon hopping is
$t^{\rm electron}_{ij}\avg{f^\dagger_if_j}$ (Sec.~\ref{sec:parton}), so
an electron model with nearest-neighbor hopping alone gives the chargon
no second-neighbor amplitude, however large the spinon $\xi_2$.  The
doped state then lies below the crossing, in the four-valley regime.

The variational optimum of Ref.~\cite{Bieri2016} for the same state is
$|\xi_2/\xi_1|=0.25/0.75\simeq0.33$ at
$(J_1,J_2,J_d)\simeq(0.63,0.13,0.24)$, that is, $J_2/J_1\simeq0.21$ and
$J_d/J_1\simeq0.38$, where $J_d$ is the hexagon-diagonal coupling that
Hu \emph{et al.}\ call $J_3$.  This optimum confirms the magnitude of
the second-neighbor amplitude but not the regime.  Its phases,
$(\varphi_1,\varphi_2)\simeq(0.08\pi,-0.61\pi)$ in the conventions
fixed by the $\theta_{abc}$ flux, coincide with the
$(J_2,J_3)=(0.2,0.4)$ optimum of Hu \emph{et al.},
$(0.0869\pi,-0.6150\pi)$.  At these phases the sectors do not cross.
The four-valley $B_-$ sector condenses for all $r_2\le0.4$ (margin
$0.28\,t_1$ at $r_2=1/3$), and a two-valley $B_-$ sector condenses
beyond.  The same holds at the other tabulated optima of
Ref.~\cite{Hu2015}.  At $r_2=1/3$ the $B_-$ sector condenses at
$(0.5,0.7)$ (four valleys, margin $0.09\,t_1$) and at $(0.2,0.3)$ (two
valleys at the Dirac momenta, margin $0.29\,t_1$).

Granting the proportionality of chargon and spinon amplitudes, and
within the nearest- and second-neighbor model at the $(0.5,0.6)$
optimum of Hu \emph{et al.}, the doped Kalmeyer--Laughlin state would
therefore be a two-valley $B_+$ chargon theory.  We use the
representative value $r_2=0.35$ [bottom $-3.308\,t_1$] for it.  The
optimized $\varphi_2$ varies by a factor of two across the chiral
phase, however, and at the other tabulated optima, including Bieri's
point, the doped state condenses the $B_-$ sector instead, with four
or two valleys as listed above.  We leave that regime aside.

Below the crossing the condensing sector is the four-valley $B_-$
sector.  A scan of $r_2$ from $0$ to $0.265$ in steps of $0.01$ at
Hu's phases, refined near $r_{2,c}$, shows that its minima stay at the
valleys $\QQ_{1\dots4}$ of Eq.~\eqref{eq:valleys}, as they do at the
crossing itself.  The four-valley construction of
Secs.~\ref{sec:psg} and \ref{sec:phases} applies to this sector once
the projective symmetry group is replaced by that of the chiral
ansatz, in which time reversal and the mirror survive only in the
combination $\sigma\Theta$, and once the $\Z_2$ pairing channels of
Sec.~\ref{sec:z2} are dropped, since the pure hopping ansatz lacks
them.  The Landau invariants of that four-valley chiral theory would
have to be rederived.

The crossing is a coincidence of two band bottoms at commensurate
momenta, the $B_+$ pair at the spinon Dirac points $(\tfrac12,\tfrac14)$
and $(\tfrac12,\tfrac34)$ in units of the magnetic reciprocal vectors
$\mathbf{G}_{1,2}$ and the $B_-$ quartet at the valleys $\QQ_{1\dots4}$
of Eq.~\eqref{eq:valleys}, so it is grid independent.  Bisection on $k$
grids of $N_g=12$, $24$, $48$, $96$ and $192$ points per magnetic
reciprocal vector returns the same value $r_{2,c}=0.2650822340$ on
every grid, and a continuum refinement of both bottoms agrees to
$10^{-10}$.

Because the phases are Hu's quoted optimum, the variations below
measure parameter sensitivity and are not an uncertainty estimate.
Scaling $\varphi_1$ or
$\varphi_2$ separately by $\pm10\%$ moves $r_{2,c}$ within
$[0.255,\,0.278]$, with
$\partial r_{2,c}/\partial\ln\varphi_1=-0.105$ and
$\partial r_{2,c}/\partial\ln\varphi_2=+0.115$.  The two nearly cancel
under a common rescaling.  Only a $+20\%$ change in
$\varphi_2$ pushes the crossing to $0.337$, marginally above $r_2^*$.
At $r_2=0.35$ the $B_+$ sector remains the condensing sector under
every one of these variations, with a sector margin of $0.174\,t_1$ at
Hu's phases.  These variations are small compared with the spread of the
optimized phases across the chiral phase of Ref.~\cite{Hu2015}
($\varphi_2$ from $-0.28\pi$ to $-0.62\pi$ over the four tabulated
optima), over which the condensing sector at $r_2=1/3$ changes as stated
above.

These variations stay within the nearest- and second-neighbor model.
The crossing is far more sensitive to the purely imaginary
intra-hexagon hopping that symmetry also permits
(Sec.~\ref{sec:ansatz}), whose phase is locked at $\beta_d=\pi/2$.  We
write its amplitude as the signed coefficient $r_d=t_d/t_1$ of that
pattern.  The pattern is fixed by the realified symmetry constraint,
whose solution is one dimensional and unimodular with bond phases
$\pm\pi/2$ (Appendix~\ref{app:csl}).  At $r_2^*=0.334$ the sector
margin is $0.155\,t_1$ for $r_d=0$.  It is linear in $r_d$ and
vanishes at $r_d=+0.0388$, while the opposite sign enlarges it beyond
$|r_d|=0.4$.  The diagonal link is a $\mathcal{T}$-odd direction in a
state that already breaks $\mathcal{T}$, so it is generated, not
tuned, and its sign is not free.  The half-filled spinon energy
responds linearly at $r_d=0$ and, at every $r_2$ we have examined,
decreases in the direction that \emph{closes} the margin.  Its
magnitude follows from the same decoupling,
$r_d=(J_d/J_1)(c_d/c_1)$, where $c_d$ and $c_1$ are the projections of
$\avg{f^\dagger_if_j}$ on the diagonal and nearest-neighbor patterns.
At $r_2^*$ this gives $|c_d/c_1|=0.200$.  Iterated to a fixed point, it
gives $r_d=0.048$, $0.107$ and $0.207$ for $J_d/J_1=0.20$, $0.38$ and
$0.60$, the last being the $J_3$ of Hu's own optimum.  All three
exceed $0.0388$.

Together with the distinction between chargon and spinon links made
above, this leaves the two-valley placement without support.  If the
chargon links are proportional to the spinon links, the same
proportionality carries the diagonal link, which closes the sector
margin.  If they are not, then in an electron model with
nearest-neighbor hopping alone the chargon acquires neither a
second-neighbor nor a diagonal amplitude, and the state lies below the
crossing.  We therefore do not claim that the doped chiral state
realizes the two-valley regime, although our analysis does not exclude
it.  Our estimate is unprojected mean field, whereas Hu's is projected
and variational, and the threshold is computed at fixed $r_2$, whereas
a complete treatment would determine $r_2$ and $r_d$ together.
Section~\ref{sec:csllandau} develops the Landau theory of the
two-valley regime, $r_2>r_{2,c}$, as a conditional result.

\subsection{Two-valley Landau theory: exact $\mathbb{CP}^1$ moduli and selection by loop terms}
\label{sec:csllandau}

We now assume that the two-valley regime applies.  The two surviving
valley fields $B_{1,2}$ carry the projective representation
$T_1=-i\sigma^2$ (valley swap), $T_2=-i\sigma^3$, and a
$1/\sqrt2$-mixing rotation with $R^6=-1$, and the antiunitary
$\sigma\Theta$ closes within the doublet.  The invariant theory of this
doublet sharpens the moduli-space theorem of Sec.~\ref{sec:moduli}:
\begin{enumerate}[label=(\roman*),leftmargin=*,itemsep=0pt,topsep=2pt,parsep=0pt,partopsep=0pt]
\item There is \emph{exactly one} quartic invariant, $(n_1+n_2)^2$, so
the entire condensate $\mathbb{CP}^1$ is degenerate at quartic order
for arbitrary couplings.
\item At sextic order the invariant space is exactly
$\mathrm{span}\{(B^\dagger B)^3,\,xyz\}$ in Bloch-sphere coordinates,
a single tetrahedral anisotropy with extrema $\pm1/(3\sqrt3)$ on two
four-state tetrahedral orbits.  The anisotropy is allowed because
$\sigma\Theta$ acts as a reflection that preserves $xyz$, whereas in
the $\mathcal{T}$-symmetric theory the $90^\circ$ rotations and the
inversion remove it.
\item The on-site quadratic and cubic moments $\sum_r\rho_r^2$ and
$\sum_r\rho_r^3$ are \emph{exactly constant} on $\mathbb{CP}^1$, so no
on-site term through sextic order can select a state, although the
site-density pattern itself varies over the sphere.  The three
$M$-point density amplitudes are the three components of the Bloch
vector $\hat n=(x,y,z)$ of the doublet,
$|\rho(M_a)|=|\hat n_a|\bar\rho/3$, where
$\rho(\mathbf{k})=N^{-1}\sum_r\rho_r\,e^{-i\mathbf{k}\cdot\mathbf{r}}$
and the three $M$ points are labeled accordingly.  The axis points are
therefore uni-$M$ stripes with three density levels
$\{0,1,2\}\bar\rho$ on $72$, $288$ and $72$ of the $432$ sites.  The
tetrahedral points $\hat n=(\pm1,\pm1,\pm1)/\sqrt3$ are triple-$M$
crystals with two levels, $(1\mp1/\sqrt3)\bar\rho=0.42265\,\bar\rho$
and $1.57735\,\bar\rho$, each on half the sites ($\sum_r\rho_r^4$
varies by $7\%$ between them).
\end{enumerate}

Intersite loop terms resolve the degeneracy.  The simplest is the
triangle-loop sextic
$\sum_\triangle\mathrm{Re}\prod_{(ij)\in\triangle}\bar B_i t_{ij}B_j$.
For a condensate,
$\prod_{(ij)\in\triangle}\bar B_it_{ij}B_j
=\big(\prod_{(ij)\in\triangle}t_{ij}\big)\prod_{i\in\triangle}\rho_i$,
so this term reduces to the flux-weighted three-site density product
$t_1^3\cos\theta_\triangle\sum_\triangle\rho_i\rho_j\rho_k\propto(1+xyz)$.
Selection in the doped chiral state is made by intersite density
correlations on the triangles, which no on-site moment through cubic
order can detect.  The two extremal states are triple-$M$ chiral
chargon crystals that are density conjugates of one another
($\rho\to2\bar\rho-\rho$, which exchanges high- and low-density sites
and keeps the same two-level histogram) with distinct loop-current
circulation patterns.  As for phase II$^-$, the sign of the sextic
determines the sign of the triple-$M$ modulation.  The weak-coupling
loop-current theory of the kagome metals also arrives at a four-state
degenerate manifold at its leading order \cite{ZhanLCO2026}.

The symmetric pair composites $B_\eta B_{\eta'}$ of the two-valley
theory exhaust the \emph{full $M$ star}, with \emph{no} uniform
($\Gamma$) channel at quadratic order.  Unlike those of the $\Z_2$
descendants, these chargon composites do not produce electron
pairing.  The chiral ansatz is a pure hopping ansatz with
invariant gauge group U(1), so it carries no spinon pairing, while the
induced electron pair amplitude is linear in the spinon pairing [the
$\bar B_+\bar B_+\Delta^{\rm sp}$ term of Eq.~\eqref{eq:pairop}, the
$B_+$-sector counterpart of Eq.~\eqref{eq:elpair}].  Every channel of
the electron pair field, uniform and finite-momentum alike, therefore
vanishes identically for the doped chiral state at this order,
provided that the second gauge sector stays gapped, as in
Sec.~\ref{sec:bands}.  The $M$-star pair composites then do not
describe a pair-density-wave superconductor.

The gap that enforces this is now the sector margin $E_--E_+$, which
plays the role of $\Delta_{\rm flat}$.  It is $0.15\,t_1$ at
$r_2^*=0.33$ and $0.17\,t_1$ at $r_2=0.35$
(Sec.~\ref{sec:cslsectors}), closes at $r_{2,c}$, and shrinks to
$0.01\,t_1$ at $r_2=0.35$ under the $+20\%$ variation of $\varphi_2$
quoted there.  At and near the crossing the two sectors form a
near-degenerate pair, and whether both condense is then determined by
the inter-sector quartic couplings, which we have not computed.  If
they do, the $\chi$ terms of Eq.~\eqref{eq:pairop} yield an electron
pair amplitude without any spinon pairing, at the momenta of the mixed
composites $\bar B_+\bar B_-$, as in the doublet mechanism of the
square and triangular lattices \cite{Christos2023,Feuerpfeil2026}.
The conclusions below therefore hold for $r_2$ well above $r_{2,c}$,
where the $B_-$ sector is gapped by the quoted margin.

The particle--hole content of the condensate survives, in the form of
the triple-$M$ chiral chargon crystals of (iii).  These form an
$M$-star charge and loop-current crystal, selected by the triangle
density correlations of the loop sextic.  At the level of the chargon
composites this parallels the triangular chiral case, where the uniform
$d\pm id$ bilinears likewise vanish for the chiral spin liquid and only
half of the $K/2$ pair-density-wave channels survive
\cite{Feuerpfeil2026}.  The two lattices differ in the finite-momentum
content of the surviving composites, which is the full $M$ star here
and the $K/2$ star there.  Uniform chiral superconductivity, the most
commonly assumed outcome of doping a Kalmeyer--Laughlin state, is thus
not generated by the chargon condensate within the Higgs framework, on
kagome as on the triangular lattice \cite{Feuerpfeil2026}.

A route to uniform chiral superconductivity outside the Higgs framework
remains a distinct, competing scenario.  The chargons may form a
bosonic integer quantum Hall state, which is itself a $d+id$
superconductor \cite{SongVishwanathZhang2021,SongZhang2023}, and anyon
superconductivity in doped chiral topological states is under active
study
\cite{Ko2009,ShiSenthil2025,Divic2025,Pichler2026,Kuhlenkamp2025,Chen2026,Feuerpfeil2026}.
Unbiased numerics support that route on another lattice.  Density-matrix
renormalization and determinantal quantum Monte Carlo on the lightly
doped triangular Hofstadter--Hubbard model at a quarter flux quantum per
triangle find a chiral superconductor with quantized spin Chern number
two when either the integer quantum Hall parent or the chiral spin
liquid parent is doped.  Its pairing correlations decay more slowly
than the period-four charge stripes that accompany them
\cite{Chen2026}, and the companion study of
Ref.~\cite{Kuhlenkamp2025} reaches the same conclusion.  The kagome
$t$--$J$ numerics of Sec.~\ref{sec:strongcoupling} instead find an
insulating holon crystal without superconductivity, but the two
settings differ in lattice, in flux and in the chirality of the
parent.

Within the Higgs framework, at this order and away from the sector
crossing, a doped kagome CSL is therefore predicted to be a
nonsuperconducting $M$-star charge and current crystal.

The emergent gauge field of the chiral state differs from the Maxwell
field of the DSL.  Both spin species occupy bands of Chern number one,
which is the Kalmeyer--Laughlin class identified in
Sec.~\ref{sec:cslansatz}, so integrating out the spinons leaves a
Chern--Simons term at level two, and a single chargon of unit gauge
charge carries statistical angle $\pi/2$.  The Landau theory of
Sec.~\ref{sec:csllandau} describes the mean-field condensation of
these semionic chargons, the chiral-metal branch of the analysis of
Song, Vishwanath and Zhang \cite{SongVishwanathZhang2021}.
The other branch of that analysis, a bosonic integer quantum Hall state
of the chargons that descends to a $d+id$ superconductor, is the
competing scenario mentioned above.  Any Higgs phase built on the
chiral state also inherits the spinon Chern number, because the
condensate locks the emergent gauge field to the electromagnetic one.
The $M$-star crystal of the two-valley regime would be a
chiral metal with an intrinsic anomalous Hall response in addition to
the loop currents of (iii).

The one existing numerical study of a doped kagome chiral spin liquid
points the same way.  Peng \emph{et al.}\
\cite{Peng2021} doped the $t$--$J$--$J_\chi$ model by density-matrix
renormalization on six- and eight-leg cylinders at $\delta=1/24$ to
$1/12$.  The model has nearest-neighbor hopping only, with $t=3$,
$J=1$ and $J_\chi=0.5$, deep in the chiral phase.  They found a stripe
crystal of spinless holons with one hole per emergent cell, no
doping-induced magnetic order, and pair correlations decaying
exponentially within one unit cell, an insulating charge crystal
without superconductivity like the doped $t$--$J$ liquids of the same
work (Sec.~\ref{sec:strongcoupling}).  The result is consistent with
the nonsuperconducting outcome derived here.  The crystal found
there is, however, a stripe and not a triple-$M$ crystal, and a
nearest-neighbor-only model gives the chargon no bare second-neighbor
hopping, so the model lies in the four-valley regime below the sector
crossing (Sec.~\ref{sec:cslsectors}), outside the two-valley theory
treated here.

\section{The doped case}
\label{sec:doped}

\subsection{Finite density: the Gross--Pitaevskii regime}
\label{sec:gp}

Doping the DSL at fixed fractionalization is expected to yield a
fractionalized Fermi liquid (FL$^*$)
\cite{SenthilSachdevVojta2003,Chatterjee2016,Bonetti2026}.  This
expectation is taken over from the square-lattice construction.  It
presupposes the binding energy of a chargon to a spinon in the presence
of the gapless compact $U(1)$ gauge field of the DSL and the scaling
dimensions of the monopoles once Fermi pockets are present, and we
establish neither here.  In this state the chargons remain gapped, and
most of the doped charge is carried by Fermi pockets of chargon--spinon
bound states, which are the electron-like quasiparticles.  The
condensate analysis of the preceding sections does not apply to it.

The Luttinger count depends on neither of these inputs.  The
pocket volume counts the doped density $p$, i.e.\ $3p$ holes per
three-site cell.  A conventional metal at the same filling would instead
have a large Fermi surface in the half-filled middle kagome band, with
$1+3p$ holes per cell ($1/3+p$ per site).  The rest of the charge is
carried by the fractionalized insulating background.

The Higgs transition out of FL$^*$ is reached when the chargon gap
closes.  At nonzero density the Landau theory acquires a first-order time
derivative \cite{Feuerpfeil2026}, while the potential of
Eq.~\eqref{eq:potential} is unchanged.  The dynamical exponent is then
$z=2$, so the two-loop check of Appendix~\ref{app:betas}, made for the
relativistic theory of the undoped transition, does not carry over.  The
protection of the Hubbard ray (Sec.~\ref{sec:hubbardray}) does carry
over, because it follows from symmetry alone.  The charge carried by the
condensate is the Noether density of the physical $U(1)$.  Besides the
usual derivative contribution, it contains a term
$\kappa\sum_\eta n_\eta$, with $\kappa$ set by the first-order time
derivative, so it is not $\sum_\eta n_\eta$ itself.  Here $\kappa$ is a
nonuniversal coefficient determined by the microscopic dynamics and, in
general, by the chemical potential.  The doping constrains the sum of the
pocket and condensate charges, not $\sum_\eta n_\eta$.

The opposite limit, in which all of the doped charge resides in the
chargon condensate (the holon picture of Ref.~\cite{Ko2009}), is
described by the lattice Gross--Pitaevskii functional of the chargon
hopping problem,
\begin{align}
E[B]&=\sum_{\langle ij\rangle}t\,s_{ij}\bigl(\bar B_{{\bm i},+}B_{{\bm j},+}
-\bar B_{{\bm i},-}B_{{\bm j},-}\bigr)+{\rm H.c.}\nonumber\\
&\quad+\frac{U}{2}\sum_r\rho_r^2\,,\qquad
\rho_r=|B_{r,+}|^2+|B_{r,-}|^2,
\label{eq:gpfunctional}
\end{align}
where the hopping term is Eq.~\eqref{eq:decoupling} with $t$ the chargon
hopping.  We minimize it on the $12\times12$ torus with the link signs
$s_{ij}$ of the $[0,\pi]$ ansatz, keeping both gauge components, at fixed
total chargon number $\sum_r\rho_r=xN$.

Fixing the total chargon number does not impose the site-resolved Gauss
law, which is then violated at the order of the modulation itself.
Imposing it suppresses the density modulation by factors of $18$ and
$44$ at $x=1/12$ and $1/6$, and the physical bond currents by the smaller
factors $1.4$ to $7.0$ across $x=1/36$ to $1/6$.  The complex valley
coefficients, and with them the point on the moduli space, are unchanged
to the figures we resolve.  At $x=1/36$ and $U_B/t_B=4$ the constrained
theory lies inside a miscibility gap, which is absent at $U_B/t_B=16$ and
at the physical $60$ to $80$.  There its homogeneous state is a
long-wavelength inhomogeneity whose root-mean-square contrast is only
$1.2$ times below the unconstrained one (Appendix~\ref{app:gausslaw}).

The symmetry classification of each candidate crystal survives this
suppression, because it follows from the projective symmetry group and
the point on the moduli space independently of the minimization.  The
densities reported in this section and the next do not survive it.  They
are properties of an unconstrained trial background and bound the
physical modulation from above, but they do not give its local values.

The functional is first order in time, so the chargon number is its
conserved charge and the time-derivative term plays no role in the
static minimum.  Rescaling
$B\to\sqrt{x}\,B$ shows that the normalized problem depends on $U/t$ and
$x$ only through $Ux/t$.  The static potential of the projected Landau
theory, Eq.~\eqref{eq:potential}, is the restriction of
Eq.~\eqref{eq:gpfunctional} to the four condensation valleys.  The full
lattice functional differs from it by the admixture of the
non-condensing chargon bands, and eliminating these bands at second
order gives the microscopic sextic of Secs.~\ref{sec:phasediagram} and
\ref{sec:phaseII}.  That admixture selects the state, because for
on-site repulsion alone the projected quartic sits on the I/II$^-$
boundary.  The sextic is larger in magnitude on the
II$^-$ level set than on the pair sphere, by the ratio $0.456N$ to
$0.088N$ of the resolvent elements, and it selects the unbalanced
four-valley $K/2$ crystal of Eq.~\eqref{eq:phase2min}.

For the charge bookkeeping, the unimodular rotor of
Sec.~\ref{sec:parton}, $|B_+|^2+|B_-|^2=1$ on every site, is relaxed
in the standard way to a soft complex doublet
\cite{Chatterjee2016,Christos2023,Feuerpfeil2026}, and $x$ is the doped
charge per site carried by that doublet.  With only $B_-$ condensed the
fusion rule of Eq.~\eqref{eq:fusion} gives
$c_{\uparrow}=\bar B_-\,f^{\dagger}_{\downarrow}$, so the electron is
the particle-hole conjugate of the spinon.  The doped charge resides in
the chargon number, $\sum_rB_r^\dagger B_r=xN$, and the Gauss law
$(n_f-1)+n_+-n_-=0$ of Sec.~\ref{sec:parton} then fixes the spinon
filling at $n_f=1+x$, or $1-x$ for the conjugate, electron-doped
quanta.  The electron-like quasiparticles carry the doped density
$1\mp x$.  When the condensate is projected onto the four low-energy
valleys this reads $\sum_\eta n_\eta=xN$, with the valley Bloch states
normalized to unity (Sec.~\ref{sec:parton}).  In the full lattice
minimization the interaction-induced higher-band admixture means that
the four-valley weight $\sum_\eta n_\eta/(xN)$ need not remain equal to
unity.

We have minimized this functional for the full chargon doublet, as a
mean-field model of the fully condensed regime, at $x=1/12$ and at
$x=1/3$ (the ``Dirac metal'' filling targeted by Ga substitution
\cite{Mazin2014,Puphal2019} and by Li intercalation \cite{Kelly2016} of
herbertsmithite), with on-site repulsion $U/t=0.5$--$16$.

(i)~Across both fillings and the full interaction range, direct
minimization of the complete chargon doublet converges to a purely $B_-$
condensate.  The same holds once the site-resolved Gauss law is imposed.
Then $\avg{B_+}=0$ is a minimum of the constrained functional, by a
margin that stays positive at every doping and coupling tested
(Appendix~\ref{app:gausslaw}), in the functional that keeps the
number-conserving part of the fusion rule alone.  For $Ux\le1/3$ the minimizer is the
sextic-selected $K/2$ crystal on the phase-II$^-$ manifold.  Its valley
projection gives $(\mathcal{P},\mathcal{X})=(\tfrac12,-\tfrac14)$ to
$10^{-4}$, while the four-valley weight decreases from $0.999$ to $0.981$
as the interaction-induced higher-band admixture grows.

For $Ux\ge2/3$, continuation from strong coupling finds pure-$B_-$
states below the $K/2$ crystal.  They lie lower marginally at $Ux=2/3$
($8\times10^{-5}\,t$ per chargon) and by $2.4\times10^{-3}$ to
$6.9\times10^{-3}\,t$ per chargon for $Ux\ge4/3$.  These states still
lie on the II$^-$ quadric
[$(\mathcal{P},\mathcal{X})=(0.5000,-0.2500)$ at $Ux=2/3$ and
$(0.5001,-0.2494)$ above], but their valley content is unbalanced, their
four-valley weight is $0.75$ to $0.78$ ($0.89$ at $Ux=2/3$), and their
density has no residual translation symmetry on the $432$-site torus.
The landscape is glassy, and global optimality is not established at any
coupling.

The minimizer landing on the II$^-$ manifold is consistent with the
quartic analysis of Sec.~\ref{sec:phasediagram} and corroborates
Sec.~\ref{sec:phaseII}.  Because the functional contains on-site
repulsion only, its projected quartic lies exactly on the I/II$^-$
boundary $\theta_U=\arctan\sqrt5$ of the pure projected on-site
representative, and the selection is made beyond quartic order.
Eliminating the non-condensing chargon bands at second order generates a
sextic whose strength on a given condensate is the quadratic form
$v^\dagger Gv$.  Here $v_r=\rho_r\psi_r$ is the density-weighted
condensate profile and $G$ the resolvent of the eliminated bands, and a
larger value means a lower energy.  On the $432$-site torus the form
equals $4(12-\sqrt6)=38.20$ on all of phase~I (the value scales with
$N$).  It is constant over the pair sphere, as the moduli theorem of
Sec.~\ref{sec:moduli} requires of any sextic, and it reaches
$\simeq197$ ($\simeq0.456N$) at the selected $K/2$ crystal on the
II$^-$ quadric.

The selection therefore depends on the interaction, and neither state is
singled out as the physical one.  For the pure projected on-site
representative, the $K/2$ crystal is selected beyond quartic order.
Once the intersite repulsion exceeds the mean-field crossover
$V_c\simeq9.3\,U^2x/t$ of Eq.~\eqref{eq:Vcrossover} (in the
$U\sum_r\rho_r^2$ normalization of Sec.~\ref{sec:phasediagram}, or
$2.3\,U^2x/t$ in the $(U/2)\sum_r\rho_r^2$ normalization of this
section), i.e.\ for $V/U$ of order $Ux/t$, the phase-I $M$ state is
selected instead.

The resulting state is a $K/2$-modulated chargon Higgs phase rather
than a superfluid.  With only $B_-$ condensed, the diagonal combination
$Q_{\rm EM}-Q_g$ of the physical electromagnetic charge $Q_{\rm EM}$
and the emergent gauge charge $Q_g$ remains unbroken.  The condensate
therefore carries charge density and loop current but no physical pair
amplitude, which would require spinon pairing (Sec.~\ref{sec:z2}).
Section~\ref{sec:electronspectrum} examines the spectrum of its
electron-like quasiparticles.

For $Ux\le1/3$ its density shares the thirty-six-site cell of the
sextic-selected crystal of Sec.~\ref{sec:phaseII}.  The functional
predicts that the modulation is progressively screened as the coupling
grows.  The density range is $\rho/\bar\rho\in(0.47,1.88)$ at
$Ux=1/24$ (the controlled point $U=0.5t$, $x=1/12$) and $(0.57,1.65)$
at $Ux\approx0.17$, and it narrows to $(0.80,1.19)$ at $Ux=4/3$ and to
$(0.93,1.06)$ at $Ux=16/3$ for the lowest states found there.  Those
stronger-coupling states break the $K/2$ superlattice ($39\%$ of their
modulation power lies off it at $Ux=16/3$), but the screening persists.
Beyond the weak-coupling points the screening is a prediction of the
mean-field functional and is not controlled (see (ii) below).

In its valley content, the minimizer at $Ux\approx0.17$ is the
sextic-selected crystal of Sec.~\ref{sec:phaseII} itself, and not just
another state on the same level set.  Its valley projection has
$\langle\rho^3\rangle/\bar\rho^{\,3}=2.1443$, which is the closed-form
value $2+1/(4\sqrt3)$ of the state selected by the negative-definite
on-site sextic.  That state is the density conjugate
$\rho\to2\bar\rho-\rho$ of the $\sum_r\rho_r^3$ minimizer on the II$^-$
level set, for which the value is $2-1/(4\sqrt3)$.  The agreement
between the lattice functional and the sextic selection holds over the
weak-coupling part of the scan, $Ux/t\le1/3$.  At stronger coupling the
landscape of Eq.~\eqref{eq:gpfunctional} is glassy
(Sec.~\ref{sec:strongcoupling}).

(ii)~\emph{The flat band is never populated in the scanned regime, in
the functional without the pairing block.}
At mean field the sign of the coupling favors this result, which
symmetry does not enforce.  Because $B_+$ and $B_-$ carry equal
electric and opposite gauge charge, every gauge- and charge-allowed local coupling
between them is a polynomial in the densities $\rho_\pm=B_\pm^\dagger
B_\pm$.  The leading one is the cross term of the on-site repulsion, and
it enters with a positive sign.  In the normalization used here the
functional is $(U/2)\sum_r\rho_r^2$, so $U\rho$ is the Hartree potential
and the cross term is $U\rho_+\rho_-$.  In the normalization
$U\sum_r\rho_r^2$ of Sec.~\ref{sec:phasediagram} the cross term reads
$2U\rho_+\rho_-$, and $Ux=16/3$ here corresponds to $Ux=8/3$ there.  A
$B_-$ condensate of density $\bar\rho$ therefore raises the $B_+$ mass
by $U\bar\rho$.

The stability gap nevertheless narrows with coupling.  The same
diagonal Hartree potential $U\rho$ acts in both gauge sectors and
raises the $B_+$ bottom, for the $K/2$ crystal by $0.96\,U\bar\rho$ at
$Ux=16/3$.  The condensate chemical potential, however, rises by at
least $U\langle\rho^2\rangle/\bar\rho\ge U\bar\rho$ for a modulated
density, and by $1.01\,U\bar\rho$ for the $K/2$ crystal at $Ux=16/3$
once the condensate's own kinetic-energy cost is included.  The
stability gap against infinitesimal $B_+$ admixture,
$\lambda_{\min}(H_++U\rho)-\mu_-$, decreases from the bare
$(\sqrt6-1)t\simeq1.449\,t$ to about $1.14\,t$ at $Ux=16/3$, the largest
coupling of the scan.  A separate continuation of the lowest state to
$Ux=40$ gives $1.09\,t$.  For the $K/2$ crystal itself, which is a
higher-lying local minimum for $Ux\ge2/3$, the corresponding values are
$1.154\,t$ and $1.10\,t$.  For a uniform condensate density the two
sectors shift rigidly by the same Hartree energy, and the gap stays at
$(\sqrt6-1)t$ exactly.  The whole decrease is therefore an effect of the
density modulation.

These values refer to the lowest states found: the $K/2$ crystal
of the II$^-$ manifold for $Ux\le1/3$ and the states described under (i)
above it.  The gap depends on the state.  A condensate constrained to
the single-valley Bloch sector of the phase-I stripe is a stationary
point of the functional but not its minimizer.  Its self-consistent gap
is $1.29\,t$ at $Ux=1/6$ and $0.30\,t$ at $Ux=16/3$.  The gap is smaller
here because the flat multiplet can occupy the stripe's low-density
sublattice, but it stays positive.  This stripe value, computed with
on-site repulsion alone, illustrates the state dependence and is not a
prediction for the gap of the intersite-selected phase-I regime.  The
gap remains positive throughout, and neither full-doublet minimization
nor fixed-$B_+$-weight scans find a competing mixed-sector state.  All of
these are Hartree--Gross--Pitaevskii statements about a flat multiplet
that is 144-fold degenerate on the 432-site torus, and they indicate
stability at mean field, within the local potential.
\begin{figure*}[t]
\includegraphics[width=0.92\textwidth]{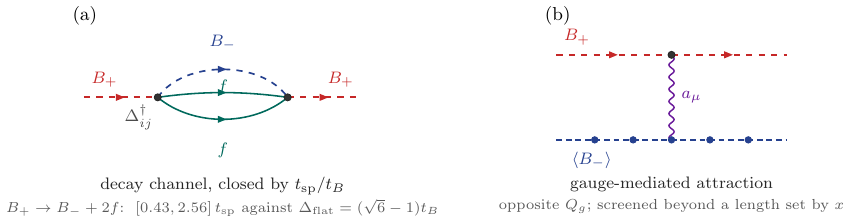}
\caption{Processes beyond the mean-field analysis in the U(1) state, where the
spinon pair amplitude vanishes.  Dashed red lines denote the flat $B_+$ multiplet, dashed indigo the
condensing $B_-$, solid teal the spinons, and the wavy violet line the emergent gauge field.
(a)~Second-order self-energy of $B_+$ through the gauge-neutral hopping term of
Sec.~\ref{sec:bands}.  The vertex is the spinon pair operator $\Delta^\dagger_{ij}$, and the
intermediate state contains one $B_-$ and two spinons, a number fixed by the Gauss law.  These
intermediate states lie at the spinon scale and $\Delta_{\rm flat}$ at the chargon scale.  Against
the bare splitting the channel is closed.
(b)~Gauge exchange between a $B_+$ quantum and the $B_-$ condensate.  The two carry opposite
emergent gauge charge, so the interaction is attractive.}
\label{fig:bplus}
\end{figure*}

Beyond the local potential, $B_+$ couples to the spinon sector through
the gauge-neutral hopping term linear in $B_+$ displayed in
Sec.~\ref{sec:bands} [Fig.~\ref{fig:bplus}(a)].  In the U(1) state this
term gives $B_+$ a second-order self-energy whose intermediate states
contain two spinons.  The number is fixed by the Gauss law of
Sec.~\ref{sec:parton}, since converting a $B_+$ quantum into a $B_-$
raises $n_f$ by two.  These intermediate states span
$[0.43,2.56]\,t_{\rm sp}$, a spinon scale, whereas
$\Delta_{\rm flat}=(\sqrt6-1)t_B$ is set by the chargon scale, and the
two are far apart.  With $t_{\rm sp}\simeq c_1J$ and $t_B=2c_1|t_c|$
(Sec.~\ref{sec:ansatz}), their ratio is $t_{\rm sp}/t_B=2|t_c|/U=0.056$
to $0.072$ over the parameter range of Sec.~\ref{sec:experiment}.  Here
the exchange is held at its measured value, so $t_c$ and $U$ are not
independent.  The continuum reaches $\Delta_{\rm flat}$ only for
$t_{\rm sp}/t_B>0.566$, eight to ten times the physical ratio, so
against the bare splitting the decay channel $B_+\to B_-+$ two spinons
is closed.  This comparison sets a dressed continuum against an
undressed level, and a fully dressed evaluation lies beyond the present
treatment.

The on-shell spectral shift and the static stability correction are
distinct, nonadditive quantities.  The spectral position of the branch
is set by the retarded self-energy on shell, at
$\omega=\Delta_{\rm flat}$, which lies above every intermediate state.
There the shift is repulsive, $+(1.5$ to $2.0)\,x\,t_c^2/t_B$ across
the $144$ states.  It moves the lowest level up from
$\Delta_{\rm flat}=1.449\,t_B$ to $1.68$ and $2.13\,t_B$ at $x=1/36$
and $x=1/12$ for the stripe, and to $1.69$ and $2.17\,t_B$ for the
$K/2$ crystal.  Whether the sector \emph{condenses}, which is what the
width of the window measures, is governed by the static limit
$\omega\to0$, where every denominator reverses sign and the shift is
attractive, $-(0.4$ to $1.2)\,x\,t_c^2/t_{\rm sp}$.  The upward shift
of the high-energy branch thus carries no gain in stability.

At the physical scales the static shift is not controlled.  With
$t_c^2/(t_{\rm sp}t_B)=2.56\,u$ and $u=U/|t_c|=28$ to $36$
(Sec.~\ref{sec:experiment}), it already amounts to $0.6$ to $2.1$ times
$\Delta_{\rm flat}$ at $x=1/36$, and to $1.6$ to $6.3$ times at
$x=1/12$.  This tracks the expansion parameter $t_cx/t_{\rm sp}$ of
Sec.~\ref{sec:electronspectrum}, which is of order one at the lowest
commensuration and larger above it.  The separation of scales closes
the decay channel against the bare splitting, but neither shift fixes
the width of the window at physical coupling.

The emergent gauge field also couples the flat sector to the condensate
directly [Fig.~\ref{fig:bplus}(b)].  Because $B_+$ and $B_-$ carry
opposite emergent gauge charge, it mediates an attraction between a $B_+$
quantum and the $B_-$ condensate, screened only beyond a length set by the
doping.  The $B_+$ states are exactly flat, so an arbitrarily weak
attraction binds.  A bound $B_+$ quantum is gauge neutral with physical
charge $2e$ (Sec.~\ref{sec:bands}), and its condensation would restore
superconductivity at the $U(1)$ level.  A bound state with binding energy
$E_B$ lies at $\Delta_{\rm flat}-E_B$, however, which is still above the
condensate unless $E_B$ exceeds $\Delta_{\rm flat}$. Moreover, once the
condensate response and the polarization of the Dirac spinons are
included, the kernel in the condensed gauge theory is not a two-body
potential.  The positive cross-quartic $U\rho_+\rho_-$ and the
Gross--Pitaevskii stability gap run against binding.  The static
spinon-continuum shift runs toward it, by an amount the second-order
treatment does not control.  A further channel exists at the same order.
The fusion rule that gives $B_+$ its hopping, applied to the electron
hopping, also pairs the spinons through a singlet field proportional to
$B_+\bar B_-$, so the spinon sector acquires a pairing block that enters
the energy at the same order in $B_+$ as the Hartree gap and is absent
from it.  Its effect requires an all-orders treatment of the spinon
sector at each chargon configuration, which lies beyond this paper.  The
conclusion that the condensed phase does not superconduct is therefore a
mean-field statement, valid at the Hartree level for $B_+$.  This
dynamical question separates the condensation route from the
variational superconducting scenario of Sec.~\ref{sec:strongcoupling}.

The functional also has a limited domain of validity.  Over the scanned
range $Ux/W_-$ runs from $0.06$ to $7.4$, where
$W_-=(\sqrt6-\sqrt3)\,t$ is the width of the lowest $B_-$ band
(Sec.~\ref{sec:bands}).  Only the $U/t=0.5$ points are therefore
controlled.  The others are properties of the mean-field functional and
say nothing about the strong-coupling problem.  Flat-band-assisted
chargon crystallization as a strong-coupling effect beyond
Gross--Pitaevskii thus remains a conjecture, suggested but not
established by these results.

(iii)~The natural commensurate fillings of the charge-crystal channels
are set by one chargon per crystal cell: $x=1/6$ for the uni-$M$ stripe
(six-site cell), $x=1/12$ for the $2\times2$ crystal (twelve sites), and
$x=1/36$ for the thirty-six-site $\sqrt{12}\times\sqrt{12}$ $K/2$ crystal
realized both by the sextic selection and by the GP minimizer.

These are commensurate fillings of the composite stars, but the
functional does not lock in to them.  The crystal cell is fixed by the
valley momenta at every $x$ and the Gross--Pitaevskii energy is
analytic in $x$, so nothing in the functional singles out one chargon
per cell.  A true lock-in, with a cusp in the energy or an insulating
plateau, would require the hard-core physics of chargons in their own
superlattice, which lies beyond the functional [see (ii)].  Throughout,
``commensurate filling'' is meant in this sense.

The $K$-star composite is not a charge channel in this construction.
Its site-density form factor and its gauge-invariant bond charge out to
third-neighbor range vanish identically at $K$ for every condensate,
although it can carry loop current (Sec.~\ref{sec:catalog}).  The
$\sqrt3\times\sqrt3$ cell therefore defines no charge-crystal endpoint
at $x=1/9$.  Special two-valley representatives of the $K/2$ family have
smaller eighteen-site cells ($x=1/18$), so the $K/2$ commensuration
depends on the representative.  For $Ux\ge2/3$ a broadened search
(continuation from strong coupling and many starts) does find pure-$B_-$
states below the $K/2$ crystal, described under (i).  No mixed-sector
state appeared in the fixed-$B_+$-weight scans or in unrestricted
minimization over the full doublet started from any of them.

At the Gross--Pitaevskii level the theory thus predicts a
$K/2$-modulated chargon Higgs phase, which can become superconducting
only through the $\Z_2$ channel $\lambda$ of Sec.~\ref{sec:z2}.  We
examine its electron-like quasiparticles next.  The crystallization seen
in strong-coupling numerics must come from physics beyond the functional
(Sec.~\ref{sec:strongcoupling}).

\subsection{Electron spectrum in the condensed phase}
\label{sec:electronspectrum}

The condensates of Sec.~\ref{sec:gp} fix the charge and current pattern
of the Higgs phase.  Whether it conducts is set by the spinon spectrum
in their presence.  The spectra of this section are computed on those
condensates and inherit their status.  The background satisfies the Gauss law on
average but not site by site, and a constraint-consistent saddle
suppresses its density modulation by up to a factor of $44$ and its
currents by up to a factor of $7$ (Appendix~\ref{app:gausslaw}).

With only $B_-$ condensed, Eq.~\eqref{eq:fusion} gives
$c_{\uparrow}=\bar B_-f^\dagger_{\downarrow}$ (Sec.~\ref{sec:gp}), so
the physical electron hopping
$-t_c\sum_{\langle ij\rangle}(c^\dagger_ic_j+\mathrm{H.c.})$ acts on the
spinons, at mean field, as the additional hopping $t_c\bar B_iB_j$.
The spinon link field becomes
$-t\,s_{ij}\,[1-(t_c/t)\,\bar B_is_{ij}B_j]$, which is the ansatz
multiplied by the gauge-invariant bond amplitude of the condensate.
Here $t$ is the spinon hopping, an exchange-scale quantity
(Sec.~\ref{sec:ansatz}).  The expansion parameter of the linear-order
statements below is therefore $t_cx/t$, not $x$.  With $t\simeq c_1J$
and the parameters of Sec.~\ref{sec:experiment} it is of order one
already at $x=1/36$, and we present the calculation, in units of the
spinon hopping, as the leading term of that expansion.

The node shifts and the inter-node coupling below are matrix elements
of $t_c\bar B_iB_j$ between nodal states and scale with $t_c$.  The
second-order gaps scale as $t_c^2/t$, and these are the quantities that
the small-ratio assumption affects.  We have also diagonalized the same
$432\times432$ problem non-perturbatively, with the spinon hopping set
to $c_1J$ and the feedback at full strength, for which $t_cx/t=0.7$ to
$8.7$ over the doping window and the parameter range of
Sec.~\ref{sec:experiment}.  This leaves the Gauss-law pocket structure
intact, with Luttinger volume $x$ and a single carrier sign at every
coupling, and it does not change the ordering of the three condensates
by spinon energy.  It does change the near-degeneracy at $n_f=1$.
There the dressed bands acquire an overlap of $0.04$ to $0.44\,c_1J$ on
the stripe, so the small $n_f=1$ gaps quoted below describe only the
weak-feedback limit.  The decoupling of Sec.~\ref{sec:parton} fixes the
sign of $t_c$ relative to $t$.  With $B_-$ the condensed component,
$t_c<0$ (the opposite combination, $t_c>0$ with $B_+$ condensed, gives
the complex-conjugate induced hopping and an identical spectrum).

We have diagonalized this Hamiltonian on the $12\times12$ torus for the
uni-$M$ stripe and the triple-$M$ state of Sec.~\ref{sec:phaseI} and
for the $K/2$ crystal selected by the on-site sextic in
Sec.~\ref{sec:phaseII}, which is the state to which the minimizer of
Sec.~\ref{sec:gp} converges at weak coupling.  We take
$\sum_r|B_r|^2=xN$ and $t_c=-t$, with the sign fixed in
Sec.~\ref{sec:parton}, for $0.005\le x\le0.17$, and minimize the direct
gap at the node energy over twisted boundary conditions.  The spinon
filling is $n_f=1+x$ for hole doping and $1-x$ for electron doping
(Sec.~\ref{sec:parton}).  The two cases are not mirror images, because
the spinon bands are not particle--hole symmetric.  On the $12\times12$
torus $N(1\pm x)/2$ is an integer at the three commensurate dopings
$x=1/6$, $1/12$ and $1/36$, and (iv) below places the Fermi level at
these dopings.

(i)~At linear order in $x$ none of the three condensates opens a
per-node Dirac mass.  The valley densities enter the spinon problem
only through the conserved $M$-point currents of
Sec.~\ref{sec:catalog}, which shift the two nodes in energy and
momentum without generating a mass term, and a site-diagonal potential
proportional to the condensate density does not generate one either.
The only linear-order gap is the inter-node one of the triple-$M$ state
below, and a $\kk\cdot\mathbf{p}$ projection of the induced
perturbation onto the four nodal states reproduces the lattice slopes
in the limit $x\to0$.

(ii)~The uni-$M$ stripe leaves both nodes gapless at every $x$ studied
(direct gap at the node energy below $10^{-6}\,t$), and its stabilizer,
of order 24, forbids a per-node Dirac mass and any inter-node coupling
at all orders.  The linear term moves the nodes to
$\varepsilon_F+1.18\,xt$ and $\varepsilon_F+0.09\,xt$, with
$\varepsilon_F=(\sqrt3-1)t$ the node energy of the half-filled parent.

(iii)~The triple-$M$ state couples the two nodes, $|V_{12}|=0.63\,xt$,
and opens an inter-node gap at the node energy,
$\Delta\simeq0.046\,xt-0.12\,x^2t$ over the scanned range (the
$\kk\cdot\mathbf{p}$ slope is $0.049\,t$), i.e.\ $3\times10^{-3}\,t$ at
its commensurate filling $x=1/12$.  The $K/2$ crystal, for which a
per-node mass is symmetry-allowed but absent from the induced
perturbation at linear order, opens a gap at the node energy only at
second order, $\Delta\simeq0.13$ to $0.14\,x^2t$ for $x\le0.04$
(falling to $0.11\,x^2t$ at $x=0.1$), i.e.\ $1.0\times10^{-4}\,t$ at
its commensurate filling $x=1/36$ and $1.1\times10^{-3}\,t$ at
$x=0.1$.
\begin{figure}[t]
\includegraphics[width=\columnwidth]{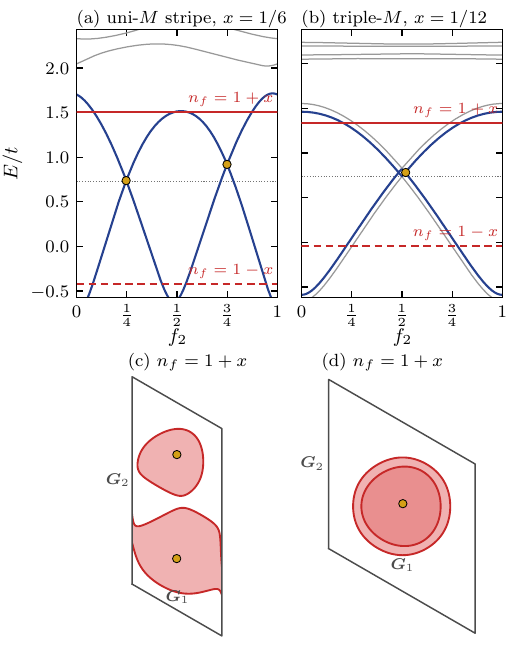}
\caption{Dressed spinon spectrum on the condensate, computed from the Bloch Hamiltonian of the
condensate cell with spinon hopping $t=1$.  (a)~Uni-$M$ stripe at $x=1/6$ (six-site cell) and
(b)~triple-$M$ state at $x=1/12$ (twelve-site cell), along the line $\bm k=\tfrac12\bm G_1+f_2\bm G_2$
of the reduced zone, which passes through both Dirac nodes of the stripe.  The two bands that meet at
the nodes are shown in blue, circles mark the node energies, the dotted line is the node energy
$(\sqrt3-1)t$ of the undoped parent, and red lines mark the Fermi level at $n_f=1+x$ (hole doping,
solid) and $n_f=1-x$ (electron doping, dashed).  (c),(d)~Fermi surface at $n_f=1+x$ in the reduced
zone spanned by $\bm G_1$ and $\bm G_2$.  Shaded regions are the occupied part of the band above the
nodes, with total volume $x$ per site, both spins.  For the stripe the Fermi level lies $0.01\,t$ above
the saddle point of that band around the lower node, so the sheet there is open along $\bm G_1$, while
the sheet around the upper node is a closed pocket.  For the triple-$M$ state the two hybridized cones
give two nested closed pockets.}
\label{fig:pockets}
\end{figure}

(iv)~Because the Gauss law places the spinons at $n_f=1\pm x$, away
from half filling, conduction is governed by the spectrum at the Fermi
level and not at the nodes.
At $n_f=1+x$ (hole doping) the Fermi level lies $0.34$ to $0.77\,t$
above both shifted nodes for the stripe, the triple-$M$ state and the
$K/2$ crystal at their commensurate dopings.  At $n_f=1-x$ (electron
doping) it lies $0.43$ to $1.35\,t$ below them.  In every case there is
no gap at the Fermi level.

All three condensates are metals at mean field
(Fig.~\ref{fig:pockets}).  Their electron-like quasiparticles form
Fermi surfaces of one carrier sign around the two nodes, with total
Luttinger volume $x$.  These are closed pockets, with one exception.
For the stripe at $x=1/6$ the Fermi level sits $0.01\,t$ above the
saddle point of the pocket band around the lower node, and that sheet
is open along $\bm G_1$.  The condensate-induced gaps of (iii) lie
$0.34$ to $0.83\,t$ from the Fermi level and do not affect conduction.
Counting gives the same answer without diagonalization.  With the
feedback included, the spinon Bloch cell coincides with the condensate
cell of six, twelve or thirty-six sites.  This cell holds $3.5$, $6.5$
or $18.5$ spinons per spin at $n_f=1+x$ ($2.5$, $5.5$ or $17.5$ at
$1-x$).  Because these numbers are half-integers, no spin-degenerate
band insulator exists at any of the three lock-ins.

An insulating gap at the commensurate fillings must come from physics
beyond mean field, such as confinement, loss of single-chargon
coherence, further-neighbor feedback of the condensate on the spinon
hopping, or interactions beyond the quadratic spinon problem.  We
compute none of these, so we do not claim an insulating gap of order
$t$ on the basis of this calculation.  Counting constrains such physics.
At $x=1/n_s$ the $n_s$-site cell holds $n_s\mp1$ electrons for hole or
electron doping, an odd number for each of the three crystals
($n_s=6$, $12$, $36$).  By the Lieb--Schultz--Mattis theorem in its
higher-dimensional form \cite{LSM1961,Oshikawa2000,Hastings2004} no
symmetric gapped insulator exists at these fillings.  An insulator there
must either break a further symmetry, for instance by doubling the
period or ordering magnetically, or carry topological order.  Spinon
pairing cannot produce topological order while the chargon stays
condensed.  Since $B_-$ carries unit gauge charge, the Higgs phase
retains no gauge structure for a charge-two spinon pair to restore, and
by Eq.~\eqref{eq:pairop} paired spinons in the presence of the
condensate give a physical charge-$2e$ superconductor rather than a
charge-conserving insulator.  A gapped $\Z_2$ liquid of the remaining
spins requires the chargons first to lose single-particle coherence,
after which paired spinons can provide it.  The same constraint applies
to the insulating holon crystals of the strong-coupling numerics
(Sec.~\ref{sec:strongcoupling}).

Together with Sec.~\ref{sec:gp}, these results characterize the
condensed phase as a chargon Higgs phase with the diagonal
U(1) $Q_{\rm EM}-Q_g$ unbroken and no superfluid response.  It carries
the charge density and loop currents of Sec.~\ref{sec:catalog}, and at
mean field its electron-like quasiparticles are metallic, with pockets
of volume $x$.  The electron spectral function is the
condensate-weighted particle--hole conjugate of the spinon spectrum, and
the doped charge resides in the condensate, as the bookkeeping of
Sec.~\ref{sec:gp} prescribes.

Because the condensing field carries unit gauge charge, the Higgs phase
has no residual gauge structure and is continuously connected to a
conventional charge-ordered metal \cite{FradkinShenker1979}.
Luttinger's theorem therefore applies to its electron-like
quasiparticles in the enlarged cell.  The $n_s(1\mp x)$ electrons per
$n_s$-site cell fill half of the reduced zone per spin, which is the
count of the parent's Dirac nodes, plus or minus a signed pocket volume
$x$ (Sec.~\ref{sec:transport}).  The theorem fixes this signed volume
only modulo filled bands.  That the pockets have one carrier sign, with
no compensating pocket elsewhere and no intervening Lifshitz transition,
is a property of the band parameters we have scanned, and we claim it
only for those crystals and values of $x$.

\subsection{Strong coupling: holon crystals and the two routes}
\label{sec:strongcoupling}

This subsection compares the symmetry and commensuration content of the
condensate theory with published strong-coupling numerics and contains
no energetics of its own.  Ko, Lee, and Wen \cite{Ko2009} already posed
the question of the strong-coupling fate of the doped kagome spin
liquid.  Their holon band coincides with our dispersive sector ($B_-$ in
the labeling of Eq.~\eqref{eq:fusion}, $B_+$ in their $t>0$ convention,
Sec.~\ref{sec:bands}), and the translation sector of its projective
representation is their Eq.~(30) (Sec.~\ref{sec:valleys}).  From the
same four minima they distinguished two states.

For the flux-free state, their Fermi-pocket state, they observe that
``by Bose statistics the holons will condense at each quadratic band
bottom'' \cite{Ko2009}.  The Landau theory of
Secs.~\ref{sec:landau} and \ref{sec:phases} tests this statement and
finds that a balanced occupation of the four band bottoms is selected in
neither limit.  On-site repulsion selects nothing among the valley
occupations at $\theta_U$, nearest-neighbor repulsion tilts the quartic
minimum to the pair condensate of phase~I, and the microscopically
generated sextic tilts it to the unbalanced four-valley $K/2$ crystal of
phase~II$^-$ (Sec.~\ref{sec:phasediagram}).

Their second state, which they argued is favored at low doping
(spinon gain $\propto$ flux$^{3/2}$ versus holon cost $\propto$
flux$^2$), arises from a different mechanism.  An emergent uniform flux
rearranges holons and spinons into Landau levels, the holons form four
$\nu=1/2$ Laughlin states, and the outcome is a time-reversal-breaking
anyon superconductor with $hc/4e$ vortices.  That construction rests on
their Landau-level energetics and assumes no symmetric condensate, and
we do not contest it on its own terms.  The present work instead
develops the condensation route systematically, with the Landau theory
in place of a symmetric valley occupation.

Strong-coupling numerics can discriminate between the two routes.
Density-matrix renormalization on the lightly doped kagome $t$--$J$
model finds an insulating crystal of \emph{spinless, unpaired} holons,
the chargon degree of freedom of the present theory, with no
superconductivity ($\xi_{\rm SC}\lesssim1.3$ lattice spacings for all
$\delta\le11\%$) \cite{Jiang2017,Peng2021}.  On the square lattice, by
contrast, the corresponding crystal consists of singlet
hole (holon) \emph{pairs} \cite{Dodaro2017,Jiang2017}, objects that in
the gauge-doublet language require both gauge components
(Sec.~\ref{sec:bands}).  The strongly renormalized,
valence-bond-dressed holon reported there also identifies the
strong-coupling ingredients (mass enhancement, induced repulsion) that
undercut the Landau-level energetics of Ref.~\cite{Ko2009} and are
absent from the Gross--Pitaevskii functional.

Whether these studies find a stripe or a crystal depends on geometry and
doping, through a competition among cells set by $1/\delta$ that lies
outside the $M$-star sector to which our moduli theorem applies.  The
tensor-network phase diagram of Ref.~\cite{XuGuYang2024} connects to our
commensurate endpoint.  At $\delta=1/6$ the $C_3$-symmetric
trihexagonal $2\times2$ crystal (two holes per cell) and a four-hole
Wigner crystal in a $2\times4$ cell are \emph{nearly degenerate}, and
the $2\times1$ stripe is close in energy.  This near degeneracy of
$C_3$-symmetric and unidirectional charge crystals parallels the
uni-$M$/tri-$M$ competition of Sec.~\ref{sec:phaseI}, but it is not the
same competition.

The doped phase diagram of Ref.~\cite{XuGuYang2024} parallels the
structure found here across its full range:
\begin{enumerate}[label=(\roman*),leftmargin=*,itemsep=0pt,topsep=2pt,parsep=0pt,partopsep=0pt]
\item insulating charge crystals for $\delta<0.27$;
\item a non-Fermi-liquid window $0.27<\delta<0.32$ in which the
\emph{extended-$s$} (our $\beta$-channel) order dies in the
extrapolation;
\item for $0.32<\delta<1/3$, a translation-invariant paired state that
the authors call a ``pair-density wave'' because the nearest-neighbor
pair amplitude alternates in sign between the up- and down-triangles of
their $1\times1$ cell,
$\Delta=(\Delta_a,\Delta_b,\Delta_a,\Delta_b,\Delta_a,\Delta_b)$ with
$\Delta_a>0>\Delta_b$ on the six bonds.
\end{enumerate}
The last is the $Q=0$, up/down-triangle sign-alternating structure of
our $\alpha$ channel \cite{XuGuYang2024}, and $d+id$ is \emph{never}
the ground state, as Sec.~\ref{sec:pairingmap} requires.  Their pairing
ansatz is uniform over a $1\times1$ cell and cannot host the $M$-star or
$K/2$-star modulations.  That work therefore neither tests nor
contradicts the finite-momentum pair content predicted here.

Adding next-nearest-neighbor couplings melts
the holon crystal into a Fermi-liquid-like state \cite{NNN2025}, which
provides a microscopic control parameter for the crystal-to-condensate
transition.  The same study finds a distinct commensurate crystal
reappearing at $\delta=1/9$ \cite{NNN2025}, the filling of one chargon
per $\sqrt3\times\sqrt3$ cell.  Its pattern has not been identified, and
it has no counterpart among the charge composites of
Table~\ref{tab:ph}.  In this construction the $K$-star composite carries
no site charge and no bond charge out to third-neighbor range, though it
can carry loop current (Sec.~\ref{sec:catalog}).  A $\sqrt3\times\sqrt3$
charge crystal at $\delta=1/9$ falls outside the charge
content of the leading condensate composites.

The most recent density-matrix renormalization results sharpen this
correspondence \cite{JiaMelting2025}.  In the nearest-neighbor $t$--$J$
model, doping alone melts the holon Wigner crystal at
$\delta\approx0.15$, just below the densest commensurate endpoint
$x=1/6$ of the crystal ladder.  The result is a uniform metallic phase
with power-law correlations and \emph{no hole pairing} out to
$\delta=0.36$.  The absence of superconductivity in the doped DSL
therefore persists on both sides of the melting transition.

In the crystal regime each emergent cell holds one hole, and stripe
patterns are reported at $\delta=1/36$--$1/18$
\cite{Jiang2017,JiaMelting2025}.  These fillings coincide with the
$K/2$-crystal commensurations of Sec.~\ref{sec:gp}.  The coincidence is
suggestive, but we do not identify the two.  The reported stripes are
unidirectional orders with emergent cells of 36 to 18 sites, whose
period depends on the cylinder geometry as well as on the filling.  They
are neither the six-site $M$ stripe of Sec.~\ref{sec:phaseI} nor the
thirty-six-site, five-level $K/2$ crystal.

Reference~\cite{JiaMelting2025} leaves open whether the melted phase is a
conventional or a fractionalized Fermi liquid.  In the chargon theory the
natural candidate is the FL* of uncondensed chargons, and the Hall number
of Sec.~\ref{sec:transport} distinguishes the two.  In a narrow window at
$\delta=1/3$ the same calculations find a pseudogap-like precursor state
with a single-particle gap and strongly enhanced pairing, adjacent to the
tensor-network $\alpha$-channel pairing window above.  It disappears on
the wider $L_y=4$ cylinder in the same work, and its fate in two
dimensions remains open \cite{JiaMelting2025}.

Variational studies point the same way.  The earliest Monte Carlo study
of the doped kagome DSL found the flux collapsing under doping toward a
zero-flux state with valence-bond order \cite{GuertlerMonien2011}.  Its
refined Star-of-David pattern is optimal at $\delta=1/12$, with one hole
per twelve-site cell and repulsive interactions
\cite{GuertlerMonien2013}.  Its cell is the $2\times2$ commensuration of
Sec.~\ref{sec:gp}, identified variationally a decade before the present
framework.

Every result from Sec.~\ref{sec:bands} onward assumes the link field of
the undoped ansatz (Sec.~\ref{sec:parton}), and the chargon theory can
estimate where that collapse sets in.  A doped carrier at the chargon
band bottom gains $(3-\sqrt6)\,t\simeq0.55\,t$ if the flux pattern
collapses to a flux-free one (band bottom $-4t$ against
$-(1+\sqrt6)\,t$), with $t$ the chargon hopping of Sec.~\ref{sec:bands}.
The half-filled spinon sector, in contrast, prefers the parent pattern
by $0.0423$ times the spinon hopping per site.  The two saddle points
therefore cross at
$x^*=\Delta E_{\rm sp}\,t_{\rm sp}/[(3-\sqrt6)\,t_B]\simeq0.003$ to
$0.007$ over the parameter range of Sec.~\ref{sec:experiment}, and at
$0.077$ if the spinon hopping is set equal to the chargon hopping.
Below $x^*$ the parent link field is stable against flux collapse, the
only competitor tested here.  Above it the flux-collapsed states are the
competitors, and the gauge-rotated states of Ref.~\cite{BogFS2021}
overtake them only at the smallest dopings for the largest $t/J$ studied
there.  The estimate is crude.  It places every carrier at the band
bottom and neglects the doping dependence of the spinon energy.  Its
message nevertheless agrees with the variational results: the
rigid-background assumption restricts the theory to the dilute end of
the doping window of Sec.~\ref{sec:experiment}.  The commensurate
fillings of that section sit at or above this range.  The lowest,
$x=1/36=0.028$, falls below $x^*$ only for $t_{\rm sp}=t_B$, which gives
$x^*=0.077$, and $x=1/12$ and $1/6$ lie above it for every choice.
The crystals at those fillings are predicted on the assumption
that the parent flux survives the doping, which the estimate does not
establish.

A later study enlarged the variational space by site-dependent SU(2)
gauge rotations of the doped ansatz \cite{BogFS2021}.  These degrees of
freedom amount to a static chargon configuration.  The study found that
this class overtakes the flux-collapsed states at low doping and selects
a $2\times2$, time-reversal- and inversion-breaking nematic pair-density
wave with a Bogoliubov Fermi surface and a coexisting $2\times2$ charge
modulation.  This combination of cell, nematicity and broken
time-reversal and inversion symmetry matches the $M$-star chargon
condensates of Sec.~\ref{sec:phases}, here realized in projected wave
functions.  That
superconducting scenario and the charge-order-first route developed
here differ in the fate of the flat sector.  A physical pair amplitude
without spinon pairing requires both gauge components, whereas the
Gross--Pitaevskii minimization above finds $\avg{B_+}=0$ protected, and
the density-matrix renormalization results, which find no hole pairing,
are consistent with it.  Whether the flat sector condenses is a
dynamical question that symmetry leaves open.

Experimentally, the realized electron-doped series, Li-intercalated
herbertsmithite \cite{Kelly2016}, remains insulating at every
composition studied, through the nominal Dirac-metal point.  The
Ga-substituted series is insulating for a different reason.  Its
Ga$^{3+}$ charge is compensated by additional OH$^-$/Cl$^-$ anions, and
x-ray absorption detects no Cu$^{1+}$, so it does not test doping of the
kagome plane \cite{Puphal2019}.  The chargon theory bears on these data
only partly and over a limited composition range.  The Li series
extends to $1.8$ Li per formula unit, $3/5$ electron per Cu
\cite{Kelly2016}, well above both the densest commensurate endpoint
$x=1/6$ of the crystal ladder and the melting doping
$\delta\approx0.15$ of the nearest-neighbor $t$--$J$ numerics above.
Its insulating behavior in that upper range is left to the extrinsic
dopant localization of Sec.~\ref{sec:materials}.

Doping a kagome DSL thus produces charge order instead of a
large-Fermi-surface metal.  At the mean-field level of
Sec.~\ref{sec:electronspectrum}, however, the condensate opens no
insulating gap.  Its electron-like quasiparticles form Fermi pockets
of total Luttinger volume $x$ in all three states, and the
condensate-induced gaps at the nodes lie $0.34$ to $0.83\,t$ away from
the Fermi level.  An insulator at the commensurate fillings therefore
requires physics beyond mean field.  The strong-coupling numerics above,
which find insulating holon crystals
\cite{Jiang2017,Peng2021,XuGuYang2024}, show that the $t$--$J$ model
contains such physics.  On that reading, insulation remains compatible
with successful doping, since the tensor-network crystals survive with
the sign of $t$ reversed, i.e.\ for electron doping
\cite{XuGuYang2024}.  Extrinsic localization remains possible
\cite{Kelly2016}.  In the weak-coupling itinerant limit the same layer
disfavors a CDW \cite{Mazin2014}, so the two limits bracket the
observations.

\subsection{Transport and finite-temperature phenomenology}
\label{sec:transport}

The transport predictions of the FL$^*$ framework
\cite{Chatterjee2016,Bonetti2026} carry over, subject to the two
assumptions flagged in Sec.~\ref{sec:gp}.  While the chargons are
uncondensed, the Hall number counts the doped density $p$, the
Luttinger volume of the pockets ($3p$ holes per three-site cell).  A
conventional Fermi liquid at the same doping would instead count the
large Fermi surface of the half-filled middle band, $1+3p$ holes per
cell (the one-band value $1+p$ does not apply to kagome).  Throughout,
Hall numbers are meant in the relaxation-time sense.  The quantity
$n_H=1/(eR_H)$ measures a carrier count only for pockets of a single
sign with comparable mobilities, because the weak-field Hall
conductivity is a property of the scattering-path geometry and not of
the carrier density alone \cite{Ong1991}.  Anisotropy, variation of the
scattering time over the pockets, and symmetry-allowed anomalous terms
all loosen the identification.  Within these limits the Hall number,
invoked in Sec.~\ref{sec:strongcoupling}, separates the
large-Fermi-surface metal from both fractionalized states.

Beyond the Higgs transition the condensed phase is the chargon Higgs
phase of Sec.~\ref{sec:electronspectrum}, with the diagonal U(1)
$Q_{\rm EM}-Q_g$ unbroken and no reconstruction to a
large-Fermi-surface metal.  At mean field its electron-like
quasiparticles form Fermi pockets of one carrier sign and total
Luttinger volume $p$ around the two shifted Dirac nodes, in all three
condensates (Sec.~\ref{sec:electronspectrum}).  On the same reading, the
Hall number of the Higgs phase therefore counts $p$, as in the FL$^*$,
but for a band-structure reason.  The parent's Dirac nodes sit at half
filling, so at electron density $1-p$ the quasiparticle Fermi surface
consists of pockets of volume $p$ rather than a large Fermi surface.

The Higgs phase and the FL$^*$ thus have the same Luttinger count and
differ in translational order and in the condensate scale $T_\rho$
(Sec.~\ref{sec:experiment}).  Nor is any
Hall-number reconstruction toward the large-Fermi-surface count implied,
because the doped charge resides in the condensate and an insulating
state at the commensurate lock-ins requires physics beyond the
quadratic spinon problem (Secs.~\ref{sec:electronspectrum} and
\ref{sec:strongcoupling}).  Above the ordering temperature, thermal
chargon fluctuations produce arc-like spectral weight of the pseudogap
type \cite{Bonetti2026}.

On kagome the leading finite-temperature scale is that of the
condensate amplitude, $T_\rho$, which lies above any superconducting
scale.  For the gauge-charged condensate this scale marks a crossover
and no true transition (Sec.~\ref{sec:experiment}).  Below it the
$M$-star charge order sets in at $T_{\rm CO}$, and no Landau cubic
forces this transition to be first order (Sec.~\ref{sec:landau}).
Where superconductivity arises through a $\Z_2$ channel, it follows at
or below $T_\rho$.  The
electromagnetic stiffness of a condensate built from a chargon
condensate and paired spinons is the series (Ioffe--Larkin)
combination of the two stiffnesses \cite{LeeNagaosaWen2006}, so
$T_c\le T_\rho$.  Here $T_c$ is set by the smaller of $T_\rho$ and a
spinon-pair scale $T_\lambda$ proportional to the Higgs amplitude
$\lambda$, which the theory does not fix (Sec.~\ref{sec:experiment}).

A doped kagome DSL should therefore show the condensate scale $T_\rho$
at or above the superconducting one, as the counterpart of the hierarchy
$T_{\rm CO}\gg T_c$ of the kagome metals \cite{RMP2026}.  The inequality
$T_\rho\ge T_c$ follows from the Ioffe--Larkin combination once the
lifted $\tau^3$ degeneracy provides a chargon stiffness, although the
band structure alone does not fix the transition temperature that this
stiffness sets.  The inequality $T_{\rm CO}<T_\rho$ depends on the
octic anisotropy, the interlayer coupling and fluctuations
(Sec.~\ref{sec:experiment}).  The full sequence is thus suggested by
the ordering of scales but is not derived, and the position of $T_c$
relative to $T_{\rm CO}$ depends on $T_\lambda$.

The charge order in this window is vestigial in the usual sense: a
composite of the chargon multiplet orders over a range where the
superconducting component does not.  Unlike the usual scenario, the
ordering $T_\rho\ge T_c$ is derived here from the Ioffe--Larkin
combination of the two stiffnesses, on the premise stated above.

Throughout, $T_{\rm CO}$ denotes the discrete lattice-symmetry-breaking
transition of the charge order.  It lies below the bare stiffness scale
$T_\rho$ of the condensate and is not fixed by $t$ and $x$ alone.
Section~\ref{sec:experiment} separates the two scales and gives the
range of $T_\rho$.

\section{Discussion: experimental signatures}
\label{sec:experiment}

We fix units once, here, for herbertsmithite.  We take the exchange
$J\simeq180$~K of ZnCu$_3$(OH)$_6$Cl$_2$, the ab initio value
\cite{Jeschke2013} quoted in Ref.~\cite{Norman2016}.  Experimental
estimates run to $17$~meV $\simeq197$~K, a $5\%$ shift that the $U$
range below absorbs.  With $J=4t_c^2/U$ for an effective one-band
Hubbard $U$ of $3$ to $5$~eV, the electron hopping is $|t_c|=108$ to
$139$~meV.  The absolute scales of this section may be too low by a
factor of order two.  The $5$ to $7$~eV of the single-band fits of
Ref.~\cite{Mazin2014} and the $6$~eV at which Ref.~\cite{Jeschke2013}
obtains $J_1=182$~K would raise the electron hopping to $139$ to
$165$~meV.  The direct density-functional hopping of about $0.3$~eV
\cite{Mazin2014} lies higher still, and with it $J=4t^2/U$ overshoots
the measured exchange several-fold.  The chargon hopping is
$t_B=2c_1|t_c|=48$ to $62$~meV at the mean-field level of
Sec.~\ref{sec:parton}, where $c_1=0.221$ is the nearest-neighbor spinon
bond amplitude of Sec.~\ref{sec:bands}.  The spinon hopping is a
different and much smaller quantity, $t_{\rm sp}\simeq c_1J=3.4$~meV.
At fixed $J$ this gives $t_{\rm sp}/t_B=0.056$ to $0.072$ and
$|t_c|/t_{\rm sp}=32$ to $41$.

Each quantity below is set by the hopping that appears in its own energy
denominators.  The chargon potential and its sextic and octic terms are
set by $t_B$, the spinon-sea and spinon-polarization responses by $t_{\rm
sp}$, and the matrix elements that couple the sectors by $t_c$.  We quote
every chargon scale as a range, from its value at $t_B$ to its value at
the bare $|t_c|$, which is an upper bound.  Neither end captures the
strong-coupling renormalization of the holon bandwidth
(Sec.~\ref{sec:strongcoupling}).  The chargon flat-band gap of
Sec.~\ref{sec:bands} is then $\Delta_{\rm flat}=(\sqrt6-1)t_B=69$ to
$201$~meV ($69$ to $89$~meV for $t_B$, $157$ to $201$~meV for the bare
$|t_c|$).

The bare phase-stiffness energy of the mean-field condensate defines the
temperature $T_\rho$.  With the isotropic valley curvature $t/(2\sqrt6)$
of Sec.~\ref{sec:bands} and an areal chargon density of $x$ per site,
$k_BT_\rho\simeq(\pi/2)\,\hbar^2n/m^*=1.11\,tx$, which runs from $17$ to
$298$~K for $x=1/36$ to $1/6$ ($17$ to $132$~K for the mean-field $t_B$).
The Berezinskii--Kosterlitz--Thouless prefactor serves here as a
normalization, and no phase transition is associated with this scale
itself.  Because $B_-$ carries the emergent gauge charge, its phase is not
gauge invariant.  In the Higgs phase the gauge field absorbs it, leaving
no Goldstone mode, no superfluid stiffness and no
Berezinskii--Kosterlitz--Thouless transition.  $T_\rho$ marks the
crossover below which the condensate amplitude, and with it the Higgs
photon mass, is established and the composites of Table~\ref{tab:ph}
become the slow variables.  It estimates the condensate scale and bounds
no ordering temperature.

The $M$-star (or $K/2$-star) charge order sets in at a discrete
lattice-symmetry-breaking transition $T_{\rm CO}$, which is not fixed by
$t$ and $x$ alone.  Because the pair sphere is degenerate through sextic
order, the discrete orbit is selected by the octic anisotropy of
Sec.~\ref{sec:phaseI}, by interlayer coupling and by fluctuations.  We
expect $T_{\rm CO}$ below $T_\rho$, although we neither derive that
ordering nor compute the scale.  If the $B_2'$ flux is ordered
throughout the upper stage described below, as symmetry allows, the
trilinear of Eq.~\eqref{eq:lock} mixes the $M$-point charge and current
triplets linearly, and the charge order and the loop-current order set
in together at $T_{\rm CO}$, in a single transition.

The octic anisotropy sets one more scale.  Its splitting per site
between the stripe and triple-$M$ orbits, $\delta e=0.351\,(Ux/t)^3\,xt$
for on-site repulsion alone [Eq.~\eqref{eq:octicderived}], is
extensive.  It becomes thermodynamically decisive only on
patches of more than $k_BT/\delta e$ sites, that is, beyond a crossover
length of order
$\ell^*=\sqrt{k_BT/(\delta e\,n_s)}=a\sqrt{k_BT/(2\sqrt3\,\delta e)}$,
with $n_s=2\sqrt3/a^2$ the kagome site density used for $T_\rho$ above.
Below $\ell^*$ the choice between the two orbits is thermally
undetermined.  Since $\delta e$ falls as $x^4$, $\ell^*$ grows rapidly
toward the dilute end.

For the illustrative choice $U=t$, two orders of magnitude below the
charging-energy scale of Sec.~\ref{sec:phasediagram} ($U$ is not
derived here), $\delta e$ stays below $0.5$~K per
site throughout $x\le1/6$, and $\ell^*$ for $T=2$ to $10$~K is one to
three lattice constants at $x=1/6$ and $280$ to $720$~\AA\ ($40$ to
$105$ lattice constants, $a=6.83$~\AA) at $x=1/36$.  These are
crossover lengths.  A domain size would additionally require
domain-wall and disorder energetics, which are beyond this work.  At the
physical charging energy the octic expansion behind $\delta e$ is outside
its range (Sec.~\ref{sec:phasediagram}), so these lengths illustrate the
$x^4$ scaling and do not predict a value.

These scales suggest a two-stage finite-temperature structure.  Below
$T_\rho$ lies an upper stage in which the condensate has stiffness but
the pair sphere is not yet resolved into a discrete orbit, so uni-$M$
and tri-$M$ correlations are short-ranged.  The $\Gamma$-point
composites that are constant over the pair sphere, such as the $B_2'$
staggered flux of Sec.~\ref{sec:catalog}, are then symmetry-allowed to
order before the $M$-triplet does, and the charge-ordering transition
follows at $T_{\rm CO}$.  We do not derive this structure, and whether
these composites actually order is a fluctuation question
outside the present theory.

Superconductivity, where the $\Z_2$ channel allows it, lies at or
below $T_\rho$.  It requires both the chargon condensate and the spinon
pairing.  The electromagnetic phase stiffness of such a two-component
condensate is the series (Ioffe--Larkin) combination $\rho_{\rm
EM}=\rho_B\rho_f/(\rho_B+\rho_f)$ \cite{LeeNagaosaWen2006} of the
chargon stiffness $\rho_B$, which sets $T_\rho$, and the stiffness
$\rho_f$ of the paired spinons.  The pair amplitude and the phase
stiffness are distinct quantities, and with the spinon Fermi surface
present at $n_f=1\pm x$ the relation between them needs a separate
response calculation.  We assume that the spinon stiffness scales with
the Higgs amplitude $\lambda$, with a coefficient of order unity that we
do not compute.  Then $T_c\lesssim\min(T_\rho,T_\lambda)$, where
$k_BT_\lambda\propto\lambda$ is an input of the theory.

The physical pair amplitude of Sec.~\ref{sec:z2spectra} is linear in
$\lambda$ and in the condensate density $x$ through the composite of
Eq.~\eqref{eq:elpair} and, for $\alpha$, $\beta$ and $\delta$, through
the $B_+$ admixture [amplitude per unit $\lambda/t$:
$1/[2(1+\sqrt6)]=0.145$ for the on-site channel of $\beta$, $\sqrt3/2$
for $\alpha$, $\sqrt6/2$ for $\delta$ and zero for $\gamma$].  This
order-parameter amplitude sets the size of the induced electron pair
field and has no direct meaning as a temperature.  The band structure
provides the energy scale $k_BT_\rho=1.11\,tx$ against which $\lambda$
may be compared, and for $\lambda/t=0.05$ to $0.3$ and $x=1/12$ to
$1/6$ the two are of the same order.  The separation between $T_\rho$
and $T_c$ is therefore not parametric, and the band structure fixes
neither $T_c$ nor $T_\rho$ as a transition temperature.

\subsection{Doped spin-liquid materials: the intrinsic scenario}
\label{sec:materials}

Herbertsmithite and its polymorphs remain the central kagome spin-liquid
materials \cite{Norman2016,Han2012}, with inelastic neutron scattering
now identifying universal spin excitations across the herbertsmithite
and Zn-barlowite families \cite{ZnBarlowite2025}.  Progress on the
charge sector has been slower.  Ga substitution was predicted to give a
correlated Dirac metal at $1/3$ doping \cite{Mazin2014}, but in the
realized series the Ga$^{3+}$ charge is anion-compensated and the kagome
plane is undoped \cite{Puphal2019}.  The only electron-doped series
realized so far, Li intercalation, in which photoemission sees
Cu$^{1+}$, remains insulating at every composition studied
\cite{Kelly2016}.  First-principles work attributes this to
dopant-induced localization and polaron formation
\cite{Polaron2018,LiDopingTheory}.

The chargon theory adds an intrinsic mechanism, independent of dopant
disorder.  \emph{Even ideal, disorder-free doping of the kagome DSL is
not expected to produce a conventional metal or a superconductor at
leading order.}  It produces instead a charge-ordered chargon Higgs
phase (Secs.~\ref{sec:phases}, \ref{sec:doped}), in which the physical
$U(1)$, the diagonal combination generated by the difference of the
electromagnetic and gauge charges, remains unbroken.

With only $B_-$ condensed, the electron is the condensate-weighted
particle--hole conjugate of the spinon (Sec.~\ref{sec:parton}), and the
condensate feeds back on the spinon hopping through the gauge-invariant
bond amplitude $\bar B_is_{ij}B_j$ (Sec.~\ref{sec:doped}).  At the
mean-field level of the quadratic spinon problem this feedback opens no
Dirac mass at linear order in $x$.  At the Gauss-law filling
$n_f=1\mp x$ the electron-like quasiparticles of all three condensates
form Fermi pockets of total volume $x$ around the Dirac nodes.  (For the
electron-doped series of this section the spinons sit at $n_f=1-x$,
with the Fermi level $0.43$ to $1.35\,t$ below the nodes.)  An
insulating gap at the commensurate fillings must therefore come from
confinement, from interactions beyond the quadratic spinon problem, or
from physics that, by the counting argument of
Sec.~\ref{sec:electronspectrum}, breaks a further symmetry or carries
topological order.  At mean field the condensate does not account for
the insulating behavior of the realized doping series, and the present
analysis does not settle whether the phase is insulating.

Translational order distinguishes intrinsic chargon crystallization
from extrinsic localization, which produces no new translational order.
Ordered dopants or a dopant-induced lattice distortion can, however,
generate satellites independently, so observed satellites are compatible
with both scenarios.  Intrinsic chargon crystallization produces
\begin{itemize}
\item superlattice Bragg satellites at the charge-carrying composite
  stars of Table~\ref{tab:ph}, $M$ ($2\times2$ or stripe) and $K/2$
  ($\sqrt{12}\times\sqrt{12}$, thirty-six-site), with commensurate
  fillings at $x=1/36$, $x=1/12$ and $x=1/6$;
\item a $C_2$-odd $2\times2$ pattern (Sec.~\ref{sec:landau});
\item a charge-ordering transition free of any symmetry-mandated
  first-order mechanism;
\item loop-current magnetism whose $M$-modulated part switches on at the
  same transition, visible to $\mu$SR and, by symmetry, open to polar
  Kerr in its multi-arm realizations (triple-$M$, $K/2$) and closed to
  it in the uni-$M$ stripe (Sec.~\ref{sec:fingerprints}).
\end{itemize}
The $K$ star is absent from the diffraction list because its
site-density form factor and its gauge-invariant bond charge out to
third-neighbor range vanish identically for every condensate, so the
$\sqrt3\times\sqrt3$ channel is a $\mathcal{T}$-odd loop-current
crystal.  Local magnetic probes can see it, whereas x-rays and
electrons cannot.

Diffraction is the most direct test of the intrinsic scenario, and the
modulation it must detect is small in absolute terms.  For the phase-I
stripe the site densities of the pure valley condensate take the three
values $x(1\mp1/\sqrt2)$ and $x$, each on one third of the sites.  This
gives a peak modulation $\delta n=x/\sqrt2=0.020$ to $0.118$ electrons
per Cu for $x=1/36$ to $1/6$.  The triple-$M$ state has $\delta
n=\sqrt{2/3}\,x$, and the $K/2$ crystal has $\delta
n=(1/\sqrt2+1/(2\sqrt3))\,x\simeq x$.

The corresponding diffraction intensities are order-of-magnitude
estimates.  We compare with the total electron count per Cu of the host
($68.3$ for ZnCu$_3$(OH)$_6$Cl$_2$, all atoms, with $q\to0$ atomic form
factors, so that $I_{\rm Bragg}$ is the strongest conceivable
fundamental reflection).  The purely electronic contribution to a
superlattice reflection is then $I_{\rm super}/I_{\rm
Bragg}=|\delta n/n_{\rm tot}|^2\simeq10^{-7}$ to $3\times10^{-6}$ for
the stripe over the same doping range.  These figures are upper bounds,
because the diffracted structure factor itself is the Fourier amplitude,
$\sqrt2\,x/3$ at $M$ for the stripe and
$(1+\sqrt2)\,x/(6\sqrt3)=0.23\,x$ per member of the $K/2$ star.  The
structure-factor ratios are $4\times10^{-8}$ to $1.3\times10^{-6}$ for
the stripe and $9\times10^{-9}$ to $3\times10^{-7}$ per $K/2$
satellite.  The estimate omits powder averaging, Debye--Waller factors
and the interlayer structure of the crystal, and it assumes full
condensation of the doped charge.

Imposing the site-resolved Gauss law cuts the modulation further, by a
factor of $44$ at $x=1/6$ in the lattice calculation of
Appendix~\ref{app:gausslaw}.  At $x=1/36$, which for $U_B/t_B=4$ alone
lies inside the miscibility gap of Appendix~\ref{app:gausslaw}, the
homogeneous state does not suppress the modulation but moves it to the
longest wavelength the torus admits.  Its root-mean-square contrast
falls only by a factor of $1.2$.  Both factors are root-mean-square
contrasts at one set of parameters, and the spatial pattern changes
between the two states.  They rescale the summed Fourier power
and not any individual satellite, and correcting a specific intensity
would need a mode-resolved calculation.  The orbital fields of the
following subsection are reduced as well, by the smaller factors $1.4$
and $7.0$ at the same two dopings.

The accompanying lattice distortion, which dominates the satellite
intensity in known charge-density-wave systems, requires an
electron--phonon coupling that the chargon theory does not contain, and
we have no estimate of its size.  The electronic channel alone is a demanding
target for x-ray diffraction, so the test falls first to probes that see
the electronic modulation directly, that is, electron diffraction and
scanned probes (tunneling where a surface conducts, noncontact force
microscopy otherwise).  Synchrotron x-ray diffraction regains its
advantage if the lattice channel is appreciable.

The natural x-ray probe of an electronic modulation of this size in a
copper compound is resonant scattering at the Cu $L_3$ edge, where the
amplitude is dominated by the $3d$ valence and where charge orders of
comparable amplitude are measured in the cuprates
\cite{CominDamascelli2016}.  A modulation $\delta n$ of
the $3d$ occupancy then gives a resonant superlattice contrast of order
$\delta n^2$ if the resonant form factor follows the occupancy linearly,
i.e.\ $4\times10^{-4}$ to $10^{-2}$ for the $\delta n=0.02$ to $0.12$
per Cu of the stripe, against the $10^{-7}$ of the non-resonant
estimate.  Both stars are within reach at the edge
($|M|=0.53$~\AA$^{-1}$ and $|K/2|=0.31$~\AA$^{-1}$ against
$2k=0.94$~\AA$^{-1}$ at $930$~eV), and the resonant channel separates
the $\mathcal{T}$-even charge composites ($F_3$) from the current
composites ($F_2'$).  Neither contrast figure is a bound.  The resonant
one assumes a sensitivity that we have not calibrated, since the
contrast depends on the occupancy derivative of the complex resonant
form factor, on photon energy and polarization, and on the reference
reflection, and energy-dependent complex amplitudes and interference can
suppress the contrast as well as enhance it.

The $K/2$ target is the six-fold star
$\{(\tfrac13,\tfrac16),(\tfrac16,\tfrac13),(-\tfrac16,\tfrac16),
(-\tfrac13,-\tfrac16),(-\tfrac16,-\tfrac13),(\tfrac16,-\tfrac16)\}$ in
reciprocal-lattice units of the kagome cell, with the $60^\circ$ axes
$\mathbf{a}_1=(1,0)$, $\mathbf{a}_2=(\tfrac12,\tfrac{\sqrt3}{2})$ of
Appendix~\ref{app:conventions}.  In the standard hexagonal setting of
$R\bar3m$ herbertsmithite, $\gamma=120^\circ$, the same star is
$(\tfrac16,\tfrac16,l)$ and its symmetry equivalents, half of
$K=(\tfrac13,\tfrac13,0)$.  Its members have $|q|=2\pi/3a=|K|/2$ along
$\Gamma$-$K$ ($0.31$~\AA$^{-1}$ for $a=6.83$~\AA) and are
reciprocal-lattice vectors of the $\sqrt{12}\times\sqrt{12}$
($R30^\circ$) cell.  The $K/2$ crystal carries no $M$ or $K$ charge
component (its $M$- and $K$-star weight is loop current,
Sec.~\ref{sec:phaseII}), and the out-of-plane index is not fixed by the
single-layer theory.  The $M$-star satellites of the stripe are
extinguished by the intra-cell structure factor in every second zone.

Absence of any satellite on either star, in a cleanly doped sample and
at a sensitivity reaching the estimates above, would exclude the
fixed-parent crystals computed here at that doping, while a less
sensitive null constrains the unestimated lattice channel alone.  A
sensitive null would still leave the intrinsic chargon mechanism as such
open, since the parent flux pattern is itself expected to collapse above
the $x^*$ of Sec.~\ref{sec:strongcoupling}, which lies below the dopings
at issue.

The prediction can be tested on the Li-intercalated series
\cite{Kelly2016}.  To our knowledge no superlattice reflection has been
reported for it, and it remains to be checked whether its structural
characterization is sensitive enough to reach the electronic estimate.
The test concerns the dilute end of that series, at and below the
densest commensurate filling $x=1/6$.  Its upper range, which reaches
$3/5$ electron per Cu, lies beyond every commensurate filling and beyond
the melting doping of the $t$--$J$ numerics
(Sec.~\ref{sec:strongcoupling}), and falls to the extrinsic scenario.

Pressure is a complementary control parameter for the parent magnet.
Y-kapellasite, which orders magnetically at ambient pressure, loses its
static magnetism under hydrostatic pressure and is fully fluctuating at
$2.3$~GPa as the anisotropy of its kagome lattice is reduced
\cite{Ykapellasite2026}.  Pressure can therefore tune a clean kagome
magnet toward the spin-liquid regime.  The experiment has no
charge-sector content, since the compound is undoped, and it does not
bear on the crystallize-or-condense alternative posed here.  The same
alternative arises for the doped cluster-Mott spin-liquid candidate
1T-TaS$_2$, where gate doping of thin samples has been proposed to evade
dopant localization \cite{LawLee2017}.

In ultracold atoms the dispersive chargon problem is realizable in
principle with artificial gauge fields, by the optically induced staggered
fluxes of Ref.~\cite{MollerCooper2010} or the laser-controlled tunneling
phases discussed for kagome optical lattices in
Ref.~\cite{HuberAltman2010}.  The emulator would be bosons hopping on the
kagome lattice with the Bloch matrix $-t\,M(\kk)$ of
Sec.~\ref{sec:ansatz}, in the $[0,\pi]$ sign pattern and with the sign of
$t$ that places the flat multiplet at the top of the spectrum, at $+2t$,
and the four isotropic, nondegenerate minima at $-(1+\sqrt6)t$.

Such an emulator can test the leading selection of
Sec.~\ref{sec:phasediagram}, together with the density and current
patterns of the selected states.  Whether the condensate occupies one
translation pair or both, i.e.\ $M$-star against $K/2$-star order, is set
by the intersite repulsion measured against the crossover
$V_c\simeq9.3\,U^2x/t$ of Eq.~\eqref{eq:Vcrossover} and, for purely
on-site interactions at small $x$, by the two-boson (ladder) correction to
the projected quartic.  This correction does not lie on the Hubbard ray.
It contributes $E_{\rm I}-E_{\rm II^-}=-0.053\,U^2x^2/t$ per site and
selects the $M$ star below $x\simeq0.036$.

The octic lifting of the pair sphere is out of reach.  Wherever the
expansion of Eq.~\eqref{eq:octicderived} is controlled, the splitting of
the uni-$M$ and triple-$M$ orbits is smaller than the quartic energy by
the factor $0.263\,\epsilon_B^2$, with $\epsilon_B=U\bar\rho/t_B$ at
filling $\bar\rho$ per site.  An emulator also carries neither the
emergent gauge field nor the $B_+$ sector.  Its condensate is an ordinary
superfluid carrying the density and current patterns of
Sec.~\ref{sec:phases} without the Higgs structure, and nothing of the
pairing analysis of Sec.~\ref{sec:z2} transfers to it.

\subsection{Fingerprints of the chargon orders}
\label{sec:fingerprints}

\emph{Charge sector.}  The charge order is $2\times2$ or stripe, with an
$F_3$ ($C_2$-odd) $M$-triplet.  It has no cubic invariant, hence no
symmetry-mandated first-order transition, and its TrH-like and SoD-like
registries are exactly degenerate.  The $C_2$ parity holds at every
order for the uni-$M$ stripe, and up to an $F_1$ admixture of relative
order $Ux/t$ for the triple-$M$ state (Sec.~\ref{sec:landau}).  Scanning
tunneling microscopy needs a conducting surface, which the insulating
doped materials realized so far do not offer.  The parity must therefore
be read from diffraction, through the zone-to-zone intensity ratios and
extinctions of the $M$-star satellites, or from noncontact force
microscopy, which images lattice and charge modulations on insulating
surfaces.  The uni-$M$/tri-$M$ landscape is soft.  It is split only at
octic order, by $1/57=1.75\%$ of the octic moment and by
$0.351\,(Ux/t)^3\,xt$ per site in the derived octic
[Eq.~\eqref{eq:octicderived}].  The selected pattern should be
extremely sensitive to strain and disorder, and the stripe should form
three $120^\circ$ domain families.

\emph{Current sector.}  The currents are staggered triangle loop
currents.  Their $M$-modulated part switches on at $T_{\rm CO}$, while
the $\Gamma$-point part is symmetry-allowed to set in already in the
upper stage described at the head of this section.  For the uni-$M$
stripe the circulations are
$\tfrac32(\sqrt3-\sqrt2)\mp\tfrac12(3-\sqrt6)=\{0.2015,\,0.7520\}\,t\bar\rho$
on up-triangles and opposite on down-triangles, in the convention
$J_{ij}=2\,\mathrm{Im}[\bar B_i s_{ij}B_j]$ of the
Fig.~\ref{fig:condensates} caption, with $\bar\rho$ the mean chargon
site density.  The pattern is the sum of a $\Gamma$-point $B_2'$
staggered flux and an $F_2'$ modulation at the activated $M$ points.
The flux is $\tfrac32(\sqrt3-\sqrt2)\,t\bar\rho$ per up-triangle on
average and is saturated by every phase-I condensate
($|\Delta N|=\sum_\eta n_\eta$ over the whole moduli sphere).  The
modulation has amplitude $\tfrac12(3-\sqrt6)\,t\bar\rho$ for the
single-arm stripe, smaller by a factor of $\sqrt3$.

For the uni-$M$ stripe both parts carry zero net moment, which its
unitary glide mirror enforces.  Local probes ($\mu$SR, NMR line
broadening) and polarized neutron diffraction see them, but the uniform
response functions (polar Kerr, zero-field anomalous Hall), which
transform as $A_2'$, do not.  The triple-$M$ state and the $K/2$
crystal are not so protected (magnetic classes $3m'$ and $6m'm'$,
Sec.~\ref{sec:phaseI}).  Their three current arms form the $A_2'$
composite $\Delta_1'\Delta_2'\Delta_3'$, so a Kerr or Hall signal is
symmetry-allowed there, cubic in the current bilinears and hence of
order $x^3$.  The uniform responses discriminate the uni-$M$
stripe from the multi-arm states.

The local fields below are order-of-magnitude, single-layer, mean-field
estimates for a stated geometry, and they vary from site to site.  They
assume full condensation ($\bar\rho=x$), treat the bond currents as
classical line currents, and fix the sign pattern only up to the overall
sense of circulation.  The current that enters them is that of the
frozen link field of the undoped ansatz, and the numbers are not
evaluated at a jointly stationary, constraint-consistent saddle.  The
feedback-dressed $\chi_{ij}(B)$ of Sec.~\ref{sec:electronspectrum}
corrects them at relative order $t_cx/t_{\rm sp}$, which is of order one
over the doping window (Sec.~\ref{sec:electronspectrum}).

With $I_{ij}=eJ_{ij}$, $J_{ij}=2\,\mathrm{Im}[\bar
B_is_{ij}B_j]\,t\bar\rho/\hbar$ (the Heisenberg current of the hopping
Hamiltonian) and $\bar\rho=x$, the largest bond current of the uni-$M$
stripe is $0.884\,etx/\hbar\simeq0.6$ to $5\,\mu$A for the bare
$t=|t_c|=108$ to $139$~meV and $x=1/36$ to $1/6$ ($0.3$ to $2\,\mu$A for
$t=t_B$).  Currents and fields scale linearly with $t$.  The site values
below are for the bare $|t_c|$ and are smaller by the factor
$t_B/|t_c|=0.44$ at the mean-field chargon hopping $t_B$.

Biot--Savart integration over the bond currents of the periodically
repeated $12\times12$ torus, treating each bond as a straight segment of
length $a/2=3.42$~\AA\ (with $a=6.83$~\AA\ for herbertsmithite), gives a
site-dependent orbital field that scales as $B\simeq10\,(36x)\,(t/108\,{\rm
meV})$~G.  In the kagome plane the field is normal to the plane.  It is
$10$ to $13$~G at a hexagon center and $6$ to $9$~G at a triangle center
for $x=1/36$, rising to $60$ to $77$~G and $35$ to $55$~G at $x=1/6$,
with opposite sign on the two hexagons of the doubled cell and on up-
and down-triangles.  At $1.5$~\AA\ above a triangle center, a plausible
muon site, it is $6$ to $9$~G at $x=1/36$ and $39$ to $55$~G at
$x=1/6$, mostly in-plane there.  It falls to about half by $3$~\AA\ and
to a tenth by $5$ to $6$~\AA\ above the plane.

Over all sites within $3$~\AA\ of the plane the range is $3$ to $20$~G
at $x=1/36$ and $18$ to $118$~G at $x=1/6$, that is $0.04$ to $1.6$~MHz
of muon precession, one to three orders of magnitude above the
characteristic fields reported by zero-field $\mu$SR at the charge-order
transitions of the kagome metals, which range from about $0.1$~G
(CsV$_3$Sb$_5$, a zero-field relaxation-rate step of
$0.0075~\mu{\rm s}^{-1}$ at $T_{\rm CDW}$ \cite{Khasanov2022}) to
$0.3$~G (KV$_3$Sb$_5$ \cite{Mielke2022}).

In herbertsmithite the muon binds to a hydroxyl group, the OH--$\mu$
complex identified by zero-field $\mu$SR \cite{Mendels2007} (with a
minority Cl site), at about $1$~\AA\ from the hydroxyl O in an unrefined
direction.  Over the possible directions the field is $6$ to $16$~G at
$x=1/36$ and $35$ to $98$~G at $x=1/6$, and the height scan brackets
the Cl site and other choices.

The two NMR ligand nuclei of herbertsmithite, $^{17}$O and $^{35}$Cl,
sit off the current paths and can be estimated the same way.  The
$^{17}$O of the hydroxyl bridging each Cu--Cu bond sits $0.86$~\AA\ out
of the plane and $0.53$~\AA\ in-plane from the bond midpoint toward the
adjacent hexagon center ($1.0$~\AA\ from the midpoint; Cu--O $1.98$~\AA,
Cu--O--Cu $119^\circ$, Cu--Cu $3.42$~\AA\ \cite{Shores2005}), where the
field is $11$ to $20$~G at $x=1/36$ and $65$ to $118$~G at $x=1/6$.
The $^{35}$Cl caps each Cu$_3$ triangle on its threefold axis
$1.94$~\AA\ above the triangle center (Cu--Cl $2.77$~\AA\
\cite{Shores2005}), where it is $5$ to $7$~G and $31$ to $44$~G (all
ranges, here and above, over the $t$ interval and the inequivalent
members of the doubled cell).  The Cu nucleus sits on the current
paths.  Its orbital hyperfine field requires the current density within
the $3d$ shell, not line currents, so we quote no number for it.

The site densities also leave a signature at the Cu nucleus that needs
no current estimate.  Through the change of the electric-field gradient
with the $3d$ occupancy, and following the density levels and
multiplicities of Secs.~\ref{sec:phaseI} and \ref{sec:phaseII}, the
inequivalent sites of each crystal split the $^{63,65}$Cu nuclear
quadrupole resonance line generically into
\begin{itemize}
\item three lines of equal weight ($1{:}1{:}1$) for the stripe;
\item three lines of weights $1{:}2{:}1$ for the triple-$M$ state;
\item five lines of weights $1{:}1{:}2{:}1{:}1$ for the $K/2$ crystal,
\end{itemize}
with the corresponding $^{17}$O and $^{35}$Cl splittings set by the bond
and triangle densities.  The NQR line multiplicity settled the
charge-order pattern in CsV$_3$Sb$_5$ \cite{NikolovNQR2025}.  The
multiplicities listed are idealized and can err in either direction.
Equal density need not mean equal gradient, which can raise the
multiplicity, while an accidental coincidence of gradients at
inequivalent densities, or disorder broadening, can lower the resolved
count.  An estimate of the size of the splitting would require the
gradient per unit occupancy.

The Cu$^{2+}$ paramagnetic and nuclear-dipolar background of
herbertsmithite is of comparable magnitude, but its static electronic
part is small.  The same zero-field $\mu$SR experiment bounds any static
electronic field at the muon site of the undoped compound at about
$0.3$~G \cite{Mendels2007}.  The nuclear-dipolar part is static but
temperature-independent, so it does not switch on at a transition.  The
discriminating signal is therefore a static component whose line shape
switches on at $T_{\rm CO}$ (or, for the $\Gamma$-point part, in the
upper stage), whatever its size.

A magnetic field couples the $F_3$ charge order linearly to the $F_4'$
flux channel, the $\mathcal{T}$-odd $M$-triplet that is odd under both
$C_2$ and $\sigma_v$ (the product of $F_3$ with the $A_2'$ of the field,
and not $F_2'$) \cite{RMP2026}.  No term linear in the field and in $F_3$
therefore selects the sense of circulation of the $F_2'$ currents, so the
field-training protocols of the $F_1$ kagome metals do not apply through
the charge order, and domain-selective protocols remain.  Uniaxial strain
selects one of the three $F_3$ stripe orientations and with it the
momentum of the accompanying $F_2'$ arm.  Polarized neutron diffraction,
at the nuclear Bragg positions for the $\Gamma$-point $B_2'$ flux and at
the $M$ and $K$ stars for the modulated parts, detects the staggered
orbital moments without training.

\emph{Pair sector.}  Superconductivity, when present, is a pair-density
wave on the phase-I and II$^-$ arcs, with no uniform charge-$2e$
component at the quartic level.  Phase~I carries $K/2$-star modulations
alone (period 3 along $\Gamma$-$K$,
$\sqrt{12}\times\sqrt{12}\,R30^\circ$), and branch~II$^-$ carries these
together with $M$-star (period-2) modulations, both with the
descendant-resolved irreps of Table~\ref{tab:pairing}.  A
first-harmonic ($\sin\phi$) Josephson null against a conventional lead
is the classic pair-density-wave diagnostic \cite{Agterberg2020}.
Because disorder, strain, surfaces and domain averaging all generate a
uniform component, a null is interpreted relative to a control junction
on a uniform superconductor.

The positive test is the charge of the lowest uniform pair
composite (Sec.~\ref{sec:pairingmap}).  The II$^-$ crystal has a
second-harmonic ($\sin2\phi$) Josephson coupling and an $h/4e$ flux
period of the vestigial phase, the triple-$M$ state has an $h/6e$
period, and the uni-$M$ stripe has no low-order uniform pair channel at
all.  The pairing channels are the $A_1$/$B_1$/$B_2$/$A_2$ assignments
of Sec.~\ref{sec:pairingmap}.  The vortex and edge features quoted
there correspond to them channel by channel and are not computed spectra.
The channels carry no uniform $d+id$ at leading order, nor at any order
on the $M$-star condensates.

\subsection{Kagome metals: convergences and discriminators}
\label{sec:kagomemetals}

The kagome metals lie outside the regime of this theory.  Their CDWs are
$F_1$-type, weakly first order, with substantial phonon involvement
\cite{RMP2026}.  The orders of the metals and of the doped insulator
are nevertheless classified by the same space-group irreducible
representations, and recent experiments make the comparison concrete.

In weak-coupling theory, sublattice interference suppresses all
conventional weak-coupling instabilities \cite{KieselThomale2012}, renders
the local repulsion irrelevant and, in a large-$V$ model at intermediate
coupling, promotes $M$-point pair-density waves without uniform
superconductivity \cite{WuThomaleRaghu2023}.  It also leads to $M$-star
loop-current order whose Landau trilinear vanishes
\cite{FuNSR2025,ZhanLCO2026}.  That order and the loop currents that the
strong-coupling chargon theory derives (Secs.~\ref{sec:catalog},
\ref{sec:phaseI}) both transform in the $F_2'$ channel, with the same
cubic-free structure.  Chargons in the doped spin liquid and electrons in
the kagome metals thus both prefer finite-momentum pairing.  Whether the
spinons of the undoped magnet do as well, as the proposed spinon
pair-density wave would imply \cite{DuricPDW2025}, is unsettled, since its
variational energies have been challenged as a sampling artifact
\cite{KamalComment2026}.  The recent developments below can be read in
chargon-theoretic terms, although shared irreps do not imply shared
mechanisms.

(i)~\emph{Inversion-breaking charge order.}  Combined STM/nc-AFM/SHG on
the $A$V$_3$Sb$_5$ family finds that in-plane inversion symmetry is
spontaneously broken in the $2\times2$ CDW state.  The authors report a
mixed-parity order with ferro-polar (dipolar) and nematic (quadrupolar)
moments whose coupling produces an electronic chirality
\cite{ShiParity2026}.  Its ferro-polar ($\Gamma$-point $E_1$) and
chiral components appear nowhere in Table~\ref{tab:ph}.  The
inversion-odd $M$-point part of such a state lies in the $C_2$-odd
site-density channel $F_3$ or $F_4$, mixed with $F_1$.  These two
differ only in mirror character, which is not reported.  The chargon
mechanism populates $F_3$ directly, whereas in the weak-coupling
theories it appears as an induced subsidiary order.  (Bulk NQR on
hole-doped CsV$_3$Sb$_{5-x}$Sn$_x$ still fits the staggered
trihexagonal $F_1$ state \cite{NikolovNQR2025}.)

Site-selective $^{121}$Sb NQR on CsV$_3$Sb$_5$ resolves a magnetic line
broadening below $T^*\simeq120$~K, consistent with spontaneous local
fields of order $1$~mT from loop currents \cite{Suetsugu2026}.  The
authors tie that onset to the nematic transition above $T_{\rm
CDW}\simeq94$~K.  Staggered fields that precede the charge order are
what the upper stage of Sec.~\ref{sec:experiment} would permit, with
$\Gamma$-point $B_2'$ flux and $E_2$ nematic composites
symmetry-allowed to order before the $M$ triplet.  The same material has
a null polar Kerr response below the charge-order transition
\cite{SaykinKerr2025}.  A $\mathcal{T}$-odd order with no net moment,
visible to a local probe and invisible to $\sigma_{xy}$, has the
symmetry of the staggered $B_2'$ and $F_2'$ channels of
Table~\ref{tab:ph} in their single-arm (uni-$M$) realization, whose
unitary glide mirror forbids the $A_2'$ composite that three arms would
form.  The observation agrees with the symmetry classification of (iii)
below, while the quantitative test of the present currents remains the
local-field estimate of Sec.~\ref{sec:fingerprints} for doped
herbertsmithite.

(ii)~\emph{Softness of the charge-order landscape.}  In
CsV$_{3-x}$Ti$_x$Sb$_5$, Ti substitution as dilute as $x=0.009$
suppresses the three-dimensional $2\times2\times4$ CDW of the pristine
compound and reduces the remaining $2\times2\times2$ order to a
quasi-two-dimensional one.  By $x=0.2$ x-ray diffraction and STM find
instead a (quasi-)one-dimensional stripe CDW with a correlation length
of only $\sim\!20$~\AA, whose intensity and correlation length grow
continuously below $\sim\!56$~K as at a \emph{second-order} transition
\cite{XiaoStripe2026}.  Short-range nematic domains outlive the
long-range CDW entirely \cite{XuNematic2025}.  Neither observation
tests the present theory, whose charge order is $F_3$ while that of the
metals is $F_1$.  They show that a stripe channel and a
second-order-like onset are available on the kagome lattice once the
first-order mechanism of the pristine $F_1$ order is disrupted, as they
would be for the cubic-free $F_3$ transition and the uni-$M$/tri-$M$
near-degeneracy described here.

(iii)~\emph{TRSB channel selection.}  High-resolution Sagnac
interferometry finds no spontaneous Kerr rotation in CsV$_3$Sb$_5$ and
ScV$_6$Sn$_6$ down to a noise floor below $\sim\!100$~nrad, including
under strain \cite{SaykinKerr2025}.  Meanwhile $\mu$SR reports
time-reversal breaking at the CDW in KV$_3$Sb$_5$ \cite{Mielke2022} and
in CsV$_3$Sb$_5$ itself \cite{Khasanov2022,Yu2021}, and
circular-dichroism ARPES reports it well above the CDW transition
\cite{TRSBaboveCDW2026}.  Staggered flux orders with zero net moment
reconcile the two sets of observations, and the uni-$M$ stripe of
phase~I carries such orders.  Its $\Gamma$-point $B_2'$ flux and its
single activated $F_2'$ arm are both staggered.  Neither couples
linearly to $\sigma_{xy}$, which transforms as the pseudoscalar $A_2'$
($C_2$-even, mirror-odd, $\mathcal{T}$-odd), and the stripe's unitary
glide mirror forbids any $A_2'$ component.

The Kerr null is consistent with the uni-$M$ stripe, and it
constrains the multi-arm states.  The triple-$M$ state ($3m'$) and the
$K/2$ crystal ($6m'm'$) carry the $A_2'$ composite
$\Delta_1'\Delta_2'\Delta_3'$ of their three current arms, which is
absent from Table~\ref{tab:ph} only because that table lists bilinears.
A Kerr rotation of order $x^3$ is symmetry-allowed in both,
with a size that the present theory does not fix.

(iv)~\emph{An intrinsic pair-density wave.}  Following the original STM
observation of a $\frac{4a}{3}$ ``roton'' pair modulation
\cite{ChenPDW2021}, phase-sensitive tunneling in KV$_3$Sb$_5$, read
through Ginzburg--Landau theory, supports a sign-changing $M$-star PDW
that is primary and not induced \cite{YanPDW2026}
(Sec.~\ref{sec:pairingmap}).  STM finds a field-trainable chiral
$2\times2$ pair modulation, interpreted as a pair-density wave of the
vanadium $d$-orbital sector, with the uniform pairing carried by the Sb
$p$-orbital sector \cite{SongPDW2025}.  Ring interferometry shows
$\pi$-shifted Little--Parks oscillations in nearly every device
\cite{WangRing2025}, indicating a sign-changing multicomponent
condensate.  Between the $\pi$- and $0$-phase regimes it also shows
$h/4e$-periodic oscillations, which the authors attribute to a
second-harmonic crossover between a dominant sign-changing and a
subdominant uniform component and not to charge-$4e$ order.

In both materials a substantial uniform pair component coexists with the
modulation.  The pure pair-density-wave arcs of the descendants, with no
uniform charge-$2e$ component at the quartic level, are not
realized there, and the agreement is confined to the $M$-star irrep
classification of the modulated part.  The topological-PDW scenario
requires the $M$ pair triplet to induce a uniform $d+id$ component at
third order \cite{YinTPDW2026}.  The same analysis expects that
component to be so small that the state is effectively gapless, with
Bogoliubov Fermi surfaces, at accessible temperatures.  An $F_3$ pair
triplet forbids this cubic, so its Bogoliubov Fermi surfaces would be
symmetry-enforced instead of approximate.  Since the two cases look
alike to spectroscopy, the discriminator between an $F_1$-type and an
$F_3$-type pair multiplet is the $C_2$ parity of the triplet itself, or
the presence or absence of any $Q=0$ pair component in a
phase-sensitive (Josephson) measurement.  For the descendants' own
$M$-star condensates no uniform component arises at any order, while on
the $K/2$ branch one is induced at the next order
(Sec.~\ref{sec:pairingmap}).

(v)~\emph{The $K$ and $K/2$ stars.}  The $C_3$-symmetric
$\sqrt3\times\sqrt3$ CDW of ScV$_6$Sn$_6$ is a $K$-point \emph{charge}
order, which the chargon $K$ channel does not produce.  Its site-density
form factor and its bond charge out to third-neighbor range vanish
identically (Sec.~\ref{sec:catalog}), and it carries instead a
$\mathcal{T}$-odd $\sqrt3\times\sqrt3$ loop-current crystal.
Superconducting CeRu$_3$Si$_2$ adds a case whose
$\mu$SR-only time-reversal breaking remains compatible with a
current-carrying PDW in place of $d+id$.  It has dual flat bands, a
dominant $M$-star charge order and a weak period-3 order.  The
satellites of the latter appear at $(\tfrac13,\tfrac13,l)$-type
positions, but the order is assigned to three single-$q$ orthorhombic
($Imma$) domains and not to a $K$-point order \cite{GerguriCeRu2026}.

The $K/2$ star remains the open channel.  The unidirectional
period-$4a$ stripe seen by STM in CsV$_3$Sb$_5$
\cite{ZhaoCascade2021,ChenPDW2021} sits at one quarter of a
reciprocal-lattice vector, $(\tfrac14,0)$ in the reciprocal-lattice
units of Sec.~\ref{sec:materials} and $(\tfrac14,-\tfrac14)$ in the
$R\bar3m$ setting, along $\Gamma$-$M$ in both.  Its wavevector, of
length $\pi/(\sqrt3\,a)$, is $13.4\%$ shorter than $|K/2|=2\pi/(3a)$ and
rotated by $30^\circ$ from the $\Gamma$-$K$ directions of the $K/2$
star, so the stripe is not a $K/2$ order.  Its wavelength
$2\sqrt3\,a=3.46a$, the $4a$ repeat along the lattice row times
$\cos30^\circ$, is correspondingly $15.5\%$ longer than the $3a$ of the
$K/2$ modulation.  That wavevector is the projection onto the $\Gamma$-$M$
axis of the two adjacent $K/2$ vectors, each of which differs from it by
a transverse component of length $\pi/(3a)$.  Detecting or excluding a
satellite at $|K|/2$ along $\Gamma$-$K$, as opposed to one at
$(\tfrac14,0)$, would establish whether any kagome material realizes the
thirty-six-site crystal.

(vi)~\emph{Parity through nonlinear transport.}  The $C_2$ parity that
separates the charge order of the metals from the chargon one is, at the
$M$ point, a statement about inversion, and nonlinear transport sorts the
two orders by it without a phase-sensitive measurement.  A second-order
response $j\propto E^2$ reverses sign under inversion while $E^2$ does
not, so it is forbidden outright for an inversion-even $F_1$ charge order
and symmetry-allowed for an inversion-odd $F_3$ one.  A third-order
response is allowed either way.  In CsV$_3$Sb$_5$ the third-order
longitudinal and Hall responses are large and survive to room temperature,
with kinks at the charge-order transition and at the $\sim\!39$~K anomaly,
while the second-order responses are reported as negligible
\cite{DaiNonlinear2026}.  That null is what the $F_1$ assignment of the
metals requires, and it fixes the measurement's sensitivity in the
reference material.  A doped kagome spin liquid ordering in $F_3$ should
carry a second-order response where the metals carry none, provided the
order also breaks threefold rotation and does not cancel between layers.
The intrinsic second-order response is governed by the Berry-curvature
dipole, a pseudovector in two dimensions, which any threefold rotation
annihilates.  The signal therefore requires the $C_3$ breaking of the
uni-$M$ stripe, which the triple-$M$ and $K/2$ crystals lack, in domains
whose average does not restore $C_3$.  In a stacked sample an in-plane
inversion-odd order can in addition cancel between layers.  A second-order
null thus excludes the inversion-odd order only where both conditions are
established.  The null reported for the metals is also in tension with
the spontaneous in-plane inversion breaking reported in (i), which was
measured in thicker bulk crystals at a higher charge-order temperature,
and the two observations have not yet been reconciled.

\section{Conclusions and outlook}
\label{sec:conclusions}

We have constructed the chargon theory of the kagome Dirac spin liquid and
of its $\Z_2$ and chiral descendants, following the analyses of the square
\cite{Christos2023} and triangular \cite{Feuerpfeil2026} lattices, and the
kagome lattice has returned an answer those lattices could not give.  The
chargon doublet that carries the doped charge is degenerate on the square
and triangular lattices, where the interactions decide between
superconductivity and density-wave order.  On kagome the lattice decides
first.  The two triangle orientations carry equal flux in the parent
ansatz, no symmetry undoes a reversal of that flux, and the doublet splits
into a component that disperses in four valleys, the holon band of
Ref.~\cite{Ko2009}, and a component whose lowest bands are flat, pinned by the line-graph structure of the lattice
(Sec.~\ref{sec:bands}).  The splitting between the two band bottoms is a
closed-form number, and over the window it defines only the dispersive
component condenses.  The doped spin liquid then forms charge and
loop-current crystals with the electromagnetic U(1) unbroken and no
superfluid response, and its electron-like quasiparticles are Fermi
pockets of Luttinger volume $x$ about the two Dirac nodes
(Sec.~\ref{sec:electronspectrum}).  Superconductivity needs the pair
operator to see both gauge components
\cite{WenLee1996,LeeNagaosaNgWen1998,LeeNagaosaWen2006}, which requires a
paired $\Z_2$ descendant as the parent, and it then arrives as a
pair-density wave.  Each descendant admits one uniform channel, none of
them $d+id$, and for the $M$-star crystals a uniform pair channel is
excluded at every order.  The insulating behaviour of the doped
herbertsmithite series is consistent with this picture only if a further
broken symmetry or topological order gaps the pockets, which the
Lieb--Schultz--Mattis count requires and the quadratic spinon problem does
not supply.

The symmetry analysis is where the results are sharpest.  The projective
action of the lattice symmetries on the chargon valleys leaves one
quadratic and three quartic invariants, all in closed form, so the whole
quartic competition reduces to a single angle.  Within the $M$-star phase
the competing charge crystals form a moduli space that stays degenerate
through sextic order, protected by an octahedral action that closes on the
full projective symmetry group (Appendix~\ref{app:theorem}), and it is
lifted only at octic order, where the term derived for on-site repulsion
selects the uni-$M$ stripe [Eq.~\eqref{eq:octicderived}].  The projected
on-site repulsion sits on the boundary between the $M$-star and $K/2$-star
condensates, a direction we call the Hubbard ray, and an enlarged symmetry
of the on-site quartic protects that ray to all orders under the gauged
renormalization group (Sec.~\ref{sec:hubbardray}).  Nearest-neighbour
repulsion moves the system off the ray toward the $M$ state
(Sec.~\ref{sec:phases}).  The chiral descendant, beyond a sector crossing,
adds a flat $\mathbb{CP}^1$ manifold of its own, resolved only by intersite
loop terms.  Such degeneracies make the selected order sensitive to strain,
doping and disorder, and they suggest a reason for the documented fragility
of kagome charge orders.

The Dirac spin liquid is thus a parent state on both sides of the Mott
transition.  Its fermion bilinears and monopoles account for the magnetic
and valence-bond orders of the insulator, and its chargon multiplet for the
superconducting, pair-density-wave, charge-ordered and current-carrying
phases of the doped system.  The chargon composites realize channels that
the insulating theory does not contain.  The charge order is a $C_2$-odd
triplet ($F_3$) with no cubic invariant, and hence no symmetry-mandated
first-order transition.  A loop-current triplet ($F_2'$) is locked to it
through the trilinear of Eq.~\eqref{eq:lock} wherever the $\Gamma$-point
flux orders.  A $K$-star loop-current channel and $K/2$-star charge
crystals appear at momenta that no operator of the insulating theory
carries.  No chargon composite transforms like a mass or a monopole of the
parent.  The weak-coupling theory of the kagome metals uses the same
$M$-star irreps and shares the loop-current channel, and it differs in the
parity of the charge order, $F_1$ against $F_3$ here, in the presence of
the Landau cubic, and in coupling strength, an itinerant metal without
local moments \cite{RMP2026} against a Mott insulator of $S=1/2$ moments.
That difference in parity is the one discriminator of
Sec.~\ref{sec:kagomemetals} that ordinary transport can settle, through the
second-order nonlinear response that inversion forbids for one order and
allows for the other, and the $K/2$ star, long discussed for kagome without
an experimental verdict, is a diffraction target.

The limits of the construction are those of a fixed parent and a
mean-field treatment.  The band structure, the projective classification
and the Landau theory hold for the nearest-neighbour ansatz and its
projective symmetry group.  The statement that the flat-band component
stays empty holds at the Hartree level.  The modulation amplitudes are
computed with the Gauss law imposed on average and shrink substantially
under the site-by-site constraint, without change to the symmetry
assignment (Appendix~\ref{app:gausslaw}).  The crystal selection at the
physical coupling is an extrapolation, since the selections beyond quartic
order rely on an expansion in the charging energy that is of order one
there (Sec.~\ref{sec:phasediagram}), and the finite-temperature transitions
are ordered in scale but not calculated.  Whether the doped chiral state
lies beyond its sector crossing is not established
(Sec.~\ref{sec:cslsectors}).

Several directions follow.  The condensation transitions, with the U(1)
gauge field and, in the chiral case, a Chern--Simons term, call for a
critical theory beyond mean field \cite{KaulSachdev2008}, and gauge
fluctuations may reweight density-wave against superconducting channels
\cite{Feuerpfeil2026}.  Whether the insulating crystals found by
strong-coupling numerics \cite{Jiang2017,Peng2021,XuGuYang2024} acquire
their gap from confinement, from loss of single-chargon coherence or from
interactions beyond the quadratic problem is open.  The leading coupling
between the chargon and monopole sectors, a singlet monopole times a term
quartic in the chargon field, remains to be evaluated.  If the undoped
ground state is the proposed spinon pair-density wave \cite{DuricPDW2025},
a proposal currently contested \cite{KamalComment2026}, the construction
extends to it and adds a $\Z_2$ Higgs channel at momentum $M$ that seeds
$M$-star electron pair-density waves upon doping.  The methods, projective
valley representations, composite catalogs and moduli-space analysis, apply
to any parton band structure that undergoes a Higgs transition, and the
maple-leaf Dirac spin liquid, whose Higgs criticality
\cite{FeuerpfeilDepleted2026} and monopole spectrum \cite{MapleLeaf2026}
have been analyzed recently, and the three-dimensional pyrochlore case are
the natural next applications.

\begin{acknowledgments}
I thank Pietro M.~Bonetti, Andreas Feuerpfeil, Subir
Sachdev and Ronny Thomale for valuable discussions on the
chargon theory and for collaborations on companion works.  The work of
Y.I.\ was performed, in part, at the Aspen Center for Physics, which is
supported by a grant from the Simons Foundation (1161654, Troyer).  This
research was also supported in part by grant NSF PHY-2309135 to the Kavli
Institute for Theoretical Physics (KITP).  Y.I.\ acknowledges support from
the Abdus Salam International Centre for Theoretical Physics (ICTP)
through the Associates Programme, from the Simons Foundation through
Grant No.~284558FY19, from IIT Madras through the Institute of Eminence
(IoE) program for establishing QuCenDiEM (Project
No.~SP22231244CPETWOQCDHOC), and the International Centre for Theoretical
Sciences (ICTS), Bengaluru for participation in the Discussion
Meeting---Fractionalized Quantum Matter (code: ICTS/DMFQM2025/07).
\end{acknowledgments}

\section*{Data availability}

The code and data that support the findings of this article are
available from the author upon reasonable request.

\appendix

\section{Gauge conventions and verification methodology}
\label{app:conventions}

\emph{Gauge.}  We use $\mathbf{a}_1=(1,0)$, $\mathbf{a}_2=(\tfrac12,
\tfrac{\sqrt3}{2})$, sublattice sites at $\{0,\mathbf{a}_1/2,
\mathbf{a}_2/2\}$, and the magnetic cell $\mathbf{A}_1=2\mathbf{a}_1$,
$\mathbf{A}_2=\mathbf{a}_2$ with the explicit sign pattern of
Fig.~\ref{fig:ansatz}.  There are twelve bond classes per magnetic
cell, every triangle product is $+1$, and every hexagon product is $-1$.
The
reciprocal vectors satisfy $\mathbf{G}_i\cdot\mathbf{A}_j=2\pi\delta_{ij}$,
and the magnetic Brillouin zone of an $L\times L$ torus (in units of
$\mathbf{a}_i$) contains $(L/2)\times L$ momenta.

\emph{Symmetries as operator identities.}  Every lattice symmetry $g$
(translations; $C_6$, $C_3$, $C_2$ and mirrors about the hexagon center;
antiunitary $\Theta$) is implemented as a site permutation combined with a
$\Z_2$ gauge transformation obtained by propagating outward from a
reference site along a spanning tree of the torus, with the gauge factor
attached at the \emph{preimage} site.  The
resulting unitaries satisfy $[U_g,H]=0$ to within numerical roundoff on the
$12\times12$ torus, and the valley representation of Table~\ref{tab:psg}
is obtained by projecting onto the four valley eigenstates.  All entries
were reproduced independently on a $24\times24$ torus.

\emph{Valley gauge.}  Individual valley eigenvectors carry arbitrary
phases set by the diagonalization.  Representations obtained from
independent diagonalizations therefore agree only \emph{modulo diagonal
valley rephasings} $B_\eta\to e^{i\alpha_\eta}B_\eta$.  The invariants
$\sum_\eta n_\eta$ and $\mathcal{P}$ are insensitive to these phases, but
the Umklapp part of $\mathcal{X}$ is not.  Under a rephasing the Umklapp
monomial of Eq.~\eqref{eq:Xinv} and its conjugate acquire the phases
$e^{\pm i(\alpha_2+\alpha_4-\alpha_1-\alpha_3)}$.  The real combination
printed there is thus the invariant only in a gauge in which that phase
is trivial, and statements about specific coefficient vectors (for
example, which balanced four-valley state is phase-locked) depend on the
gauge in the same way.  We fix the gauge instead of working in
rephasing-invariant form.  Requiring the translation generators $T_1$ and
$T_2$ to take the printed form of Table~\ref{tab:psg} forces
$\alpha_1=\alpha_2$ and $\alpha_3=\alpha_4$ (the printed $T_1$ carries the
same phase in both directions of each exchange).  This leaves one common
phase per translation pair, and the Umklapp term is invariant under it.
In this gauge the Umklapp coefficient is real and the closed forms of
Eqs.~\eqref{eq:Pinv}--\eqref{eq:Xinv} hold as printed.  All quoted
matrices and coefficient vectors refer to this gauge.

\emph{Flux checks.}  We checked every gauge-invariant loop product on the
torus.  For the DSL and $\Z_2$ states these are all up- and
down-triangles and all hexagons.  For the chiral state we also checked
\emph{all four} oriented fluxes of Eq.~\eqref{eq:cslfluxes}, including
the mixed triangle $\theta_{abc}$.  Without that flux the global sign of
the second-neighbor pattern, which the symmetry nullspace fixes only up
to sign, would remain undetermined (Appendix~\ref{app:csl}).

\emph{Invariant averaging.}  The invariant counts of
Sec.~\ref{sec:landau} follow from an explicit group average.  The unitary part of the projective symmetry group has 144
elements modulo the overall sign, or 288 as $4\times4$ matrices, and the
scalar $-1$ acts trivially on every composite.  (The 288-element group
used in Appendix~\ref{app:theorem}, which contains both unitary and
antiunitary elements, is counted modulo the same sign.)  The Reynolds
average of the induced action over the 144 unitary elements has trace $1$
on the 16-dimensional space of quadratic tensors $\bar B_\eta B_{\eta'}$.
On the 100-dimensional space of Hermitian quartics it has trace $3$ and
rank $3$, which is the count of Eq.~\eqref{eq:potential}.  The same
average over the translation subgroup alone (24 of the 288 matrices) has
trace $10$, over $\langle T_1,T_2,C_3\rangle$ (72 matrices) trace $4$,
and over $\langle T_1,T_2,C_6\rangle$ (144 matrices, the index-two
subgroup without the mirror) already trace $3$.  Time reversal is
therefore not needed for this count, which the 144-element unitary group
produces on its own.

\emph{Numerical settings.}  Unless stated otherwise, lattice results
refer to the $12\times12$ torus of crystallographic cells, with $N=432$
sites, $864$ nearest-neighbor bonds and $72$ magnetic cells, periodic
boundary conditions and the gauge above.  The operator identities behind
Table~\ref{tab:psg} and the flux checks were repeated on the
$24\times24$ torus ($N=1728$).  Extrema of the Bloch bands are located by
Nelder--Mead refinement from a $25\times25$ grid of seeds covering the
magnetic Brillouin zone, converged to $10^{-12}$ in momentum and
$10^{-14}$ in energy.  The chiral sectors of Sec.~\ref{sec:cslsectors}
are scanned on $N_g\times N_g$ grids up to $N_g=192$ and refined in the
same way (Appendix~\ref{app:csl}).  Because $N_g$ is a multiple of 12,
these grids contain every sector minimum exactly.  Landau functionals on
the condensate manifold are minimized by multistart BFGS (gradient
tolerance $10^{-12}$, 40 to 60 random starts) and by Nelder--Mead
(40 starts, tolerances $10^{-10}$ in the variables and $10^{-14}$ in the
value).  The condition for the quartic to be bounded below on the
condensate manifold was checked by direct minimization over the sphere
for several hundred random couplings.  The lattice Gross--Pitaevskii
functional of Sec.~\ref{sec:gp} is minimized on the full 432-site doublet
by projected gradient descent at fixed chargon number
$\sum_rB_r^\dagger B_r=xN$.  We use an adaptive step and random restarts,
continue from neighboring couplings where available, and retain the
lowest energy found.  The stability gaps quoted there are exact
eigenvalues of the lattice Hartree problem at the converged density.  The
direct gap of the spinon problem in the condensed phase
(Sec.~\ref{sec:doped}) is minimized over twisted boundary conditions,
using a $10\times10$ grid of twist angles followed by Nelder--Mead
refinement.

\emph{Tolerances.}  Operator identities such as $[U_g,H]=0$, the algebra
of Eq.~\eqref{eq:psgalg}, the entries of Table~\ref{tab:psg}, and the
closed forms quoted for the projected functionals, the density moments
and the loop-current circulations all agree with their torus evaluation
to relative residuals of $10^{-13}$ or below.  Quantities described as
constant on the pair sphere vary by less than $10^{-14}$, the crossing
of Eq.~\eqref{eq:crossing} is reproduced to ten digits across grids, and
``gapless'' in Sec.~\ref{sec:doped} means a twist-minimized direct gap
at the node energy below $10^{-6}\,t$.

\section{The local Gauss law at a constrained saddle}
\label{app:gausslaw}

The Gauss law $(n_f-1)+n_+-n_-=0$ of Sec.~\ref{sec:parton} holds on
every site, but the calculations of Secs.~\ref{sec:gp} and
\ref{sec:electronspectrum} impose only its spatial average: the
Gross--Pitaevskii functional is minimized at fixed total chargon number
$\sum_r\rho_r=xN$, and the spinon problem is filled to the global
$n_f=1+x$.  With $B_-$ alone condensed the site-resolved statement is
$\avg{n_{f,{\bm i}}}=1+\rho_{\bm i}$, and the density contrast is itself
of order $x$, so the violation is not parametrically small.  We measure that violation below, then set up and solve the constrained
saddle, and identify which results of the main text survive it.

\emph{The size of the violation.}  We evaluate it on the $432$-site
torus, using the unconstrained condensate of Sec.~\ref{sec:gp} and the
dressed spinon problem of Sec.~\ref{sec:electronspectrum} at $n_f=1+x$,
with the scale ratio $t_{\rm sp}/t_B=0.093$, marginally above the
physical range of Sec.~\ref{sec:experiment}.  The residual
$\avg{n_{f,{\bm i}}}-1-\rho_{\bm i}$ has zero mean by construction,
because the global filling \emph{is} enforced, so only its site-to-site
spread is a test.  That spread is $3.5\times10^{-2}$ at
$x=1/12$, $U/t_B=4$, against a density contrast of $6.9\times10^{-2}$,
and $6.0\times10^{-2}$ against $9.4\times10^{-2}$ at $x=1/6$, or half to
two thirds of the modulation under study.

\emph{The constrained problem.}  Adding
$\sum_{\bm i}a_{\bm i}\big[\avg{n_{f,{\bm i}}}-1-\rho_{\bm i}\big]$
places $+a_{\bm i}$ in the spinon Hamiltonian and $-a_{\bm i}$ against
the chargon density, and the saddle requires stationarity in
$a_{\bm i}$, in the spinon state and in $B$.  The functional requires some care, because the electron hopping admits
two mean-field decouplings of one and the same energy.  With
$t_c<0$ and $c_\uparrow=\bar B_-f^\dagger_\downarrow$, the
particle--hole conjugation in the fusion rule carries a sign, and the
dressed spinon link is
$-t_{\rm sp}s_{ij}+t_c\bar B_{\bm i}B_{\bm j}$ with $t_c/t_B=-1/2c_1$.
Freezing $\avg{f^\dagger_{\bm i}f_{\bm j}}$ at the parent value
$c_1s_{ij}$ turns that term into the chargon hopping $B^\dagger H_BB$
of Eq.~\eqref{eq:decoupling}; carrying the spinon state instead turns
it into the $B$ dependence of the spinon free energy $F_{\rm sp}$.
Keeping both would count the same hopping energy twice, so the
constrained functional carries it once, through $F_{\rm sp}$:
\begin{equation}
F_{\rm eff}[B,a]=\tfrac12U\sum_{\bm i}\rho_{\bm i}^2
+F_{\rm sp}[B,a]-\sum_{\bm i}a_{\bm i}\bigl(1+\rho_{\bm i}\bigr)\,,
\label{eq:feff}
\end{equation}
with $\rho=|B|^2$, $F_{\rm sp}$ the free energy of the dressed spinon
Hamiltonian at filling $n_f=1+x$, and no separate chargon kinetic term.
We verify that $F_{\rm sp}$ reproduces that kinetic term.  The ratio
$\bigl(F_{\rm sp}[B]-F_{\rm sp}[0]\bigr)/B^\dagger H_BB$ is $1.0015$ for
$|B|^2\le0.1$, to be compared with the $1.0013$ that follows from
rounding $c_1$ to three figures, and the gradient of
Eq.~\eqref{eq:feff} reproduces finite
differences to a relative $3\times10^{-8}$.  The spinon spectrum adds a
difficulty beyond the density constraint.  At $n_f=1+x$
the highest occupied and lowest empty levels of the dressed spinon
problem are degenerate to $10^{-11}$, with eighteen levels inside the
smearing window.  At sharp filling the density is not
differentiable in $a$, and the residual stalls near $6\times10^{-3}$.
The residual carries no weight in the near-null subspace of
$\chi_{\bm{ij}}=\partial\avg{n_{f,{\bm i}}}/\partial a_{\bm j}$, so nothing obstructs the constraint in principle, and
fractional occupation of the Fermi shell removes the obstruction.

\emph{Solving the constrained saddle.}  An alternating solution of the
two conditions does not converge, so we nest them instead.  Solving the
constraint for $a(B)$ to convergence at each $B$ makes every iterate
satisfy the local Gauss law.  Since $\partial F/\partial a$ vanishes
there, the gradient of $F_{\rm eff}[B,a(B)]$ is the $B$ gradient at
fixed $a$.  At $U/t_B=4$, $R=t_{\rm sp}/t_B=0.0718$ and smearing
$T=2\times10^{-3}t_B$ on the $12\times12$ torus, both residuals then
reach numerical zero together, at $x=1/12$ and at $x=1/6$, with
$\max|\avg{n_f}-1-\rho|$ at $10^{-14}$ or below and
$\max|\partial F_{\rm eff}/\partial B|_\perp$ at $10^{-9}$ to
$10^{-8}$.

\begin{figure*}[t]
\includegraphics[width=\textwidth]{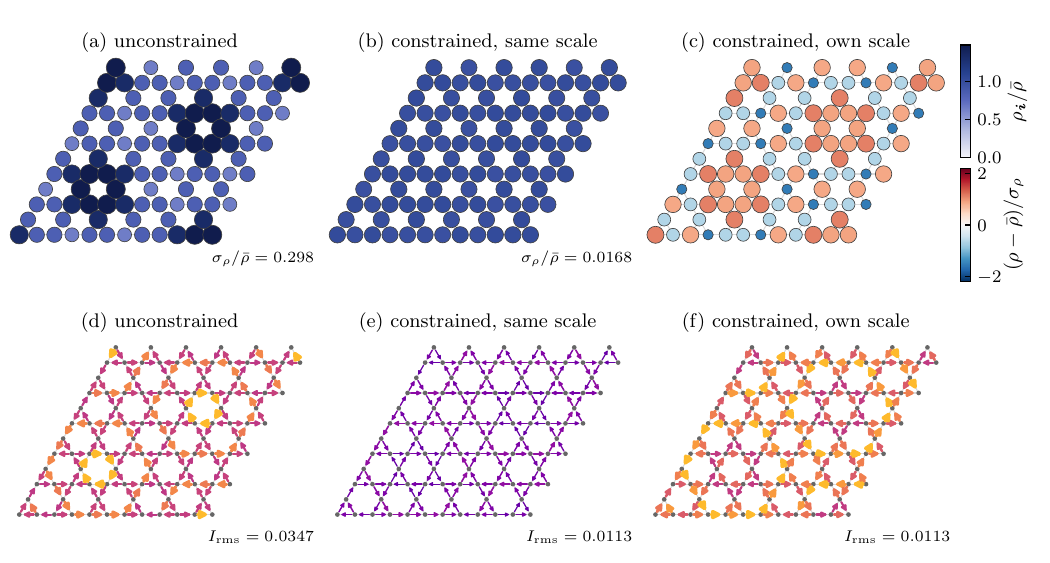}
\caption{The unconstrained Gross--Pitaevskii minimizer of Sec.~\ref{sec:gp} against the constrained saddle
of Eq.~\eqref{eq:feff} at $x=1/12$, $U_B/t_B=4$, $R=0.0718$, on a $6\times6$-cell window of the $12\times12$
torus.  Top: chargon density $\rho_{\bm i}/\bar\rho$, (a)~unconstrained and (b)~constrained on the same
color scale, with marker area following the density.  The contrast $\sigma_\rho/\bar\rho$ falls from
$0.298$ to $0.0168$.  (c)~The constrained density on its own scale, $(\rho-\bar\rho)/\sigma_\rho$, showing
five levels with the multiplicities of the unconstrained state on a partly rearranged pattern.  Bottom: the physical bond
current $I_{ij}=2\,{\rm Im}[\bar B_iB_j\avg{f^\dagger_if_j}]$ in units of $t_B$, with the frozen parent
correlator $c_1s_{ij}$ for the unconstrained state (d) and the dressed correlator for the constrained
saddle, on the unconstrained scale (e) and on its own (f).  Arrows give the direction and their width and
color the magnitude.  Bonds carrying less than $4\%$ of the maximum are omitted.  The staggered triangle pattern is
kept (correlation $0.97$, every bond keeping its sign) and the root-mean-square current falls by a
factor $3.1$.}
\label{fig:constrained}
\end{figure*}

\emph{Amplitude and pattern.}  Imposing the constraint flattens the
condensate [Fig.~\ref{fig:constrained}(a)--(c)].  At $x=1/12$ the
density contrast $\sigma_\rho/\bar\rho$ falls from $0.298$
unconstrained to $0.0168$, with $\rho$ confined to $[0.0812,0.0849]$
where it had run over $[0.0549,0.1234]$.  At $x=1/6$ it falls from
$0.204$ to $0.0046$.  The ordering-star weights fall with it, $K$ and
$K/2$ from $0.0049$ to $0.0002$ at $x=1/12$ and from $0.0065$ to below
$10^{-4}$ at $x=1/6$.  The flattened saddle lies well away from a
uniform reference with $|B_{\bm i}|=\sqrt{x}$, the phases of the
unconstrained minimizer, and $a$ solved to the same tolerance.  The
reference has gradient $2\times10^{-2}$
to $4\times10^{-2}$, four to five orders above the converged value, and
its $F_{\rm eff}$ lies above the saddle by $0.018$ at $x=1/6$ and $0.047$
at $x=1/12$.  We have not minimized over the phases of a uniform-density
state, so this test excludes that one configuration, not the
uniform-density family as a whole.

The condensate's position on the moduli space, by contrast, is
unchanged.  Populations alone would not settle this, since the invariant
$\mathcal{X}$ of Eq.~\eqref{eq:Xinv} depends on the relative phases.  For
example, the two normalized vectors $(\sqrt a,\sqrt b,\mp\sqrt b,\sqrt
a)$ with $a=(1-1/\sqrt3)/4$ and $b=(1+1/\sqrt3)/4$ share the populations
quoted below but have $\mathcal{X}=-1/4$ and $+1/12$.  The first lies on
the II$^-$ minimum manifold, and the second lies off the quadric
$Q=B_1B_3+B_2B_4=0$ altogether.  We therefore compare the complex
projected coefficients in the fixed gauge of
Appendix~\ref{app:conventions}.  They agree in modulus and in phase.  The
populations are $0.1057$, $0.3943$, $0.3943$, $0.1057$ with arguments
$(-0.4814,-0.2314,+0.2686,-0.9814)\pi$ at $x=1/12$, and $0.1057$,
$0.3943$, $0.3943$, $0.1057$ with $(-0.4832,-0.2332,+0.2668,-0.9832)\pi$
at $x=1/6$, for the unconstrained minimizer and the constrained saddle
alike, to the figures shown.  The invariants follow as
$\mathcal{P}=1/2$, $\mathcal{X}=-1/4$ and $Q=0$ to $10^{-8}$ for all
four states, which places them at the II$^-$ vertex of the
sextic-selected crystal of Sec.~\ref{sec:phaseII}.  The condensate
flattens its density by admixing states outside the four minima while
holding the ratios among the minima fixed.  The admixed weight grows from
$0.019$ to $0.076$ at $x=1/12$ and from $0.034$ to $0.078$ at $x=1/6$.
The density profile keeps five distinct levels with multiplicities
$72{:}144{:}72{:}72{:}72$ in both states and at both dopings, and at
$x=1/6$ the levels agree in shape once the amplitude is divided out, to
about $0.1\sigma_\rho$.  The real-space pattern is only partly
preserved: maximized over the $1728$ elements of the space group on the
torus, the normalized modulations overlap by $0.78$ at $x=1/12$ and
$0.45$ at $x=1/6$, and the dominant wavevector moves from $K$ to $K/2$
at $x=1/6$.  At a contrast of half a percent these last two figures
compare structure close to the threshold of the calculation, and we
attach little weight to them.  The valley weights are properties of $B$
itself and are not subject to this reservation.

The loop currents keep their pattern and lose part of their magnitude.
The physical bond current is the Peierls derivative of
Eq.~\eqref{eq:feff}, $I_{ij}\propto{\rm Im}\bigl[t_c\bar
B_iB_j\avg{f^\dagger_if_j}\bigr]$, which reduces to the frozen-link form
$2\,{\rm Im}[\bar B_is_{ij}B_j]$ only when the correlator is replaced by
$c_1s_{ij}$.  The constrained saddle carries a dressed complex
correlator, and we evaluate $I_{ij}$ with it.  The resulting currents are
divergence free, as stationarity requires, with $\max_i|\sum_jI_{ij}|$
below $10^{-9}$ of the root-mean-square bond current.  Their pattern
survives [Fig.~\ref{fig:constrained}(d)--(f)].  At both dopings they
correlate with the unconstrained currents at $0.97$ without any
translation search, and every bond on the torus keeps its sign.  The
root-mean-square physical current nevertheless falls by a factor of
$1.4$ at $x=1/36$, $3.1$ at $x=1/12$ and $7.0$ at $x=1/6$, against
density-modulation factors of $18$ and $44$ at the two higher dopings
and $1.2$ at $x=1/36$, where the
constrained minimum is not a crystal (see below).  Where a crystal
survives, the charge modulation is the more strongly suppressed of
the two.  The orbital-field estimates of Sec.~\ref{sec:experiment}, which
use the frozen-link currents, should be divided by the current factors.

\emph{Landscape and stability.}  We searched the landscape
systematically, with starts spanning the unconstrained minimizer, the
Landau candidates built from the four band minima, random valley vectors,
random full-lattice vectors that probe outside the four-minimum
subspace, and perturbations of the best state found so far.  Anchoring
the infeasibility barrier on the unconstrained minimizer from the first
evaluation removes the runaways discussed below, and all $26$ solves
converged.  At $x=1/12$ fourteen converged starts reach nine distinct
minima, and at $x=1/6$ twelve reach three.  At both dopings the lowest is
the minimum reached from the unconstrained minimizer, with
$F_{\rm eff}=-176.827474$ (reached by three starts) and $-278.785823$
(by four).  Every state at those energies carries $\mathcal{P}=1/2$,
$\mathcal{X}=-1/4$ and $Q=0$, the II$^-$ vertex.  The higher minima lie
away from this vertex.  One at $x=1/12$, for example, sits at
$(\mathcal{P},\mathcal{X},|Q|)=(0.6045,+0.1331,0.4067)$.  Basin size and
depth are anticorrelated here, which a small multistart search would
miss.  The single-valley state, with $\mathcal{P}=1$ and three density
levels, is the most frequently reached basin at both dopings (seven of
twelve starts at $x=1/6$), yet it lies $2.11$ above the minimum.  On the
constraint surface the multiplier term cancels and $F_{\rm eff}$ reduces
to $\tfrac12U\sum\rho^2+\avg{H_0(B)}-TS$, which is bounded below on a
finite lattice at fixed $\sum_r\rho_r$.  The runaways therefore occur
only once an outer step leaves that surface, along the multiplier
direction, in which the canonical spinon free energy is concave.  The
constraint also requires $\rho_{\bm i}\le1$, since
$\avg{n_{f,{\bm i}}}=1+\rho_{\bm i}$ cannot exceed $2$, but this bound
did not cause the runaways.  Over $200$ random valley starts at each
doping, the largest $\rho_{\bm i}$ reached $0.28$ at $x=1/12$ and $0.56$
at $x=1/6$.

Both solutions are minima.  We differentiate the reduced gradient of
$F_{\rm eff}[B,a(B)]$, re-solving the constraint at every displacement so
that the response $\partial a/\partial B$ is included, and obtain the low
spectrum of the resulting Hessian by reflecting about its top,
$\lambda_{\max}=11.36$ at $x=1/12$ and $13.39$ at $x=1/6$.  Two
eigenvalues vanish, corresponding to the scale direction of the
parameterization and to the global phase, each with
$|H m|\le6\times10^{-9}$.  All others are positive.  The lowest nonzero
eigenvalue is threefold degenerate at both dopings, $0.0416$ at $x=1/12$
and $0.0227$ at $x=1/6$, and the twelve lowest show the same pattern of
degeneracies at both.  The whole low spectrum softens by a common factor
$0.547\pm0.003$ between the two dopings.  ``Lowest'' means lowest found
over the starts described, with the smearing, the system size, $U/t_B$
and $R$ held fixed throughout, and implies no claim of global
optimality.

\emph{The lowest doping.}  The statements above concern $x=1/12$ and $1/6$.
At $x=1/36$ the constrained theory instead phase separates.  Started from
the unconstrained minimizer, the constrained solve reaches a state with
valley invariants $\mathcal{P}=1/2$, $\mathcal{X}=-1/4$, $Q=0$, five
density levels and a contrast $3.0$ times below the unconstrained one.
This state is a stationary point of $F_{\rm eff}$ but not a minimum.  Its
Hessian on the norm sphere, formed by differentiating the analytic
gradients with the constraint re-solved at every displacement, has six
degenerate directions of curvature $-1.1\times10^{-3}$.  Displacing along
one of them and re-minimizing lowers $F_{\rm eff}$ by $0.0168$, from
$-96.165672$ to $-96.182423$, or $1.4\times10^{-3}$ per chargon.  The
state reached is a minimum (softest curvature $+0.046$ after two exact
zero modes).  Its density runs from $0.001$ to $0.041$ per site, with the
dominant Fourier component at the smallest wavevector the torus admits,
$|q|=2\pi\cdot2/(12\sqrt3\,a)$.  A search over eighteen starts of the
kinds described above finds nothing lower on this torus.  On a
$24\times24$ torus we started the same minimization either from the
tiled minimum perturbed at the new smallest wavevector or from a
half-torus puddle.  Both starts approach the same stripe (density
modulations correlated at $0.98$, energies within $0.002\,t_B$),
although neither run reached the convergence of the $12\times12$
minimizations.  It
consists of a region covering two thirds of the system at $0.038$
chargons per site, an empty region covering a fifth, and a wall between
them, and its energy is $8.0160\,t_B$ per chargon against $8.0152$ for
the $12\times12$ minimum.  The wavelength follows the system size.

A Maxwell construction settles the interpretation.  We computed the
lowest constrained state on the $12\times12$ torus at $x=0$, $1/72$,
$1/48$, $1/36$, $1/24$, $1/18$, $1/12$ and $1/6$ (energies per site
$-0.127129$, $-0.174736$, $-0.198631$, $-0.222644$, $-0.270640$, $-0.318116$,
$-0.409323$, $-0.645338$).  The lower convex hull runs straight from
$x=0$ to $x=1/24$, and the three points below $1/24$ lie above it by
$1.6$ to $2.5\times10^{-4}$ per site.  Every point from $1/24$ upward lies
on the hull; at $1/24$ and $1/18$ three starts of each kind converge to
the same state, while at $1/12$ and $1/6$ the three-start check was made
at $U_B/t_B=16$, $60$ and $80$.  At $U_B/t_B=4$ the constrained theory has a
miscibility gap for $0<x<1/24$, between the undoped spin liquid and a
charge-ordered condensate at $x\approx1/24$, which is the density of the
dense region in every inhomogeneous state found.  The homogeneous
$x=1/36$ state whose properties are quoted in this paper is metastable
within that gap.  Its valley invariants are those of the dense phase, so
the symmetry classification is unaffected.  Its root-mean-square contrast
($1.2$ times below the unconstrained one at the minimum, $3.0$ at the
stationary point), its physical currents (suppressed by $1.4$, correlated
at $0.81$ with the unconstrained pattern) and its stability gap
($\Delta_+=1.39$ at the minimum, $1.44$ at the stationary point),
however, are properties of the metastable state.

The miscibility gap is specific to this weak coupling.  At $U_B/t_B=16$
the same construction gives a hull that is convex at every point, with
$e(1/36)$ lying $2.1\times10^{-3}$ per site below the chord from $0$ to
$1/24$.  The Gross--Pitaevskii start reaches the homogeneous state at all
seven dopings with no depleted site anywhere, and all three starts agree
at $1/18$, $1/12$ and $1/6$.  At the physical charging energy,
$U_B/t_B=60$ and $80$, all eight points lie on the
lower convex hull, and each interior point lies below the chord through
its two neighbours by $1.3$ to $74\times10^{-3}$ per site at $60$ and
$2.4$ to $97\times10^{-3}$ at $80$.  The start-to-start spread, which
sets the noise floor of the search, is about $10^{-4}$.  The local
curvature therefore changes sign between $U_B/t_B=4$ and $16$ and grows
by orders of magnitude beyond it.  This conclusion errs in the safe
direction, since a missed lower interior minimum would only push that
point further below the chord.  The $x=1/36$ state of this paper is thus
metastable at the coupling where the constrained calculation was
validated and not phase separating at the physical one.  The amplitudes quoted for it
are those of the $U_B/t_B=4$ point.  Unlike at $U_B/t_B=4$, the
strong-coupling condensate is nearly uniform at every doping, with of
order five per cent of its weight outside the four band minima, and
$e(x)$ turns over near $x=1/18$ and rises to $+0.13$ and $+0.41$ per
site by $x=1/6$, against the undoped $-0.127$, so beyond a filling of
about one twentieth the charging cost of the condensate dominates.

\emph{The $B_+$ sector at the constrained saddle.}  The same functional
determines whether $\avg{B_+}=0$ survives the constraint, once the
doublet is restored and the Gauss law is written as
$\avg{n_{f,{\bm i}}}=1+\rho_{-,{\bm i}}-\rho_{+,{\bm i}}$ at fixed
$n_++n_-=xN$.  The two components dress the spinon link with opposite
sign, as the $\tau^3$ of Eq.~\eqref{eq:decoupling} requires, and the
counting check used above confirms this.  At fixed filling $F_{\rm sp}$
reproduces each component's kinetic term with ratio $1.0013$.  The
$1.4\%$ difference that appears when the filling is tracked equals the
chemical-potential term $\mu\,\partial N/\partial\nu$ to five figures.
Because every term in $\partial F/\partial\bar B_+$ carries a factor
$B_+$, the point $B_+=0$ is exactly stationary, and the question is
whether it is a minimum.  Expanding to second order at fixed total
chargon number gives the constrained counterpart of the Hartree
stability gap, $\Delta_+=\lambda_{\min}(M_+)-\mu_-$ with $M_\pm=U\,{\rm
diag}(n)\pm{\rm diag}(a)\pm t_c(\!\operatorname{ADJ}\rho_f^{T})\mp\mu$.
This gap is positive throughout, $1.39$ at $x=1/36$, $1.28$ at $x=1/12$
and $0.65$ at $x=1/6$, against $\Delta_{\rm flat}=1.4495$.  (At $x=1/36$
the stationary point discussed above gives $1.44$.)  The analytic
gradient agrees with finite differences to $10^{-8}$.  Doubling $U/t_B$
from $4$ to $8$ moves $\Delta_+$ by under $0.2\%$ at every doping,
although $\lambda_{\min}$ and $\mu_-$ each shift by several tenths, so
the extrapolation in $U$ is mild.  Direct minimization confirms the
result.  We used seventeen starts at $x=1/12$: the pure-$B_-$
saddle, seeded doublets with $B_+$ carrying $1$, $5$, $15$, $35$ and
$50\%$ of the chargon weight, and ten random configurations drawn
independently in valley space and on the full lattice with $B_+$
fractions up to $80\%$, together with basin hopping.  All converge, and
\emph{none} retains any $B_+$ in this functional.  The largest converged $n_+/n$ is zero to
six figures, and no state lies below the pure-$B_-$ saddle.  Several
land in higher constrained minima, by $0.07$ to $1.20$, and expel $B_+$
there as well, so the expulsion is a property of the functional and not
of the basin.  The gap $\Delta_+$ narrows with doping, falling by
$55\%$ between the lowest and highest doping studied while
$\Delta_{\rm flat}$ does not move, and the search above was made at one
doping and one coupling.

The symmetry classification of the main text is unaffected by the
constraint.  It depends
only on the point of the moduli space, which the valley weights fix, and
the constraint leaves those weights unchanged to four figures.  The
projective symmetry group, the moduli-space theorem and the composite
catalog of Sec.~\ref{sec:catalog} therefore stand as derived, together
with whatever part of the selection among candidate crystals is settled
by valley content.  The amplitudes are affected.  Every density and
diffraction intensity in Secs.~\ref{sec:gp} and
\ref{sec:electronspectrum} is computed on the unconstrained background.
Its modulation exceeds the constraint-consistent one by factors of $18$
and $44$ at $x=1/12$ and $1/6$, and the orbital fields of
Sec.~\ref{sec:experiment} exceed theirs by factors of $1.4$, $3.1$ and
$7.0$ at the three dopings.  At $x=1/36$ and this coupling the
homogeneous constrained state is metastable within a miscibility gap
that is absent by $U_B/t_B=16$, and its modulation differs altogether
from the unconstrained one, at comparable contrast.  These suppression
factors refer to the two sampled parameter points $U_B/t_B=4$,
$t_{\rm sp}/t_B=0.0718$, $T=2\times10^{-3}t_B$, $L=12$, and are not
bounds.  A lower-lying constrained state, or the physical $U_B/t_B$ of
$63$ to $81$, could carry either a larger or a smaller modulation.  Nor
does a single factor rescale every satellite.  The root-mean-square
contrast fixes only the sum of Fourier powers, whereas each intensity
follows its own amplitude and form factor, and at $x=1/6$ the dominant
wavevector moves between the two states.  The site-by-site densities of
the unconstrained background are therefore not estimates of the physical
densities.

\section{Proof details of the moduli-space theorem}
\label{app:theorem}

\emph{Antecedent.}  The mechanism has a direct antecedent on the square
lattice.  For two boson fields transforming under the $\pi$-flux magnetic
translation algebra, the quartic potential leaves the relative phase of
the two components undetermined, and that residual degeneracy is lifted
only at eighth order \cite{Lannert2001,SachdevPark2002,Balents2005}.  The
kagome case differs in its in-pair group, which is octahedral, as we now
show.

\emph{In-pair generators.}  Restricted to the pair $(B_1,B_2)$ and acting
on the Bloch vector $\hat n$, the generators of Table~\ref{tab:psg} act
as follows.  $T_2$ is a $\pi$ rotation about $\hat z$, $T_1$ is a $\pi$
rotation about $\hat x$, and $C_3=C_6^2$, which maps the pair to itself,
is a $2\pi/3$ rotation about the body diagonal
$\tfrac{1}{\sqrt3}(1,-1,-1)$ of the frame they define.
These generate the tetrahedral rotation group $\mathrm{T}$.  The product
$\sigma C_6$ also maps the pair to itself, and its induced action is a
$\pi$ rotation about the $(110)$ axis, which lies outside $\mathrm{T}$.
Together with $\mathrm{T}$ it generates the octahedral rotation group
$\mathrm{O}$ of order 24, which contains the $\pi/2$ rotations about the
coordinate axes.  The antiunitary $\Theta C_2$ also maps the pair to
itself.  Writing its in-pair action as an antiunitary $2\times2$ matrix
and evaluating $\hat n\mapsto\hat n'$ on a basis of states gives
$\hat n'=-\hat n$.  The induced action is the inversion, which
completes $\mathrm{O}_h=\mathrm{O}\times\{1,-1\}$.  We verified the
closure on the full 288-element projective symmetry group.  Of its
elements, 144 (72 unitary and 72 antiunitary) preserve the pair subspace
$\{(B_1,B_2,0,0)\}$.  They induce 48 distinct $O(3)$ elements on the
Bloch sphere, which are closed under multiplication, contain the
inversion and divide into 24 proper and 24 improper elements.  The proper
part has the octahedral angle multiset
$\{0\!:\!1,\ 90\!:\!6,\ 120\!:\!8,\ 180\!:\!9\}$ and comes from the
unitary elements alone.  In the chiral theory the mirror is a symmetry
only in combination with time reversal.  The $\pi$ rotation about
$(110)$ is then absent, and the unitary in-pair group is again
$\mathrm{T}$.  The in-pair antiunitary is instead $\sigma\Theta$, whose
induced action is a reflection.  The harmonic $xyz$ is even under this
reflection, so the $\mathbb{CP}^1$ is lifted at sextic order instead of
at octic order.
Table~\ref{tab:inpair} collects the axis, angle and determinant of the
induced action of the in-pair generators and products.  The
corresponding table for the chiral doublet is
Table~\ref{tab:inpairchiral} of Appendix~\ref{app:csl}.
\begin{table}[t]
\caption{Induced action on the Bloch sphere $\hat n$ of the pair
$(B_1,B_2)$, with $\hat x,\hat y,\hat z$ the Pauli components in the
valley basis and gauge of Table~\ref{tab:psg}: rotation axis, angle in
degrees and determinant of the induced $O(3)$ element for every in-pair
generator and named in-pair product.  Elements that exchange the two
pairs ($C_6$, $\sigma$, $\Theta$, $C_2$) are not listed, since their
axes depend on the identification of the $(B_3,B_4)$ sphere with the
$(B_1,B_2)$ sphere.  The unitary in-pair elements generate the
octahedral rotation group $\mathrm{O}$ (24 elements, angle multiset
$\{0\!:\!1,\,90\!:\!6,\,120\!:\!8,\,180\!:\!9\}$).  With the inversion
the full in-pair group is $\mathrm{O}_h$ (48 elements: 24 rotations, the
inversion, nine reflections, six $S_4$ and eight $S_6$ rotoreflections).
Character sums for $\ell=0,\dots,6$: $n_\ell=(1,0,0,0,1,0,1)$ over the
24 rotations and over all 48 elements alike, and $(1,0,0,1,1,0,2)$ over
the tetrahedral subgroup $\langle T_1,T_2,C_3\rangle$.}
\label{tab:inpair}
\begin{ruledtabular}
\begin{tabular}{lllcc}
element & type & axis & angle & det \\
\colrule
$T_1$ & unitary & $\hat x$ & 180 & $+1$ \\
$T_2$ & unitary & $\hat z$ & 180 & $+1$ \\
$C_3=C_6^{\,2}$ & unitary & $(1,-1,-1)/\sqrt3$ & 120 & $+1$ \\
$\sigma C_6$ & unitary & $(1,1,0)/\sqrt2$ & 180 & $+1$ \\
$C_6\sigma$ & unitary & $(1,0,1)/\sqrt2$ & 180 & $+1$ \\
$\Theta C_2$ & antiunitary & inversion & & $-1$ \\
\end{tabular}
\end{ruledtabular}
\end{table}

\emph{Character counting.}  The character sum, with $\chi_\ell$ as in
Eq.~\eqref{eq:charsum} and taken over the classes of $\mathrm{O}$ (the
identity, six $\pi/2$ rotations, eight $2\pi/3$ rotations and nine $\pi$
rotations),
$n_\ell^{\rm inv}=\tfrac{1}{24}\big[(2\ell{+}1)+6\,\chi_\ell(\pi/2)
+8\,\chi_\ell(2\pi/3)+9\,\chi_\ell(\pi)\big]$, gives
$n_\ell^{\rm inv}=(1,0,0,0,1,0,1)$ for $\ell=0,\dots,6$, and the same
sequence results when the sum is extended over all 48 elements of
$\mathrm{O}_h$ (Table~\ref{tab:inpair}).  The tetrahedral sum alone would
give $(1,0,0,1,1,0,2)$.  The $\ell=3$ invariant $xyz$ is odd under each
of the six $\pi/2$ rotations of $\mathrm{O}$ and is removed by them.  The
inversion induced by $\Theta C_2$ is therefore not needed for the
sextic-order degeneracy, and the $\ell=4$ cubic harmonic remains the
first nontrivial invariant.  Numerically, all four symmetry-allowed
sextic invariants are constant on the pair sphere, with spreads
$\sim10^{-15}$ against a natural scale $\sim10^{-2}$, and the on-site
moments $\sum_r\rho_r^2$ and $\sum_r\rho_r^3$ are constant to the same
precision.  Only $N^{-1}\sum_r\rho_r^4$ (with $\sum_r\rho_r=N$) varies.
It takes its minimum, $19/6$, on the uni-$M$ stripe orbit and its
maximum, $29/9$, on the triple-$M$ orbit, a spread of exactly
$1/57=1.75\%$.

\emph{The $(J_1,J_2)$ frame.}  The phase-diagram angle $\theta$ of
Fig.~\ref{fig:wheel} is defined in an orthonormal frame fixed by a
\emph{deterministic} construction.  Each quartic is represented by its
coefficient tensor $Q_{ab,cd}$, symmetrized separately in $(a,b)$ and in
$(c,d)$, with $V=\sum Q_{ab,cd}\,\bar B_a\bar B_bB_cB_d$, and the
Frobenius inner product $\langle Q,Q'\rangle=\sum\bar Q_{ab,cd}Q'_{ab,cd}$
is used throughout.  In this norm $\|J_0\|=\sqrt{10}$ for
$J_0=(\sum_\eta n_\eta)^2$, $\|\mathcal{P}\|=\sqrt6$,
$\|\mathcal{X}\|=\sqrt3$, $\langle\mathcal{P},J_0\rangle=6$ and
$\langle\mathcal{X},J_0\rangle=\langle\mathcal{X},\mathcal{P}\rangle=0$.
The frame consists of $\hat J_0=J_0/\sqrt{10}$ together with $J_1$ and
$J_2$, the Gram--Schmidt orthonormalizations of $\mathcal{P}$ and
$\mathcal{X}$ against $\hat J_0$ (in that order), with the
sign convention that the $J_1$ coefficient of $\mathcal{P}$ is negative
and the $J_2$ coefficient of $\mathcal{X}$ is positive.  All three of
$\hat J_0$, $J_1$ and $J_2$ have unit Frobenius norm.  The dictionary to
the closed-form basis of Eqs.~\eqref{eq:Pinv}--\eqref{eq:Xinv} is then
exact,
\begin{equation}
\begin{aligned}
\mathcal{P}&=\tfrac{6}{\sqrt{10}}\,\hat J_0-\sqrt{\tfrac{12}{5}}\,J_1\,,\\
\mathcal{X}&=\sqrt{3}\,J_2\,,
\end{aligned}
\label{eq:dictionary}
\end{equation}
with no residual mixing.  The coefficients are
$\langle\mathcal{P},\hat J_0\rangle=6/\sqrt{10}$ and
$(6-\tfrac{36}{10})^{1/2}=\sqrt{12/5}$.  They refer to the normalized
$\hat J_0$, not to $J_0$ itself, against which the residual
$\|\mathcal{P}-\tfrac{6}{\sqrt{10}}J_0\|$ would be $4.3854$.  In this
frame the three phase-diagram vertices
are $(\mathcal{P},\mathcal{X})=(1,0)$ (phase I) and
$(\tfrac12,\mp\tfrac14)$ (II$^\mp$), and the boundary geometry gives the
phase-I window and the Hubbard direction in closed form:
$\theta_{\rm I/II^\mp}=\pm\arctan\sqrt5=\pm0.36614\pi$,
the II$^-$/II$^+$ boundary at exactly $\theta=\pi$, and
$\theta_U=\arctan\sqrt5$ [Eq.~\eqref{eq:hubbardquartic}, which the
numerical fit of the projected on-site functional reproduces as
$\tan\theta_U=\sqrt5$ to $5\times10^{-16}$], with $\theta_V=0.32362\pi$.
All boundaries were confirmed by bisection on the energy criterion
``global minimum equals the (exactly flat) pair-sphere value,'' with the
branch read off from the sign of $\mathcal{X}$ at the minimizer.  Every
angle quoted here refers to this frame.

\section{Two-loop check of the ray protection}
\label{app:betas}

We check the all-orders statement of Sec.~\ref{sec:hubbardray}
explicitly at two loops.  The check is made in the relativistic theory
(dynamical exponent $z=1$).  This theory describes the transition of the
undoped insulator tuned at fixed half filling by the chargon gap (for
instance under pressure, Sec.~\ref{sec:materials}), where the Landau
functional has a second-order time derivative.  These beta functions do
not describe the doped Higgs transition of Sec.~\ref{sec:gp}, which has
a first-order time derivative ($z=2$) and Fermi pockets.  Only the
symmetry statement, the invariance of the Hubbard ray under any
symmetric renormalization, carries over to that case.  The continuum
theory contains the four complex valley fields with the quartic of
Eq.~\eqref{eq:potential}, minimally coupled with unit charge to the
emergent $U(1)$ gauge field.  The gauge field also couples to the $N_f$
two-component Dirac spinons of the parent spin liquid, which has two
Dirac cones for each spin (Sec.~\ref{sec:ansatz}), so $N_f=4$ on kagome.
No Yukawa coupling between the charged valley fields and the spinons is
gauge invariant.  The gauge-invariant mixed quartic
$\bar B_\eta B_{\eta'}\psi^\dagger\Gamma\psi$ is marginal in $d=3$ and
lies outside the scope of the ray theorem, and we do not include it.  We
work in $d=4-\epsilon$ dimensions and continue the spinons as $N_f/2$
four-component fermions of unit charge.  The valley fields are
canonically normalized, with kinetic term
$|(\partial_\mu-ieA_\mu)B_\eta|^2$, where $A_\mu$ is the emergent gauge
field.  On the eight real components $B_a=(\phi_a+i\phi_{a+4})/\sqrt2$
the quartic $u_0\rho^2$ then reads $\tfrac14u_0(\phi^2)^2$.  The beta
functions are those of minimal subtraction and do not depend on the
gauge parameter.  With $f=e^2/16\pi^2$ and the quartic couplings measured
in units of $16\pi^2$, the two-loop beta functions follow by
specializing the general two-loop results of Luo, Wang, and Xiao
\cite{LuoWangXiao2003} to the quartic tensor of Eq.~\eqref{eq:potential}
on the eight real components of the valley fields.

The result agrees with known limits to within numerical roundoff.  At
$f=0$ the one- plus two-loop terms reproduce the beta functions of the
ungauged four-boson theory, obtained independently from the general
two-loop functional of a scalar theory whose two two-loop coefficients
are fixed by the single-field and $O(N)$ results.  For the
$U(n)$-symmetric quartic with $n=1,\dots,4$, once the two normalization
factors are fixed by one-loop matching, the two-loop terms reproduce
all four structures $u^3$, $fu^2$, $f^2u$ and $f^3$ of the
quartic beta function of the abelian Higgs model at two loops
\cite{Ihrig2019} with their full $n$ dependence.  The fate of the
antipodal ray is determined by the two-loop term $\Lambda^{2g}$ of
Eqs.~(44) and (52) of Ref.~\cite{LuoWangXiao2003}, in the notation of
that paper.  It appears with the same coefficient in the
re-examination of the general two-loop equations in
Ref.~\cite{Schienbein2019} [Eqs.~(5.14) and (A.19) there].  The
corrections that reference makes to the earlier results
\cite{MachacekVaughn1985,LuoWangXiao2003} concern the off-diagonal scalar
wave-function renormalization and do not arise here.  The eight real
components of the valley fields form an irreducible representation of
the projective symmetry group together with the $U(1)$, so the scalar
anomalous-dimension matrix is proportional to the identity.  The beta
functions read
\begin{align}
\beta_f&=-\epsilon f+\tfrac43(N_f+2)f^2+(4N_f+32)f^3,
\label{eq:betaf}\\
\beta_{u_0}&=-\epsilon u_0+32u_0^2+24u_0u_P+6u_X^2-12fu_0+6f^2\nonumber\\
&\quad-456u_0^3-528u_0^2u_P-120u_0u_P^2-204u_0u_X^2\nonumber\\
&\quad-120u_Pu_X^2+208fu_0^2+192fu_0u_P+12fu_X^2\nonumber\\
&\quad+\big(\tfrac{742}3+\tfrac{20}3N_f\big)f^2u_0+120f^2u_P\nonumber\\
&\quad-\big(\tfrac{292}3+\tfrac{32}3N_f\big)f^3,
\label{eq:betau0}\\
\beta_{u_P}&=-\epsilon u_P+24u_0u_P+24u_P^2-4u_X^2-12fu_P\nonumber\\
&\quad-488u_0^2u_P-816u_0u_P^2+96u_0u_X^2-312u_P^3\nonumber\\
&\quad+92u_Pu_X^2+96fu_0u_P+144fu_P^2+4fu_X^2\nonumber\\
&\quad+\big(\tfrac{142}3+\tfrac{20}3N_f\big)f^2u_P,
\label{eq:betauP}\\
\beta_{u_X}&=u_X\Big[-\epsilon+24u_0+8u_P-12f-488u_0^2-432u_0u_P\nonumber\\
&\qquad\quad-56u_P^2+28u_X^2+96fu_0+64fu_P-48fu_X\nonumber\\
&\qquad\quad+\big(\tfrac{142}3+\tfrac{20}3N_f\big)f^2\Big].
\label{eq:betauX}
\end{align}
Writing $\beta_P$ and $\beta_X$ for $\beta_{u_P}$ and $\beta_{u_X}$,
on $u_X=-2u_P$ one finds $\beta_X+2\beta_P=0$ identically, with all
gauge terms included, whereas on $u_X=+2u_P$,
$\beta_X-2\beta_P=-384fu_P^2$.  The asymmetry between the two rays has
a single source.  The map $g$ of Sec.~\ref{sec:hubbardray} requires
$\beta_{u_0}$ and $\beta_{u_P}$ to be even and $\beta_{u_X}$ to be odd
in $u_X$, and only one monomial of the system violates this, the term
$-48fu_X^2$ of Eq.~\eqref{eq:betauX}, which is even in $u_X$.  The
covariant remainder contributes $+192fu_P^2$ to $\beta_X+2\beta_P$ on
the Hubbard ray and $-192fu_P^2$ to $\beta_X-2\beta_P$ on its antipode,
and the term $-48fu_X^2$ contributes $-192fu_P^2$ on both rays, so that
the sum vanishes on the Hubbard ray and equals $-384fu_P^2$ on the
antipode.  At $f=0$ the covariance is exact and both rays are
preserved.  At one loop both rays are preserved also at $f\neq0$,
because the one-loop gauge contributions to $\beta_{u_P}$ and
$\beta_{u_X}$ are $-12fu_P$ and $-12fu_X$.  This common rescaling
preserves every ray through the origin of the $(u_P,u_X)$ plane, and the
remaining one-loop gauge terms, $-12fu_0$ and $6f^2$, enter
$\beta_{u_0}$ alone.  The two rays are therefore first distinguished at
two loops, where the gauge vertices separate the charged pair channel
$Q$ from the neutral channel $\tilde Q$.  The same conclusion
holds in the full $330$-dimensional space of quartic couplings of eight
real fields.
At points of the Hubbard plane $u_X=-2u_P$, the one- and two-loop beta
functions lie in that plane and in the three-dimensional PSG-invariant
subspace, to within numerical roundoff.  On the antipodal plane
$u_X=+2u_P$ the two-loop beta function stays in the PSG-invariant
subspace but leaves the plane, with $\beta_X-2\beta_P=-384fu_P^2$ to the
same accuracy.

Equations~\eqref{eq:betaf}--\eqref{eq:betauX} also bear on the order of
the transition, which Sec.~\ref{sec:landau} leaves open.  We record what
they imply without pursuing the question further.  At one loop, with
$f^*=3\epsilon/[4(N_f+2)]$, a real quartic fixed point on the
$U(4)$-symmetric line $u_P=u_X=0$ requires $N_f\ge-11+12\sqrt3=9.78$
[the discriminant of Eq.~\eqref{eq:betau0} restricted to that line is
proportional to $N_f^2+22N_f-311$], and at $N_f=4$ the one-loop system
has no real charged fixed point anywhere in $(u_0,u_P,u_X)$.  The
ungauged theory has no stable one-loop fixed point either.  Its $O(8)$
point is unstable to both PSG anisotropies, each of which is relevant
there with renormalization-group eigenvalue $\epsilon/4$.
Within the $\epsilon$ expansion the four-valley theory therefore shows
the Halperin--Lubensky--Ma tendency toward a fluctuation-induced
first-order transition \cite{Halperin1974}.  Thresholds of this kind are
unreliable in $d=3$.  For the abelian Higgs model the one-loop
$n_c\approx183$ falls to $n_c\approx12.2\pm3.9$ at four loops, and the
route matters.  That figure comes from a dimensional interpolation
between $d=2+\epsilon$ and $d=4-\epsilon$ rather than from resumming the
$4-\epsilon$ series, whose Pad\'e and Pad\'e--Borel approximants at four
loops still range from $-14$ to $130$ and whose optimized estimate is
$-52\pm45$ \cite{Ihrig2019}.  Here the two-loop terms of
Eqs.~\eqref{eq:betaf}--\eqref{eq:betauX} are not small at $\epsilon=1$,
and the Dirac sea screens the photon through the $N_f$ terms above.  We
therefore claim neither a continuous nor a first-order transition.

\section{The two off-ray quartics}
\label{app:offray}

This appendix gives the two second-order quartics of Sec.~\ref{sec:phasediagram},
drawn in Fig.~\ref{fig:quartics}, in the form in which they were evaluated.

\emph{Normalization.}  Write the condensate as
$B_{\bm r}=\sqrt{xN}\sum_\eta c_\eta\phi_\eta({\bm r})$ with
$\sum_\eta|c_\eta|^2=1$ and $\phi_\eta$ the exact valley Bloch states of
Sec.~\ref{sec:bands}.  Both terms below are quartic in $B$ and therefore enter
the Landau functional of Sec.~\ref{sec:landau} in the same form, contributing a
per-site energy $x^2\big[u_0J_0+u_P\mathcal{P}+u_X\mathcal{X}\big]$ with
\begin{align}
J_0&=\Big(\sum_\eta n_\eta\Big)^{2},\qquad
\mathcal{P}=(n_1{+}n_2)^2+(n_3{+}n_4)^2\,,\nonumber\\
\mathcal{X}&=(n_1{-}n_2)(n_3{-}n_4)\nonumber\\
&\quad+2\,{\rm Re}\big[\bar c_1\bar c_3c_2c_4+\bar c_2\bar c_4c_1c_3\big]\,,
\end{align}
and $n_\eta=|c_\eta|^2$.  Each term is evaluated on eighty random valley vectors
and fitted to these three invariants.  The residual of the fit, which is at the
level of the arithmetic, checks that no fourth invariant appears.  The
coefficients are quoted as $N(u_0,u_P,u_X)$, the combination in which the
per-site energy is independent of torus size.  The energy difference between
the competing orders is
\begin{equation}
E_{\rm I}-E_{\rm II^-}=\tfrac12u_P+\tfrac14u_X=\tfrac14\big(u_X+2u_P\big)\,,
\label{eq:splitdef}
\end{equation}
using $(\mathcal{P},\mathcal{X})=(1,0)$ on phase~I and
$(\tfrac12,-\tfrac14)$ on branch~II$^-$.  It vanishes identically on the Hubbard
ray $u_X=-2u_P$, so a term lying on that ray leaves the two degenerate at any
magnitude.

\emph{(a) The spinon-polarization quartic.}  The condensate dresses the spinon
link field into $-t_{\rm sp}s_{ij}+t_c\bar B_{\bm i}B_{\bm j}$
(Sec.~\ref{sec:electronspectrum}), and the filled Dirac sea responds at second
order,
\begin{equation}
\Delta E_{\rm sp}=2\!\!\sum_{a\,{\rm occ}}\sum_{b\,{\rm emp}}\!
\frac{\big|\langle b|\,t_c\bar B_{\bm i}B_{\bm j}\,|a\rangle\big|^2}
{\varepsilon_a-\varepsilon_b}\,,
\label{eq:spquartic}
\end{equation}
the matrix element taken between eigenstates of the undoped ansatz and the factor
two coming from spin.  Equation~\eqref{eq:spquartic} is quartic in $B$ and carries
spinon particle--hole denominators, so its scale is $t_c^2/t_{\rm sp}$.  Evaluated
on the $432$-site torus in the three antiperiodic sectors of
Appendix~\ref{app:conventions}, which gap the Dirac nodes, and averaged over them,
\begin{equation}
N(u_0,u_P,u_X)=(-0.1374,\,+0.0635,\,+0.1096)\,t_c^2/t_{\rm sp}\,,
\end{equation}
so that $u_X/u_P=+1.73$ against the $-2$ of the Hubbard ray.  The term therefore
lies far off that ray, at $\theta=0.652\pi$, inside phase~II$^-$.  It is constant on the phase-I pair
sphere and on the II$^-$ level set to the precision of the evaluation, as the
moduli theorem of Sec.~\ref{sec:moduli} requires of any function of $\mathcal{P}$
and $\mathcal{X}$ alone, and Eq.~\eqref{eq:splitdef} gives
$E_{\rm I}-E_{\rm II^-}=+0.0592\,x^2t_c^2/t_{\rm sp}$ per site.

\emph{(b) The two-boson ladder.}  For a dilute Bose gas the effective coupling
between two condensed chargons is the two-body $T$ matrix rather than the bare
$U_B$.  At second order,
\begin{align}
\delta U_{abcd}&=-\sum_{(m,n)}
\frac{U_{ab,mn}\,U_{mn,cd}}{\varepsilon_m+\varepsilon_n-2E_0}\,,\nonumber\\
U_{ab,cd}&=U_B\sum_{\bm r}\bar\phi_a\bar\phi_b\phi_c\phi_d({\bm r})\,,
\label{eq:ladder}
\end{align}
the sum running over intermediate pairs not both in the fourfold band bottom and
$E_0$ being the bottom energy.  The denominators are chargon two-particle
energies, so the scale is $U_B^2/t_B$.  The individual coefficients grow
logarithmically with system size, as a two-dimensional ladder at threshold does,
but the growth is entirely along the Hubbard ray, the one direction on which
Eq.~\eqref{eq:splitdef} vanishes, so the off-ray part converges:
\begin{equation}
E_{\rm I}-E_{\rm II^-}=-0.061,\ -0.054,\ -0.053\ \ U_B^2x^2/t_B
\end{equation}
per site at $L=12$, $24$ and $36$.

\emph{Comparison.}  The two carry different denominators and different couplings,
so comparing them requires the scale relations of Sec.~\ref{sec:experiment}.  With
$t_{\rm sp}=c_1J$, $J=4t_c^2/U$ and $t_B=2c_1|t_c|$, the ratio is
$t_{\rm sp}/t_B=2|t_c|/U$, so the single variable $u=U/|t_c|$ controls both once
the soft-chargon coupling is identified with the electronic one, $U_B=U$, as the
charging-energy estimate of Sec.~\ref{sec:phasediagram} does.  In common units of
$x^2t_c^2/t_B$,
\begin{align}
E_{\rm I}-E_{\rm II^-}&=+0.0296\,u &&\text{(spinon)}\,,\nonumber\\
&=-0.0525\,u^2 &&\text{(ladder)}\,,
\end{align}
which cross at $u=0.563$.  Section~\ref{sec:phasediagram} quotes this comparison
and discusses the physical range of $u$ and the control of each expansion.

\section{The four $\Z_2$ projective symmetry groups}
\label{app:z2psg}

All four descendants share $g_{\mathcal T}=i\tau^1$,
$g_{T_1}=g_{T_2}=\tau^0$ and $\eta_{\mathcal T}=\eta_{12}=\eta_{C_6T_1}
=-1$, and differ by the uniform $i\tau^3$ dressings of
Eq.~\eqref{eq:etatriples} [Ref.~\cite{Lu2011}, Eqs.~(C6)--(C10), with
$\eta_{\sigma C_6}$ defined through the relation
$\sigma^{-1}C_6\sigma C_6=e$ of Eq.~(A14) there].  We
constructed each state explicitly on the torus and validated the number of
symmetric pairing patterns at every range (on-site, NN, 2NN, and both
3NN classes) against Ref.~\cite{Lu2011}, Eqs.~(C16)--(C19).  All twenty
independent counts are reproduced (Table~\ref{tab:z2content}).  The
unique symmetric $\pm1$ second-neighbor pattern $\nu_{ij}$ (serving both
$\chi_2$ and $\Delta_2$) matches Eq.~(10) of Ref.~\cite{Lu2011}.  The
valley matrix of the corresponding form factor has eigenvalue exactly
$-2$ on all four valleys in the flux-canonical sign convention, giving
$\lambda_{\rm eff}=\lambda-2\Delta_2$ [Eq.~\eqref{eq:lambdaeff}].
The real-space form factors quoted in Sec.~\ref{sec:fourZ2} are, for
$\alpha$, $\mathrm{sgn}(\Delta/\chi)=+1$ on up- and $-1$ on down-triangles,
for $\gamma$ the analogous alternation on the inscribed second-neighbor
triangles, and for $\beta$, $\Delta_2\propto\nu_{ij}$.  They are fixed
uniquely by the symmetry conditions,
whose nullspace is one-dimensional in each case, and agree with those of
Ref.~\cite{Lu2011}.  The spinon-pairing channels that connect the U(1)
DSL to proximate $\Z_2$ spin liquids on the kagome lattice, and the Higgs
criticality of the resulting transitions, are analyzed in
Ref.~\cite{FeuerpfeilDepleted2026}.

\section{Chiral ansatz verification}
\label{app:csl}

The chiral ansatz is built directly on the torus from Hu's oriented
convention \cite{Hu2015}, which we make explicit here so that
Sec.~\ref{sec:csl} can be reproduced from the paper.  In
Eq.~\eqref{eq:cslfluxes}, $a$, $b$, $c$ are three consecutive vertices of
a hexagon and $d$ is the vertex opposite $b$, so that $ab$ and $bc$ are
nearest-neighbor bonds while $ac$, $cd$ and $da$ are second-neighbor
chords of the same hexagon; $\theta_\triangle$ and $\theta_{\hexagon}$
are the fluxes through a triangle and a hexagon, $\theta_{abc}$ that
through the mixed triangle (two bonds and one chord) and $\theta_{acd}$
that through the inscribed chord triangle.  Every flux is the phase of
the product of directed hopping amplitudes around the loop traversed
counterclockwise, with $t_{ji}=\bar t_{ij}$.  Each nearest-neighbor bond
belongs to a single triangle and each chord to a single hexagon, which
fixes the orientation conventions.  The bond $ij$ carries
$t_1s_{ij}e^{\pm i\varphi_1}$, with the upper sign when $i\to j$ runs
counterclockwise around its triangle, and the chord $ij$ carries
$t_2\nu_{ij}e^{\pm i\varphi_2}$, with the upper sign when $i\to j$ runs
counterclockwise around its hexagon.  Because the edges of a hexagon are
traversed clockwise with respect to their own triangles, these
conventions give $\theta_\triangle=3\varphi_1$,
$\theta_{\hexagon}=\pi-6\varphi_1$, $\theta_{acd}=3\varphi_2$ and
$\theta_{abc}=\pi-2\varphi_1-\varphi_2$, the two $\pi$'s coming from the
sign patterns.  Here $s_{ij}$ is the pattern of Fig.~\ref{fig:ansatz}
and $\nu_{ij}=\pm1$ is the unique symmetric second-neighbor pattern of
Appendix~\ref{app:z2psg}, defined on the twelve chord classes of the
magnetic cell (six chords in each of its two hexagons).  The symmetry
nullspace over all second-neighbor bonds is one-dimensional and fixes
$\nu_{ij}$ only up to a global sign.  Attaching the oriented phase
$\varphi_2$ makes that sign physical, and the mixed-triangle flux pins
it.  At the real reference point ($\varphi_1=\varphi_2=0$) the canonical
state has $abc$ loop product $-1$ and $acd$ product $+1$.  In closed
form the sign-fixing condition reads $\nu_{ij}=-s_{ik}s_{kj}$, with $k$
the common nearest neighbor of $i$ and $j$.  Every mixed triangle then
has real loop product $-1$.  The three chords of a chord triangle have
the six edges of its hexagon as their nearest-neighbor legs, and the
hexagon flux of the DSL is $\pi$, so every chord triangle has product
$-(-1)=+1$.  The reference point is therefore $[0,\pi;\pi,0]$.  Our
implementation
reproduces all four fluxes of
Eq.~\eqref{eq:cslfluxes} at general angles on every triangle, hexagon,
mixed triangle and chord triangle of the torus.  We establish each
residual symmetry of the chiral state by exhibiting the accompanying
U(1) gauge transformation explicitly, built as in
Appendix~\ref{app:conventions}, which confirms the class
$(\tau_\sigma,\tau_R)=(1,0)$.
For the diagonal bond, the constraint relating $u_{ij}$ to $u_{ji}$ is
antilinear.  Realifying it over $(\mathrm{Re}\,\xi_d,\mathrm{Im}\,\xi_d)$
leaves a one-dimensional solution with phase locked to $\beta_d=\pi/2$ in
the chiral state, and no solution in the $\mathcal{T}$-symmetric DSL.

The sector data of Sec.~\ref{sec:cslsectors} come from Brillouin-zone
scans of both chargon sectors on $N_g\times N_g$ grids, with $N_g$ a
multiple of 12 so that the grids contain every sector minimum exactly,
followed by Nelder--Mead refinement of the continuum band bottoms.  In
these scans the $B_\pm$ sectors are exactly
degenerate at the NN-only point.  There the $B_-$ hopping, minus the
complex conjugate of the $B_+$ hopping, carries the same fluxes
$[\pi/2,0]$, and the two sectors are U(1) gauge copies (the gauge
transformation is constructed as above).  The crossing of
Eq.~\eqref{eq:crossing} occurs at Hu's phases and is reproduced to ten
digits from $N_g=12$ to $192$ and by the continuum refinement.  Beyond
the crossing the $B_+$ minimum is two-valley.  The estimate
$r_2^*\simeq0.334$ is obtained by inverting Hu's quoted mean-field gap,
converted from their $\mathrm{Re}\,t_1=1$ normalization to units of
$|t_1|$, with our exact bands.  Grid-resolved readings of the same
quantity give larger values, and every reading exceeds $r_{2,c}$ at the
$(J_2,J_3)=(0.5,0.6)$ phases of Hu \emph{et al.}  Those scans hold the
diagonal amplitude at zero.  With the diagonal amplitude switched on in
the realified pattern above, the sector margin at $r_2^*$ is linear in
$r_d$ and changes sign at $r_d=+0.0388$, which the self-consistent value
estimated in Sec.~\ref{sec:cslsectors} exceeds.  The margin quoted there
at $r_d=0$ is $0.155\,t_1$ at $r_2^*=0.334$ and $0.174\,t_1$ at
$r_2=0.35$.

\emph{Induced action on the doublet.}  Table~\ref{tab:inpairchiral}
gives the induced action of the projective generators on the condensate
Bloch sphere of the two-valley theory of Sec.~\ref{sec:csllandau}, with
axes referred to the doublet basis of that section.  This basis differs
from the four-valley frame of Appendix~\ref{app:theorem}, so the axes in
Tables~\ref{tab:inpair} and \ref{tab:inpairchiral} do not correspond
element by element.  $T_1$, $T_2$ and $R=C_6$ generate the
tetrahedral rotation group $\mathrm{T}$ of order 12, with angle multiset
$\{0\!:\!1,\,120\!:\!8,\,180\!:\!3\}$.  Since $C_2=R^3$ is a pure phase
on the doublet, $R$ itself is the threefold generator there.  In the
$\mathcal{T}$-symmetric theory, by contrast, $C_6$ exchanges the pairs
and only $C_3=C_6^2$ is in-pair.  The antiunitary $\sigma\Theta$ acts as the
reflection $(x,y,z)\mapsto(-z,y,-x)$ in the plane perpendicular to
$(1,0,1)$, and the full magnetic in-pair group is $\mathrm{T}_d$ (order
24: twelve rotations, six reflections and six $S_4$ rotoreflections,
containing no inversion).  The character sums are
$n_\ell=(1,0,0,1,1,0,2)$ under $\mathrm{T}$ and $(1,0,0,1,1,0,1)$ under
$\mathrm{T}_d$ for $\ell=0,\dots,6$.  The $\ell=3$ harmonic $xyz$
therefore survives and is the unique sextic anisotropy of
Sec.~\ref{sec:csllandau}.  The table was computed at the representative
point $r_2=0.35$ of Sec.~\ref{sec:cslsectors}.  The group structure does
not depend on the point within the two-valley regime.
\begin{table}[t]
\caption{Induced action of the projective symmetry group of the chiral
two-valley theory on the condensate Bloch sphere of the $B_+$ doublet
(Sec.~\ref{sec:csllandau}), in the doublet basis of that section: axis
(for the reflection $\sigma\Theta$, the normal of the mirror plane),
angle in degrees and determinant of the induced $O(3)$ element.  Here
$C_2=R^3$ is a pure phase on the doublet.  The unitary elements form
the tetrahedral rotation group $\mathrm{T}$ (order 12, angle multiset
$\{0\!:\!1,\,120\!:\!8,\,180\!:\!3\}$) and the full magnetic group is
$\mathrm{T}_d$ (order 24).  The character sums for $\ell=0,\dots,6$ are
$n_\ell=(1,0,0,1,1,0,2)$ under $\mathrm{T}$ and $(1,0,0,1,1,0,1)$ under
$\mathrm{T}_d$.}
\label{tab:inpairchiral}
\begin{ruledtabular}
\begin{tabular}{lllcc}
element & type & axis & angle & det \\
\colrule
$T_1$ & unitary & $\hat y$ & 180 & $+1$ \\
$T_2$ & unitary & $\hat z$ & 180 & $+1$ \\
$R=C_6$ & unitary & $(1,-1,-1)/\sqrt3$ & 120 & $+1$ \\
$C_3=R^2$ & unitary & $(-1,1,1)/\sqrt3$ & 120 & $+1$ \\
$C_2=R^3$ & unitary & identity & 0 & $+1$ \\
$\sigma\Theta$ & antiunitary & $(1,0,1)/\sqrt2$ (normal) & & $-1$ \\
\end{tabular}
\end{ruledtabular}
\end{table}

\section{Composite tables}
\label{app:composites}

The sixteen particle--hole bilinears $\bar B_\eta B_{\eta'}$ decompose,
with all projective offsets included, into $\Gamma\,(\times2)$
[$n_{\rm tot}$ and the valley polarization $\Delta N$], the full $M$ star
($\times2$ per member, one $\mathcal{T}$-even and one $\mathcal{T}$-odd
Hermitian combination), the full $K/2$ star ($\times6$), and $\pm K$.  The
ten symmetric pairs decompose into $\Gamma\,(\times1)$
[$(B_1B_3+B_2B_4)/\sqrt2$], the $M$ star ($\times3$), and the $K/2$ star
($\times6$).  The extended-point-group analysis uses the explicitly
constructed 48-element group $C_{6v}'''$ (space group modulo $2\times2$
translations, with the affine parts referred to the hexagon center), with class
structure and character table as in Refs.~\cite{Venderbos2016,RMP2026} and
in Appendix~C of Ref.~\cite{Hermele2008}.
We \emph{calibrate} the $\sigma_v$-versus-$\sigma_d$ labeling instead of
assuming it.  Interchanging the two mirror families interchanges $F_3$ with
$F_4$ (and $B_1$ with $B_2$) and fixes the remaining irreps, so the
twelve-site site-permutation representation,
$A_1\oplus E_2\oplus F_1\oplus F_3\oplus F_4$ under either labeling, does
not discriminate between them and serves only as a check of the class
structure (no $F_2$).  The calibration rests on the twelve-plaquette flux
(pseudoscalar) representation, whose $M$ content must be of $F_2$ and
$F_4$ type only \cite{RMP2026}.  This holds for only one of the two
labelings ($2F_2\oplus F_4$, against $2F_2\oplus F_3$ for the other).  It
fixes $\sigma_v$ as the mirror family perpendicular to the lattice
vectors and yields the assignments of Tables~\ref{tab:ph} and
\ref{tab:pairing}.  In the same calibration the $\Gamma$-sector staggered
triangle flux is $B_2$, confirming the $B_2'$ assignment of $\Delta N$.
As for microscopic activation, on a $T_1$-pair condensate
$(B_1,B_2,0,0)$ the three $M$ amplitudes are proportional to the three
components of the Bloch vector $\hat n$.  The condensate therefore
activates one, two or three arms of the $M$ star according to how many
components of $\hat n$ are nonzero.  The one-arm states are the
six coordinate-axis points of the pair sphere only (one orbit of
$\mathrm{O}_h$, containing the single-valley states), the body diagonal
$\hat n=(1,1,1)/\sqrt3$ is the triple-$M$ state, and the whole sphere is
the moduli space of Sec.~\ref{sec:moduli}.  On a one-arm state the
$F_3$ (charge/bond) and $F_2'$ (current) composites have equal magnitude
$|\Delta|=\tfrac12$.  Within the balanced four-valley manifold, distinct
phase lockings place the $M$ weight entirely in the $F_3$ or entirely in
the $F_2'$ channel.  Every invariant cubic vanishes identically
on all condensates, consistent with the absence of invariant cubics in the
$(F_3\oplus F_2')$ space.

\bibliography{references}

\end{document}